\documentclass[reprint,superscriptaddress,preprintnumbers,longbibliography,
amsmath,amssymb,aps,superscriptaddress,floatfix,tightenlines,pra
]{revtex4-2}
\usepackage{subfigure}
\usepackage{graphicx}% Include figure files
\usepackage{dcolumn}% Align table columns on decimal point
\usepackage{bm}% bold math
\usepackage{color}
\usepackage[colorlinks=true,
citecolor=green,
linkcolor=red,
anchorcolor=black,
urlcolor=blue]{hyperref}
\usepackage{overpic}
\usepackage{braket}

\renewcommand{\figurename}{Fig.}
\begin{document}
	
\title{Simulating Chern insulators on a superconducting quantum processor}

\author{Zhong-Cheng Xiang}
\thanks{These authors contributed equally to this work.}
\affiliation{Institute of Physics, Chinese Academy of Sciences, Beijing 100190, China}

\author{Kaixuan Huang}
\thanks{These authors contributed equally to this work.}
\affiliation{Beijing Academy of Quantum Information Sciences, Beijing 100193, China}
\affiliation{Institute of Physics, Chinese Academy of Sciences, Beijing 100190, China}
\affiliation{Key Laboratory of Weak Light Nonlinear Photonics, Ministry of Education, Teda Applied Physics Institute and School of Physics, Nankai University, Tianjin 300457, China}

\author{Yu-Ran Zhang}
\thanks{These authors contributed equally to this work.}
\affiliation{School of Physics and Optoelectronics, South China University of Technology, Guangzhou 510640, China}
\affiliation{Theoretical Quantum Physics Laboratory, Cluster for Pioneering Research, RIKEN, Wako-shi, Saitama 351-0198, Japan}
\affiliation{Center for Quantum Computing, RIKEN, Wako-shi, Saitama 351-0198, Japan}

\author{Tao Liu}
\affiliation{School of Physics and Optoelectronics, South China University of Technology, Guangzhou 510640, China}

\author{Yun-Hao Shi}
\affiliation{Institute of Physics, Chinese Academy of Sciences, Beijing 100190, China}

\author{Cheng-Lin Deng}
\affiliation{Institute of Physics, Chinese Academy of Sciences, Beijing 100190, China}

\author{Tong Liu}
\affiliation{Institute of Physics, Chinese Academy of Sciences, Beijing 100190, China}

\author{Hao Li}
\affiliation{Institute of Physics, Chinese Academy of Sciences, Beijing 100190, China}

\author{Gui-Han Liang}
\affiliation{Institute of Physics, Chinese Academy of Sciences, Beijing 100190, China}

\author{Zheng-Yang Mei}
\affiliation{Institute of Physics, Chinese Academy of Sciences, Beijing 100190, China}

\author{Haifeng Yu}
\affiliation{Beijing Academy of Quantum Information Sciences, Beijing 100193, China}

\author{Guangming Xue}
\affiliation{Beijing Academy of Quantum Information Sciences, Beijing 100193, China}

\author{Ye Tian}
\affiliation{Institute of Physics, Chinese Academy of Sciences, Beijing 100190, China}

\author{Xiaohui Song}
\affiliation{Institute of Physics, Chinese Academy of Sciences, Beijing 100190, China}

\author{Zhi-Bo Liu}
\affiliation{Key Laboratory of Weak Light Nonlinear Photonics, Ministry of Education, Teda Applied Physics Institute and School of Physics, Nankai University, Tianjin 300457, China}

\author{Kai Xu}
\email{kaixu@iphy.ac.cn}
\affiliation{Institute of Physics, Chinese Academy of Sciences, Beijing 100190, China}
\affiliation{Beijing Academy of Quantum Information Sciences, Beijing 100193, China}
\affiliation{CAS Centre for Excellence in Topological Quantum Computation,
	UCAS, Beijing 100190, China}
\affiliation{Songshan Lake Materials Laboratory, Dongguan 523808, China}

\author{Dongning Zheng}
\affiliation{Institute of Physics, Chinese Academy of Sciences, Beijing 100190, China}
\affiliation{CAS Centre for Excellence in Topological Quantum Computation,
	UCAS, Beijing 100190, China}
\affiliation{Songshan Lake Materials Laboratory, Dongguan 523808, China}

\author{Franco Nori}
\email{fnori@riken.jp}
\affiliation{Theoretical Quantum Physics Laboratory, Cluster for Pioneering Research, RIKEN, Wako-shi,
	Saitama 351-0198, Japan}
\affiliation{Center for Quantum Computing, RIKEN, Wako-shi, Saitama 351-0198, Japan}
\affiliation{Physics Department, University of Michigan, Ann Arbor, Michigan 48109-1040, USA}

\author{Heng Fan}
\email{hfan@iphy.ac.cn}
\affiliation{Institute of Physics, Chinese Academy of Sciences, Beijing 100190, China}
\affiliation{Beijing Academy of Quantum Information Sciences, Beijing 100193, China}
\affiliation{CAS Centre for Excellence in Topological Quantum Computation, UCAS, Beijing 100190, China}
\affiliation{Songshan Lake Materials Laboratory, Dongguan 523808, China}

\begin{abstract}
	\noindent The quantum Hall effect, fundamental in modern condensed matter physics, continuously inspires 
	new theories and predicts emergent phases of matter. Here we experimentally demonstrate three types of Chern 
	insulators with synthetic dimensions on a programable 30-qubit-ladder superconducting processor. We directly measure 
	the band structures of the 2D Chern insulator along synthetic dimensions with various configurations of 
	Aubry-André-Harper chains and observe dynamical localisation of edge excitations. With these two signatures of topology, 
	our experiments implement the bulk-edge correspondence in the synthetic 2D Chern insulator. Moreover, we simulate two 
	different bilayer Chern insulators on the ladder-type superconducting processor. With the same and opposite periodically 
	modulated on-site potentials for two coupled chains, we simulate topologically nontrivial edge states with zero Hall 
	conductivity and a Chern insulator with higher Chern numbers, respectively. Our work shows the potential of using 
	superconducting qubits for investigating different intriguing topological phases of quantum matter.
\end{abstract}

\maketitle
\noindent\textbf{\Large{}Introduction}{\Large\par}
\noindent Topological phases of matter \cite{Hasan2010,Qi2011}, classified beyond  Landau's symmetry-breaking theory, have been attracting growing interest in recent decades. It started with the
discovery of the two-dimensional (2D) integer quantum Hall effect (QHE) \cite{Klitzing1980},
which arises from the topological nature of Bloch bands characterised by
the Chern number \cite{Thouless1982}. Topological band theory provides a
direct link between theory and experiments, which are successful for
identifying salient characteristics of topological states and predicting new classes of
topological phases \cite{Bansil2016}. The existence of robust edge states
is deeply related to the topology of gapped bulk band structures,
which is the ubiquitous bulk-edge correspondence in topological systems.

Since exploring higher-dimensional physics requires an exponentially growing number of qubits,
there is growing interest in creating synthetic dimensions to construct a higher-dimensional
lattice in a lower-dimensional system with its internal degrees of freedom
\cite{Yuan2018,Ozawa2019,Dutt2020,Leefmans2022}. Moreover, a lower-dimensional topological charge pump
shares the same topological origin as higher-dimensional topological physics, e.g., a
one-dimensional (1D) Thouless pump for the 2D integer QHE \cite{Thouless1983} and a 2D topological
pump for a four-dimensional quantum Hall system \cite{Lohse2018,Zilberberg2018}.

In addition to the 2D electron gas, the integer QHE and Chern insulators (i.e., a lattice version of the QHE)
have also been observed in other physical platforms,
including ultra-cold atoms in optical lattices \cite{Jotzu2014,Aidelsburger2015,Chalopin2020},
photonic systems \cite{Wang2009,Hafezi2013,Khanikaev2017},
etc. \cite{Ningyuan2015,Klembt2018}. However,
it still remains very challenging to synthesise topological quantum phases and to further
demonstrate the bulk-edge correspondence in quantum simulation platforms
\cite{Georgescu2014,Satzinger2021,Semeghini2021}.

Here we observe  several topological signatures of
2D  and bilayer
Chern insulators
with synthetic dimensions on a programmable 30-qubit-ladder superconducting processor.
We experimentally measure the band structure of the 2D Chern insulator along a synthetic dimension by analysing the temporal frequency of the system's
response to local perturbations. By monitoring quantum walks (QWs) of a single
excitation initially prepared at an edge qubit, we demonstrate
dynamical localisation of the topologically protected edge states.
The measured band structure and the dynamical signatures of topological edge states
together demonstrate the {bulk-edge correspondence}.
Furthermore, we synthesise two different bilayer Chern insulators
on the  30-qubit-ladder processor.
Given the same periodically modulated on-site potentials of two coupled chains, we probe the nontrivial topological edge states
with {zero} Hall conductivity.
When on-site potentials of two coupled chains have opposite signs,
a Chern insulator with {higher} Chern numbers is probed.
Our results show that superconducting simulation platforms are capable for
studying  different intriguing topological phases of quantum matter.

\vspace{.5cm}
\noindent\textbf{\Large{}Results}{\Large\par}
\noindent\textbf{\large{}Experimental setup and model Hamiltonians}{\large\par}
\noindent Our experiments are performed on a superconducting circuit \cite{Gu2017} consisting of 30 transmon qubits
(Q$_{j,s}$, with $j$ varied from 1 to 15 and pseudo-spin $s\in\{\uparrow,\downarrow\}$) \cite{You2007,Koch2007},
which constitute a two-legged qubit ladder \cite{Elbio1996}, see Fig.~\ref{fig:1}\textbf{a}.
By setting $\hbar=1$, the system's
Hamiltonian can be written as \cite{Roushan2017,Ye2019,Yan2019}
\begin{eqnarray}
	H&=&J_{\parallel}\sum_{j,s}(\hat{c}_{j,s}^\dag\hat{c}_{j+1,s}+\textrm{H.c.})+J_{\perp}\sum_{j}(\hat{c}_{j,\uparrow}^\dag\hat{c}_{j,\downarrow}+\textrm{H.c.})\nonumber\\
	&&+\sum_{j,s}V_{j,s}\hat{c}_{j,s}^\dag\hat{c}_{j,s},\label{eq1}
\end{eqnarray}
where  $\hat{c}^{\dag}$  ($\hat{c}$) is the hard-core bosonic creation (annihilation) operator
with $(\hat{c}^{\dag})^2=\hat{c}^{2}=0$,  and $[\hat{c}^{\dag}_{i,s},\hat{c}_{j,r}]=\delta_{ij}\delta_{sr}$.
Here, $J_{\parallel}/2\pi\simeq 8$~MHz and $J_{\perp}/2\pi\simeq 7$~MHz denote
the nearest-neighbour (NN) hopping between nearby qubits on the same leg and on the same rung,
respectively, and $V_{j,s}$ is the tunable on-site potential.
Experimental details of our system are described in the Supplementary Note 1 and Supplementary Note 2.

With a dimensional reduction procedure \cite{Kraus2012a},
a 2D integer quantum Hall system can be mapped to a 1D model with a periodic parameter
as a synthetic dimension. In this context, we
experimentally simulate a 2D integer QHE associated with Chern insulator using 15 qubits on one leg of the ladder
(i.e., qubits labelled by Q$_{1,\uparrow}$ to Q$_{15,\uparrow}$),
where the on-site potential $V_{j,\uparrow}$ of each qubit is periodically modulated.
This system can be described by the 1D Aubry-Andr\'{e}-Harper (AAH) model \cite{Harper1955,aubry1980}
with a tight-binding Hamiltonian
\begin{equation}
	H_\textrm{AAH}=J_{\parallel}\sum_{j=1}^{14}(\hat{c}_{j}^\dag
	\hat{c}_{j+1}+\textrm{H.c.})+\Delta\sum_{j=1}^{15}\cos(2\pi b j+\phi)\hat{c}_{j}^\dag
	\hat{c}_{j},
\end{equation}
where the second index $\uparrow$ is omitted. Here, $b$ determines the modulation periodicity, and
the modulation phase $\phi$ corresponds to the momentum in a synthetic dimension  \cite{Kraus2012a}, see Fig.~\ref{fig:1}\textbf{b}.
Note that  there exists
hopping between next-nearest neighbour (NNN) qubits $H'=J'_\parallel\sum_{j=1}^{13}(\hat{c}_{j}^\dag\hat{c}_{j+2}+\textrm{H.c.})$  on the same leg with a strength $J'_\parallel\simeq0.1J_\parallel$, see the Supplementary Note 1.
This model is topologically equivalent to the lattice model of the 2D integer QHE, proposed by Hofstadter \cite{Hofstadter1976},
where electrons hop within the 2D lattice, subjected to a perpendicular magnetic field with $b$ being the magnetic flux (normalised to the magnetic flux quantum) threading each unit cell. In our experiments, we fix $b=\frac{1}{3}$, and
vary $\phi$ from $0$ to $2\pi$ to obtain various instances of AAH models with potentials
by
tuning the qubits' frequencies
as $\omega_j=\omega_0+\Delta\sum_{j=1}^{15}\cos(2\pi b j+\phi)$, with a reference frequency $\omega_0/2\pi=4.7$~GHz (Fig.~\ref{fig:1}\textbf{c}).

\vspace{.5cm}
\noindent\textbf{\large{}Band structure spectroscopy}{\large\par}
\noindent Band structures play an essential role in characterising topological phases of matter
and discovering novel classes of intriguing topological materials \cite{Bansil2016}.
Here, we directly measure the topological band structure of the integer quantum Hall system
along the synthetic dimension using a dynamic spectroscopic technique applied in refs.~\cite{Senko2014,Jurcevic2015,Roushan2017}.
This method detects quantised eigenenergies of the quantum system from the Fourier transformation (FT)
of the subsequent response of the system given local
perturbations. The experimental sequence of pulses of the band structure spectroscopy
are shown in Fig.~\ref{fig:1}\textbf{d}. With 15 qubits initialised at their idle points, we place
one target qubit Q$_j$ in the superposed state $|+_j\rangle=(|0_j\rangle+|1_j\rangle)/\sqrt{2}$,
using a Y$_{\frac{\pi}{2}}$ pulse. Then, all qubits are detuned to their corresponding
frequencies for the quench dynamics with a time $t$ before the readout of the
Q$_j$ at its idle point in the $\hat{\sigma}^x$ and $\hat{\sigma}^y$ bases. For each
$\phi$, time evolutions of $\langle\hat{\sigma}_j^x\rangle$ and
$\langle\hat{\sigma}_j^x\rangle$ are recorded when choosing  a target qubit, e.g., Q$_8$ in Fig.~\ref{fig:2}\textbf{a}.
Figure~\ref{fig:2}\textbf{b} shows the squared FT magnitude $|A_j|^2$ of the response function
$\chi_j(t)\equiv\langle\hat{\sigma}_j^x(t)\rangle+i\langle\hat{\sigma}_j^y(t)\rangle$
for each qubit with $\phi=\frac{2\pi}{3}$ and $\Delta/2\pi =12$~MHz.
With the summation of the squared FT magnitudes of all selected qubits $I_\phi\equiv\sum_{j}|A_j|^2$,
the positions of its peaks clearly indicate the eigenenergies $E/2\pi$ of the system for
each $\phi$ (Fig.~\ref{fig:2}\textbf{c}).

\vspace{.5cm}
\noindent\textbf{\large{}Simulating 2D Chern insulators with a synthetic dimension}{\large\par}
\noindent The topological nature of the 1D AAH model and the 2D integer QHE associated with Chern insulator can be identified from its
band structure. When setting $\Delta/2\pi=12$~MHz,  we plot in Fig.~\ref{fig:2}\textbf{d} the band
structure of the 1D AAH model along %the synthetic dimension of
$\phi$ with $N=15$ and open boundary conditions along $x$-direction.
As $\phi$ evolves, the gapless edge states (red curves) appear in two gaps between
three ``bulk band'' regimes
(dense blue curves). We experimentally map out this band structure by measuring $I_\phi$ for $\phi\in[0,2\pi]$
(Fig.~\ref{fig:2}\textbf{e}),  which also agrees well with the numerical result by simulating the system dynamics (Fig.~\ref{fig:2}\textbf{f}).
Two gapless Dirac bands within two band gaps are clearly observed in Fig.~\ref{fig:2}\textbf{e}, and
each gap is related to a quantised Hall conductance $\sigma =
({e^2}/{h})\mathcal{C}$,
with the integer  $\mathcal{C}$ being determined by the Chern number \cite{Thouless1982}.

Moreover, the topological edge states predicted in the band structure can be verified in real space
by observing localisation of an edge excitation during its QWs on the 15-qubit chain \cite{Cai2019}.
After the system initialisation, we excite one qubit with a X$_\pi$ pulse,
tune all qubits
to their corresponding frequencies and measure them at a time $t$ after their
free evolutions (Fig.~\ref{fig:1}\textbf{e}). %for experimental waveform sequences.
For the topologically
trivial case with $\Delta/2\pi=0$~MHz,  the measured density distributions $P_j(t)$
of single-excitation QWs initialised at any position of the qubit chain
show a light-cone-like propagation
of information with  boundary reflections \cite{Yan2019}.

When we set $b=\frac{1}{3}$ and $\Delta/2\pi=12$~MHz, the $P_j(t)$ for the excitation initialised at a boundary qubit
(Q$_1$ or Q$_{15}$) exhibits localisation for $\phi=\frac{2\pi}{3}$, where the edge states appear in
the middle of a band gap (Fig.~\ref{fig:3}\textbf{a1},~\textbf{a3}).
This localised dynamical behaviour is due to the fact that the edge-excitation modes
have a main overlap with the in-gap edge states which are topologically protected by the band gaps.
This is also verified by the squared FT magnitudes $|A_1|^2$ and $|A_{15}|^2$  for edge qubits
(Fig.~\ref{fig:3}\textbf{b1}, \textbf{b3}), which mainly contain information of the in-gap edge states.
The discontinuity of the FT signals for edge qubits results from
the second and first band gaps, respectively, because the edge states localised at two boundaries have opposite
propagation directions, like in the standard QHE. When a qubit away from either end, e.g. Q$_8$,
is excited, we observe the propagation of the excitation in Fig.~\ref{fig:3}\textbf{a2} due to the absence of
topological protection, and its FT signal $|A_8|^2$ merely shows partial information of the bulk
band (Fig.~\ref{fig:3}\textbf{b2}).
Thus, by directly measuring the band structure  and observing dynamical localisation
of edge excitations, our experiments demonstrate \textit{the bulk-edge correspondence
	in the synthetic 2D Chern insulator}.

In addition, a topological charge pump, entailing the charge transport in a 1D time-varying
potential driven in adiabatic cycles \cite{Thouless1983,Kraus2012}, provides an alternative way to explore the 2D integer QHE.
The charge transported in a pumping cycle is determined
by the Chern number \cite{Niu1984}, which is defined over a 2D Brillouin zone with one spatial and one
temporal dimensions. We experimentally simulate the charge pump by adiabatically varying $\phi$
in a $2\pi$ period starting from $\phi_0=\frac{5\pi}{3}$ with $\Delta/2\pi =36$~MHz.
Figure~\ref{fig:3}\textbf{c1}, \textbf{c3} plot the evolutions of distributions  $P_j(t)$ of an excitation
initialised at the central qubit Q$_8$ for the forward and backward pumping schemes, respectively.
In one pumping cycle, the excitation propagates through an integer number of unit cells, in this  case three qubits,  which
is determined by the Chern number.

Following Laughlin's argument \cite{Laughlin1981,Fabre2022}, the role of
the threading magnetic flux is played by the adiabatic variation of $\phi$, leading to the excitation's transport.
The  displacements of the centre of mass (CoM) $\delta_x$ are shown in  Fig.~\ref{fig:3}\textbf{d},
of which the slight deviations
from the Chern numbers $\pm1$ may result from the boundary reflection on the finite-size 1D qubit chain and the small fraction of excitations for the higher bands.
{A brief discussion of  fast pumping \cite{Fedorova2020} is given in the Supplementary Note 3.}
In comparison, there exists no excitation transport when $\phi$ is not pumped  (Fig.~\ref{fig:3}\textbf{c2},~\textbf{d}).

\vspace{.5cm}
\noindent\textbf{\large{}Simulating bilayer Chern insulators}{\large\par}
\noindent Next, using all thirty qubits, we simulate two different bilayer Chern insulators.
The Hamiltonian of the 30-qubit-ladder quantum processor is given in equation~(\ref{eq1}),
where the on-site potentials on the qubits in two legs ($s\in\{\uparrow,\downarrow\}$) are modulated as
\begin{equation}
	V_{j,s}=\Delta_s\cos(2\pi b j  +\phi).
\end{equation}
with $b=\frac{1}{3}$.
For instances of two coupled AAH chains by varying $\phi$ with the same $\Delta_{\uparrow(\downarrow)}/2\pi=12$~MHz
and opposite  $\Delta_\uparrow/2\pi=-\Delta_\downarrow/2\pi = 12$~MHz, we effectively simulate two different bilayer quantum
systems~\cite{Rozhkov2016}, where the magnetic fluxes threading two layers are the same and have a difference of $\pi$, respectively.
Using the above spectroscopic technique, we
measure the band structures for these two cases (Figs.~\ref{fig:4}\textbf{a},~\textbf{b}),
which agree well with the theoretical predictions (dashed curves) using the experimental parameters in the Supplementary Note 1.

For $\Delta_{\uparrow(\downarrow)}/2\pi=12$~MHz with  $J_\perp$ being
comparable to $J_\parallel$, gapless edge states are experimentally observed
(Fig.~\ref{fig:4}\textbf{a}), even when the Chern number for half filling  is predicted to be zero (see Methods).
This novel topological phase results from the existence of a pair of counter-propagating chiral edge states in the second gap (Fig.~\ref{fig:4}\textbf{c}),
which are protected by the {inversion symmetry} and lead to a {zero} Hall conductivity $\sigma=0$ \cite{Lau2015}.
When we excite a corner
qubit (e.g., Q$_{1,\uparrow}$) and monitor the excitation's QWs for $\phi=\frac{2\pi}{3}$, the time-evolved distribution
$P_{j,s}$ shows
oscillations between two corner states localised at the same boundary rung of the qubit
ladder (Fig.~\ref{fig:4}\textbf{f}).
This dynamical behaviour can be further understood from the measured
squared FT magnitudes $|A_{1,\uparrow}|^2$ and $|A_{1,\downarrow}|^2$ with respect to the corner
qubits Q$_{1,\uparrow}$ and Q$_{1,\downarrow}$, respectively, which share information of the same
edge states (Fig.~\ref{fig:4}\textbf{c}).

By modulating opposite on-site potential fields with
$\Delta_\uparrow/2\pi=-\Delta_\downarrow/2\pi = 12$~MHz,
we obtain a different band structure (Fig.~\ref{fig:4}\textbf{b}) from the one
with two layers having the same magnetic flux.
In this case, the magnetic fluxes threading two layers have a difference of $\pi$, and
the gapless edge states are characterised by higher Chern numbers  (see Methods).
The QWs of the excitation initialised at a corner qubit
(e.g., Q$_{1,\uparrow}$ or Q$_{1,\downarrow}$) for $\phi=\frac{2\pi}{3}$ show dynamical localisation at the initial
qubit (Fig.~\ref{fig:4}\textbf{g}), and the measured squared FT magnitudes of two corner qubits show
information of different edge states in different gaps
(Fig.~\ref{fig:4}\textbf{d}).
Thus, this synthetic bilayer quantum system with two layers having a $\pi$-flux difference
presents a Chern insulator with chiral edge states, identified by {higher} Chern numbers.
Since high-Chern-number insulators without the formation of Landau levels
have attracted increasing attentions \cite{Ge2020,Chen2020,Zhao2020}, our work provides a new perspective for exploring these emergent topological phases. In comparison, for $\Delta_{\uparrow(\downarrow)}/2\pi=0$, the QWs of the corner
excitation first show linear propagation and then indicate thermalisation on the qubit ladder as a non-integrable system \cite{Ye2019,Zhu2022}.

\vspace{.5cm}
\noindent\textbf{\Large{}Discussion}{\Large\par}
\noindent In summary, we simulate 2D and bilayer Chern insulators
with synthetic dimensions on a programmable
30-qubit-ladder superconducting processor.
By measuring the band structures and monitoring dynamical localisation of edge
excitations, we implement the bulk-edge correspondence in the 2D Chern insulator.
In addition, we synthesise two different bilayer Chern insulators  with two layers
having the same magnetic flux and a $\pi$-flux difference, respectively, and simulate distinct
and novel topological phases. Our experiments, using a relatively large number of superconducting
qubits with long coherence times and accurate readouts, show
the future potential of using superconducting simulating platforms for investigating
intriguing topological quantum phases and quantum many-body physics.
For instance, by upgrading the ladder-type superconducting processor with tunable interactions between 
two qubit chains, we could observe the quantum phase transitions between 
topological phases with zero Chern number (Fig.~\ref{fig1chain}b)  and 
with higher Chern numbers (Fig.~\ref{fig1chain}c). In addition, when  
tuning weak couplings between two chains and modulating $b$ either positive or negative for two chains, 
respectively, the quantum spin Hall effect 
could be demonstrated by simulating bilayer quantum Hall systems with opposite Chern numbers 
\cite{Kane2005}. 
%In addition, quantum many-body physics could  
%be experimentally studied on the ladder-type superconducting processor with more controllability, e.g., 
% Bose-Mott insulators \cite{Zhu2013} and hole pairing in ladder systems \cite{Hirthe2023}.}

\renewcommand{\figurename}{Figure}
\setcounter{figure}{0}  % reset counter

\vspace{.5cm}
\noindent\textbf{\Large{}Methods}{\Large\par}

\noindent\textbf{\large{}System Hamiltonian}{\large\par}

\noindent Our superconducting quantum processor can be described as a Bose-Hubbard ladder
with a Hamiltonian ($\hbar=1$) \cite{Roushan2017,Ye2019}
\begin{eqnarray}
	H_{\textrm{BH}}&=&J_{\parallel}\sum_{j,s}(\hat{a}_{j,s}^\dag\hat{a}_{j+1,s}+\textrm{H.c.})
	+\sum_{j,s}\frac{\eta_{j,s}}{2}\hat{n}_{j,s}(\hat{n}_{j,s}-1)\nonumber\\
	&& +J_{\perp}\sum_{j}(\hat{a}_{j,\uparrow}^\dag\hat{a}_{j,\downarrow}+\textrm{H.c.})
	+\sum_{j,s}V_{j,s}\hat{n}_{j,s},\label{BHladder}
\end{eqnarray}
where $\hat{a}^{\dag}$ ($\hat{a}$) is the bosonic creation
(annihilation) operator, and $\hat{n}\equiv \hat{a}^{\dag}\hat{a}$ is the number operator.
Here, $J_{\parallel}/2\pi\simeq 8$~MHz and $J_{\perp}/2\pi\simeq 7$~MHz denote
the nearest-neighbour (NN) hopping between nearby qubits on the same leg and on the same rung,
respectively. Also, $\eta$ is the on-site nonlinear interaction,
and $V_{j,s}$ is the tunable on-site potential.

Our device is designed to fulfil the hard-core limit $|\eta/J|\gg1$, and thus,
the highly occupied states of transmon qubits are blockaded, which represents fermionisation of strongly
interacting bosons \cite{Yan2019}. The system Hamiltonian can then be simplified as
\begin{eqnarray}
	H&=&J_{\parallel}\sum_{j,s}(\hat{c}_{j,s}^\dag\hat{c}_{j+1,s}+\textrm{H.c.})
	+\sum_{j,s}V_{j,s}\hat{c}_{j,s}^\dag\hat{c}_{j,s}\nonumber\\
	&&+J_{\perp}\sum_{j}(\hat{c}_{j,\uparrow}^\dag\hat{c}_{j,\downarrow}+\textrm{H.c.}),
\end{eqnarray}
where $\hat{c}^{\dag}$  ($\hat{c}$) is the hard-core bosonic creation (annihilation) operator
with $(\hat{c}^{\dag})^2=\hat{c}^{2}=0$,
and $[\hat{c}^{\dag}_{j,s},\hat{c}_{i,r}]=\delta_{ji}\delta_{sr}$.

Note that in addition to the hopping between nearest-neighbour (NN) qubits, there also exist
the hopping between next-nearest neighbour (NNN) qubits on  different legs:
\begin{equation}
	J_\times\sum_j(\hat{c}^\dag_{j,\uparrow}\hat{c}_{j+1,\downarrow}+\hat{c}^\dag_{j,\downarrow}\hat{c}_{j+1,\uparrow}+\textrm{H.c.}),
\end{equation}
and the hopping between third-nearest-neighbour (TNN) qubits on the same leg:
\begin{equation}
	J'_\parallel\sum_{j,s}(\hat{c}^\dag_{j,s}\hat{c}_{j+2,s}+\textrm{H.c.}).
\end{equation}

\noindent\textbf{\large{}Mapping the 2D quantum Hall model to various instances of Aubry-Andr\'{e}-Harper chains}{\large\par}
\noindent An electron moving in a 2D lattice with a perpendicular magnetic field $b$
is described by the integer quantum Hall model,
\begin{eqnarray}
	H_{\textrm{IQH}}&=&t_x\sum_{j=1}^{N-1}\sum_{m=1}^{M}(\hat{c}_{j,m}^\dag
	\hat{c}_{j+1,m}+\textrm{H.c.})\nonumber\\
	&&+t_y\sum_{j=1}^{N}\sum_{m=1}^{M}(\hat{c}_{j,m}^\dag
	\hat{c}_{j,m+1}e^{i2\pi b  j}+\textrm{H.c.}),
\end{eqnarray}
where $t_x$ and $t_y$ are hopping strengths along the $x$ and $y$-axes,
respectively.
We consider a periodic boundary condition in the $y$-direction and introduce the
Fourier transformation (FT):
\begin{equation}
	\hat{c}_{j,k_y}=\sum_{m=1}^{M}e^{-ik_ym}\hat{c}_{j,m},\hspace{0.2 in}  \hat{c}_{j,k_y}^\dag=\sum_{m=1}^{M}e^{ik_ym}\hat{c}_{j,m}^\dag.
\end{equation}
Then, the Hamiltonian transforms to $H_{\textrm{IQH}}=\sum_{k_y}H_{k_y}$, with $k_y$ being the quasi-momentum in the $y$-direction,
%\begin{align}
%  H&=\sum_{qq'}\sum_{m=1}^{M}\left[t_x\sum_n^{N-1}(\hat{a}_{n,q}^\dag
%   \hat{a}_{n+1,q'}e^{-i(q-q')m}+\textrm{H.c.})+t_y\sum_n^{N}(\hat{a}_{n,q}^\dag
%    \hat{a}_{n,q'}e^{i2\pi b n}e^{-i(q-q')m}e^{iq'}+\textrm{H.c.})\right]\\
%    &=\sum_{q}\left[t_x\sum_n^{N-1}(\hat{a}_{n,q}^\dag
%     \hat{a}_{n+1,q}+\textrm{H.c.})+2t_y\sum_n^{N}cos(2\pi b n+q)\hat{a}_{n,q}^\dag
%      \hat{a}_{n,q}\right]\\
%  &=\sum_qH_q,
%\end{align}
where
\begin{eqnarray}
	H_{k_y}&=&t_x\sum_j^{N-1}(\hat{c}_{j,k_y}^\dag
	\hat{c}_{j+1,k_y}+\textrm{H.c.})\nonumber\\
	&&+2t_y\sum_n^{N}\cos(2\pi b j+k_y)\hat{c}_{j,k_y}^\dag
	\hat{c}_{j,k_y}.
\end{eqnarray}
By replacing $t_x$, $2t_y$, and $q$ with  $J$, $\Delta$, and $\phi$, we obtain
the 1D Aubry-Andr\'{e}-Harper (AAH) model
\begin{equation}
	H_\textrm{AAH}=J\sum_j^{N-1}(\hat{c}_{j}^\dag
	\hat{c}_{j+1}+\textrm{H.c.})+\Delta\sum_j^{N}\cos(2\pi b j+\phi)\hat{c}_{j}^\dag
	\hat{c}_{j},\label{AAH}
\end{equation}
where the second index has been omitted in the 15-qubit experiment.

Since we simulated  a 1D tight-binding  fermionic Hamiltonian (\ref{AAH})  in our experiments,
to investigate the behaviour of one excitation (a hard-core boson) effectively
captures the topological property of the system.
Furthermore, in the case when we only excite one qubit, the interaction term in equation~(\ref{BHladder})
is zero, and the behaviour of the quantum system we simulated is not affected
by the statistics of the particle (fermionic or bosonic).
Since our device is designed to fulfil the hard-core limit $|\eta/J|\simeq25\gg1$,
the highly occupied states of transmon qubits are blockaded, which represents fermionisation of strongly
interacting bosons \cite{Yan2019}.
Thus, the dynamical behaviour of a multiple-excitation system in the 15-qubit-chain experiment
is similar as a 1D fermionic system, as demonstrated in ref.~\cite{Yan2019}.

\vspace{.5cm}
\noindent\textbf{\large{}Band structure spectroscopy and effect of decoherence}{\large\par}

\noindent After initialising the selected qubits at their idle points, we prepare
one target qubit Q$_{j,s}$ in the superposed state
$|+_{j,s}\rangle=(|0_{j,s}\rangle+|1_{j,s}\rangle)/\sqrt{2}$,
using a Y$_{\frac{\pi}{2}}$ pulse. Then, all qubits are tuned to their corresponding
frequencies for the quench dynamics, and at a time $t$, we
measure the Q$_{j,s}$ at its idle point in the $\hat{\sigma}^x$ and $\hat{\sigma}^y$ bases.
For each $\phi$, time evolutions of $\langle\hat{\sigma}_{j,s}^x(t)\rangle$ and
$\langle\hat{\sigma}_{j,s}^y(t)\rangle$ are monitored. Then,
we calculate the squared Fourier transformation (FT) magnitude $|A_{j,s}|^2$ of the response function \cite{Roushan2017}
\begin{equation}
	\chi_{j,s}(t)\equiv\langle\hat{\sigma}_{j,s}^x(t)\rangle+i\langle\hat{\sigma}_{j,s}^y(t)\rangle.
\end{equation}
With the summation of the squared FT magnitudes of all selected qubits $I_\phi\equiv\sum_{j,s}|A_{j,s}|^2$,
the positions of its peaks clearly indicate the eigenenergies $\{E_n\}$ of the system for
each $\phi$.

Given $|\Psi_{j,s}(0)\rangle=\sum_n c_{n}^{({j,s})}|\psi_n\rangle$, with $H|\psi_n\rangle=E_n|\psi_n\rangle$
and $c_{n}^{({j,s})}\equiv \langle\psi_n|\Psi_{j,s}(0)\rangle$, and the evolution of the state
can be obtained by solving the Schr\"{o}dinger equation as
\begin{equation}
	|\Psi_{j,s}(t)\rangle=e^{-iHt}|\Psi_{j,s}(0)\rangle=\sum_n c_{n}^{({j,s})}e^{-itE_n}|\psi_n\rangle.
\end{equation}
We obtain the response function of the system after the initial perturbation as
\begin{equation}
	\chi_{j,s}(t)\sim2\langle\Psi_{j,s}(0)|\Psi_{j,s}(t)\rangle-1=2\sum_n |c_{n}^{({j,s})}|^2 e^{-itE_n}-1,
\end{equation}
and the FT of the response function is calculated as
\begin{eqnarray}
	\tilde{\chi}_{j,s}(\omega)&\sim&\frac{1}{2\pi}\int\!d\omega\;\left[2\sum_n |c_{n}^{({j,s})}|^2 e^{-it(E_n-\omega)}-e^{i\omega t}\right]\nonumber\\
	&=&2\sum_n |c_{n}^{({j,s})}|^2\delta(\omega-E_n)-\delta(\omega),
\end{eqnarray}
where the eigenenergies $\{E_n\}$ are indicated by the peaks of the FT signals.

Furthermore, when considering decoherence with a decaying rate $\gamma$ in a form
\begin{eqnarray}
	\chi_{j,s}(t)&\sim&2\langle\Psi_{j,s}(0)|\Psi_{j,s}(t)\rangle-1\nonumber\\
	&=&\left[2\sum_n |c_{n}^{({j,s})}|^2\delta(\omega-E_n)-\delta(\omega)\right]e^{-\gamma t},
\end{eqnarray}
the FT of the response function is calculated as
\begin{eqnarray}
	\tilde{\chi}_{j,s}(\omega)&\sim&\frac{1}{2\pi}\int\!d\omega\;\sum_n |c_{n}^{({j,s})}|^2 e^{-it(E_n-\omega)}\nonumber\\
	&=&\sum_n |c_{n}^{({j,s})}|^2\frac{4\gamma}{\gamma^2+(\omega-E_n)^2}-\frac{2\gamma}{\gamma^2+\omega^2},
\end{eqnarray}
which indicates that the presence of decoherence increases the width of the peaks of the FT signals,
and the locations of the peaks can still indicate the values of the eigenenergies.

There also exists the unwanted zero frequency signal of the FT.
In our experiments, to eliminate the unwanted zero frequency signal of the FT
we calculate the FT of oscillations of the response function
$\delta\chi(t)\equiv\chi(t)-\overline{\chi(t)}$, i.e.,
the response function minus the its average over the time interval.

\vspace{.5cm}
\noindent\textbf{\large{}Avoiding rung-pair excitations in strongly interacting Bose-Hubbard ladders}{\large\par}
\noindent In the 30-qubit-ladder experiment, we excited either one corner qubit or a bulk qubit
and monitored the quantum walks of the excitation, to study the topologically protected
edge states of the bilayer quantum Hall systems. Localisation of the excitation
initialised at a corner qubit and the propagation of the excitation initialised
at the bulk qubit indicate the existence of a topological edge state protected by
the topology of the bulk structure.

However, we do not consider to excite two corner qubits at the same edge simultaneously, because it has been experimentally and theoretically shown in refs.~\cite{Ye2019,Li2020} that the dynamics of single- and double-excitation states have very distinct behaviours. Specifically, in the hard-core limit, there exists rung-pair localisation at the edges
even for the topologically trivial case without the modulation of the on-site potentials, $\Delta_{\uparrow}/2\pi=\Delta_{\downarrow}/2\pi=12$~MHz.
In the centre-of-mass frame, the two-particle system can be mapped into an
effective single-particle Hamiltonian, and there exists a zero-energy
flat band in the hard-core limit, which is the origin of the localisation \cite{Li2020}. Therefore, in the 30-qubit-ladder experiment, we avoid exciting two qubits on the same rung simultaneously, when investigating the topologically-protected edge states.

%\subsection{Discussions of the jump structure of the edge band structure}
\vspace{.5cm}
\noindent\textbf{\large{}Quantum charge pumping}{\large\par}
\noindent In addition to the study of 2D topological systems, topological charge pumping provides an alternative way to explore the quantised transport with topological protection in a dynamical 1D system. The concept of topological charge pumping was first proposed by D. J. Thouless \cite{Thouless1983}, and recognised a topological quantisation of charge
transport in a 1D time-varying potential driven in adiabatic cycles. The charge transported
in a pumping cycle is determined by the Chern number \cite{Thouless1982},
which is defined over a 2D Brillouin zone with one spatial dimension and one temporal dimension.

{We experimentally simulate the quantum charge pump by adiabatically varying $\phi$ in a $2\pi$
	period starting from $\phi_0=\frac{5\pi}{3}$ with $\Delta/2\pi= 36$~MHz.
	After initialising 15 qubits in the state $|0\rangle^{\otimes 15}$ at their idle frequencies, we prepared the
	central qubit, Q$_{8, \uparrow}$, in the $|1\rangle$ state.
	Then, we set the frequencies of qubits as
	$\omega_{j}(\phi) = \omega_0 + \Delta\sum_{j=1}^{15}\cos(2 \pi bj + \phi)$, with
	$\Delta/2\pi = 36$~MHz and an initial phase $\phi_0 = \frac{5\pi}{3}$.
	In this case, the single-excitation initial state has the minimum energy, and
	with a high value of $\Delta$, the initial excitation stabilises at Q$_{8, \uparrow}$.
	
	The  frequencies for all qubits are calibrated using the frequency calibration procedure
	as shown in the Supplementary Information.
	Then, by modulating the frequencies of all 15 qubits simultaneously,
	we slowly vary $\phi$ from $\phi_0$ to $\phi_0\pm2\pi$ in {1,100}~ns with a relatively slow speed
	$\sim{1.8}\pi/\mu$s for the backward and forward  pumping schemes, respectively.
	For the case of no pump, $\phi$ is fixed at $\phi_0$ during the time evolution.
	
	In our experiments, we measured the $|1\rangle$-state occupation probability of each qubit at its
	idle point in the $\hat{\sigma}^z$ basis for  each evolution time $t$. To reduce the effect of the
	stochastic fluctuations, we maintained a fixed sample of {4,000} single-shot readouts and repeated
	the measurement procedure 10 times %(16 times for the no pump case) 
	for estimating the mean values
	and standard deviations at each evolution time $t$. We obtained the average displacements of the centre of
	mass (CoM) as
	\begin{equation}
		\delta x(t) = \sum_{j=1}^{15}{P_{j,\uparrow}(t) (j-8)}/{3},
	\end{equation}
	where $P_j(t)$ is the $|1\rangle$-state occupation probability of the Q$_{j,\uparrow}$.
	Note that the CoM is divided by 3, because we set $b=\frac{1}{3}$
	and there are 3 qubits in one unit cell.
	The experimental results can be found in Fig.~3 in the main text.}

\vspace{.5cm}
\noindent\textbf{\large{}Characterisation of bilayer Chern insulators}{\large\par}
\noindent The Hamiltonian of the 30-qubit ladder reads
\begin{eqnarray}\label{realspace}
	H&=&\sum_{j,s}(J_{\parallel}\hat{c}_{j,s}^\dag\hat{c}_{j+1,s}+J'_{\parallel}\hat{c}_{j,s}^\dag\hat{c}_{j+2,s}+\textrm{H.c.})\nonumber\\
	&& +J_{\perp}\sum_{j}(\hat{c}_{j,\uparrow}^\dag\hat{c}_{j,\downarrow}+\textrm{H.c.})\nonumber\\
	&&+J_{\times}\sum_{j}(\hat{c}_{j,\uparrow}^\dag\hat{c}_{j+1,\downarrow}+\hat{c}_{j,\downarrow}^\dag\hat{c}_{j+1,\uparrow}+\textrm{H.c.})\nonumber\\
	&&+ \sum_{j,s}\Delta_s\cos\left(2\pi b_s j+\phi \right)  \hat{c}_{j,s}^\dag\hat{c}_{j,s} ,
\end{eqnarray}
where $s\in\{\uparrow,\downarrow\}$, and $b=\frac{1}{3}$ determines the modulation
periodicity. The typical hopping strengths for our sample are $J_{\parallel}/2\pi=8$~MHz,
$J_{\perp}/2\pi=7$~MHz, $J'_{\parallel}\simeq 0.1J_{\parallel}$, and
$J_{\times}\simeq 0.2J_{\parallel}$. The AAH ladder can be mapped
into two coupled  Hofstadter lattices, subjected to the same effective magnetic fields
for each layer.

We now proceed to discuss the topological properties in two cases,
i.e.,
\begin{equation}
	\Delta_\uparrow=\Delta_\downarrow=\Delta,\textrm{ and }\Delta_\uparrow=-\Delta_\downarrow =\Delta.\nonumber
\end{equation}
They correspond to the study of two types of bilayer quantum systems with two layers having the
same magnetic flux and a $\pi$-flux difference, respectively.
In our experiments, we set $\Delta/2\pi =12$~MHz.\\

\noindent \textbf{Two qubit chains with the same periodically modulated on-site potentials.}
For $\Delta_\uparrow=\Delta_\downarrow = \Delta $, two identical AAH chains are
coupled to form an AAH ladder. In Fig.~\ref{fig1chain}, we plot the band structures for different
inter-chain hopping strengths $J_\perp$.
Figure~\ref{fig1chain} shows the energy spectra of different topological phase regimes,
showing that topological phase transitions occur as $J_\perp$ varies.

We can  identify the topologically nontrivial and trivial bands by using the Chern number of each band \cite{Thouless1982}, which is defined as
\begin{equation}\label{H1}
	\mathcal{C}_n=\frac{1}{2\pi} \int_{0}^{2\pi} \!\!\!\!\! dk \int_{0}^{2\pi}\!\!\!\!\! d \phi\; \left[\partial_k \mathcal{A}_\phi^n-\partial_\phi \mathcal{A}_k^n \right].
\end{equation}
The $n$th band Berry connection $\mathcal{A}^n$ is written as
\begin{equation}\label{H2}
	\mathcal{A}_\gamma^n = i \langle\varphi_n(k, \phi)|\partial_\gamma|\varphi_n(k, \phi)\rangle,
\end{equation}
where $|\varphi_n(k, \phi)\rangle$ is the  $n$th band's wavefunction, and $\gamma=k,\phi$. Then, the Hall conductivity reads
$\sigma=\sum_{n} \mathcal{C}_n$,
with $n$ being summed over the occupied bands.

When the inter-chain hopping strength $J_\perp$ is much smaller (Fig.~\ref{fig1chain}\textbf{a})
or much larger (Fig.~\ref{fig1chain}\textbf{c}) than the intra-chain hoping strength $J_\parallel$,
the topological boundary states are well characterised by the Chern number.
However, when $J_\perp$ is comparable to $J_\parallel$, topological edge states with zero Hall
conductivity appear (Fig.~\ref{fig1chain}\textbf{b}) for the half filling.

The zero Hall conductivity results
from the contribution of a pair of counter-propagating chiral edge states. This novel topological
phase was also studied in a dimerised Hofstadter model \cite{Lau2015} and has never been
experimentally observed before.
Figure~\ref{intensity} shows the density distribution of a mid-gap state for the half filling, where
the in-gap state occupies the end sites of both chains.

The topological edges with zero Chern number (i.e., zero Hall conductivity) is protected by the inversion symmetry. After Fourier transforming equation~(\ref{realspace}), we have the following momentum-space Hamiltonian
\begin{widetext}
	\begin{equation}\label{H2momentum}
		H(k) = \sum_\mathbf{k} \Phi^\dagger_\mathbf{k} \left(\begin{matrix}
			\Delta \cos(\frac{2 \pi}{3} + \phi) & ~~J_\parallel+J'_{\parallel} e^{-i k} &~~J_\parallel e^{-i k} + J'_{\parallel} & ~~J_\perp &   J_{\times}  & ~~J_{\times}  e^{-i k} &  \\
			J_\parallel + J'_{\parallel} e^{i k}  & ~~	\Delta \cos(\frac{4 \pi}{3} + \phi) &~~J_\parallel +J'_{\parallel} e^{-i k}& ~~ J_{\times}    & J_\perp & ~~  J_{\times}    &  \\
			J_\parallel e^{i k}+J'_{\parallel}  & ~~J_\parallel+J'_{\parallel} e^{i k} &~~	\Delta \cos(\phi) & ~~ J_{\times} e^{i k} &  J_{\times}    & ~~J_\perp &  \\
			J_\perp & ~~J_{\times}    &~~J_{\times}  e^{-i k} & ~~	\Delta \cos(\frac{2 \pi}{3} + \phi) &  J_\parallel +J'_{\parallel} e^{-i k}& ~~J_\parallel e^{-i k}+J'_{\parallel}  &  \\
			J_{\times}    & ~~J_\perp &~~ J_{\times}    & ~~J_\parallel + J'_{\parallel} e^{i k} &  	\Delta \cos(\frac{4 \pi}{3} + \phi) & ~~J_\parallel + J'_{\parallel} e^{-i k} &  \\
			J_{\times} e^{ i k} & ~~ J_{\times}    &~~J_\perp & ~~J_\parallel e^{i k} + J'_{\parallel}  &  J_\parallel + J'_{\parallel} e^{i k} & ~~	\Delta \cos(\phi) &
		\end{matrix}\right) \Phi_\mathbf{k},
	\end{equation}
\end{widetext}
where $\Phi_\mathbf{k} = (\hat{c}_{k,1,\uparrow},~\hat{c}_{k,2,\uparrow},~\hat{c}_{k,3,\uparrow},~\hat{c}_{{k},1,\downarrow},~\hat{c}_{k,2,\downarrow},~\hat{c}_{k,3,\downarrow} )^T$, with $\hat{c}_{k,l,s}$ ($l=1,2,3$ and $s\in\{\uparrow,\downarrow\}$) being the annihilation operator of a hardcore boson at momentum $k$, sublattice $l$, and pesudospin $s$.

The Hamiltonian $H(k)$ in equation~(\ref{H2momentum}) has the inversion symmetry
\begin{equation}
	\hat{\mathcal{P}} H(k) \hat{\mathcal{P}}^{-1} = H(-k),
\end{equation}
for $\phi=\frac{2\pi}{3}$ and $\phi=\frac{5\pi}{3}$, where the inversion symmetry
operator $\hat{\mathcal{P}}$ is
\begin{equation}\label{H3}
	\hat{\mathcal{P}} = \left(\begin{matrix}
		0 & ~~0 & ~~0 & ~~0 &  ~~0 & ~~1 &  \\
		0 & ~~0 & ~~0 & ~~0 &  ~~1 & ~~0 &  \\
		0 & ~~0 & ~~0 & ~~1 &  ~~0 & ~~0 &  \\
		0 & ~~0 & ~~1 & ~~0 &  ~~0 & ~~0 &  \\
		0 & ~~1 & ~~0 & ~~0 &  ~~0 & ~~0 &  \\
		1 & ~~0 & ~~0 & ~~0 &  ~~0 & ~~0 &
	\end{matrix}\right).
\end{equation}
Then, we define an integer invariant $\mathcal{N}$ to characterize this novel topological phase,
which is expressed as \cite{Lau2015}
\begin{equation}\label{invariant}
	\mathcal{N} \equiv |N_1-N_2|,
\end{equation}
where $N_1$ and $N_2$ are the number of negative parities (by applying the inversion symmetry operator to eigenstates) at the high symmetry points $k=0$ and $k=\pi$, respectively.

Figure~\ref{bloch} shows the Bloch band structures for $\phi=\frac{2\pi}{3} $ and
$\phi=\frac{5\pi}{3}$, respectively, and the topological invariant $\mathcal{N}$ at each
band gap is indicated. For $\phi=\frac{2\pi}{3} $, we find $\mathcal{N}=1$ for $\frac{1}{4}$-filling and
half-filling, indicating a pair of edge states pinned at $\phi=\frac{2\pi}{3} $.
For $\phi=\frac{5\pi}{3}$, we have $\mathcal{N}=1$ for $\frac{3}{4}$-filling and half-filling,
indicating a pair of edge states pined at $\phi=\frac{5\pi}{3}$. Therefore, two pairs of
edge states appear at half-filling, which propagate in opposite directions at each edge.\\

\noindent \textbf{Two qubit chains with opposite periodically modulated on-site potentials.}
For $\Delta_\uparrow=-\Delta_\downarrow = \Delta$, the AAH ladder shows two typical topological band structures, as shown in Fig.~\ref{fig2chain}. This bilayer structure, with two layers having a $\pi$-flux difference, shows different topological features from the bilayer structure with two layers having the same magnetic flux.  Furthermore, the in-gap states mainly occupy the end site of one of the chains, as shown in Fig.~\ref{intensity2}.
Thus, this synthetic bilayer quantum system, with two layers having a $\pi$-flux difference, presents an integer quantum Hall effect with chiral edge states, identified by higher Chern numbers.

\vspace{.5cm}
\noindent\textbf{\large{}Data availability}{\large\par}
\noindent The source data underlying all figures are available at \url{https://doi.org/10.6084/m9.figshare.23925009}.~Other data are available from the corresponding author upon request.

\vspace{.5cm}
\noindent\textbf{\large{}Code availability}{\large\par}
\noindent The codes are available upon reasonable request from the corresponding author.

\vspace{.5cm}
\noindent\textbf{Acknowledgments}
We thank Hongming Weng for stimulating discussions.
This research was supported by the NSFC (Grants No.~T2121001, No.~11934018, No.~11904393,
No.~92065114, No.~12174207, No.~12274142), the CAS Strategic Priority Research Program
(Grant
No.~XDB28000000, No.~YJKYYQ20200041), the Beijing Natural Science Foundation
(Grant No.~Z200009),
the State Key Development Program for Basic Research of China (Grant
No.~2017YFA0304300), the Key-Area Research and Development Program of Guangdong Province
China (Grant No.~2020B0303030001).
F.N. is supported in part by: Nippon Telegraph and Telephone Corporation (NTT) Research, the Japan Science and Technology Agency (JST) 
[via the Quantum Leap Flagship Program (Q-LEAP), and the Moonshot R$\&$D Grant Number JPMJMS2061], 
the Asian Office of Aerospace Research and Development (AOARD) (via Grant No. FA2386-20-1-4069),
and the Foundational Questions Institute Fund (FQXi) via Grant No. FQXi-IAF19-06.\\

\noindent\textbf{Author contributions}
Y.-R.Z., Tao Liu, K.X., F.N., and H.F. conceived the research; Y.-R.Z., Tao Liu., Z.-C.X., K.H., K.X., F.N., and
H.F. designed the experiment; Z.-C.X. designed and fabricated the device with the help of G.-H.L., Z.-Y.M., X.S. and D.Z.; K.H. and C.-L.D performed the experiment supervised by K.X., Z.-B.L. and H.F.; Tong Liu, H.L and Y.T. helped to build up the experimental setup supervised by K.X.; K.H., Y.-R.Z., and Y.-H.S. performed numerical simulations; K.H., Y.-R.Z., and Z.-C.X. analysed experimental results. Tao Liu.; Y.-R.Z., F.N., and H.F. performed theoretical explanations; G.X and H.Y. provide the Josephson Parametric Amplifiers; All authors contributed to the discussions of the results and the development of the manuscript; F.N. and H.F. supervised the whole project.\\

\noindent\textbf{Competing interests} The authors declare no competing interests.

%\bibliography{Manuscript.bib}

\begin{thebibliography}{99}%
	\makeatletter
	\providecommand \@ifxundefined [1]{%
		\@ifx{#1\undefined}
	}%
	\providecommand \@ifnum [1]{%
		\ifnum #1\expandafter \@firstoftwo
		\else \expandafter \@secondoftwo
		\fi
	}%
	\providecommand \@ifx [1]{%
		\ifx #1\expandafter \@firstoftwo
		\else \expandafter \@secondoftwo
		\fi
	}%
	\providecommand \natexlab [1]{#1}%
	\providecommand \enquote  [1]{``#1''}%
	\providecommand \bibnamefont  [1]{#1}%
	\providecommand \bibfnamefont [1]{#1}%
	\providecommand \citenamefont [1]{#1}%
	\providecommand \href@noop [0]{\@secondoftwo}%
	\providecommand \href [0]{\begingroup \@sanitize@url \@href}%
	\providecommand \@href[1]{\@@startlink{#1}\@@href}%
	\providecommand \@@href[1]{\endgroup#1\@@endlink}%
	\providecommand \@sanitize@url [0]{\catcode `\\12\catcode `\$12\catcode
		`\&12\catcode `\#12\catcode `\^12\catcode `\_12\catcode `\%12\relax}%
	\providecommand \@@startlink[1]{}%
	\providecommand \@@endlink[0]{}%
	\providecommand \url  [0]{\begingroup\@sanitize@url \@url }%
	\providecommand \@url [1]{\endgroup\@href {#1}{\urlprefix }}%
	\providecommand \urlprefix  [0]{URL }%
	\providecommand \Eprint [0]{\href }%
	\providecommand \doibase [0]{https://doi.org/}%
	\providecommand \selectlanguage [0]{\@gobble}%
	\providecommand \bibinfo  [0]{\@secondoftwo}%
	\providecommand \bibfield  [0]{\@secondoftwo}%
	\providecommand \translation [1]{[#1]}%
	\providecommand \BibitemOpen [0]{}%
	\providecommand \bibitemStop [0]{}%
	\providecommand \bibitemNoStop [0]{.\EOS\space}%
	\providecommand \EOS [0]{\spacefactor3000\relax}%
	\providecommand \BibitemShut  [1]{\csname bibitem#1\endcsname}%
	\let\auto@bib@innerbib\@empty
	%</preamble>
	\bibitem [{\citenamefont {Hasan}\ and\ \citenamefont {Kane}(2010)}]{Hasan2010}%
	\BibitemOpen
	\bibfield  {author} {\bibinfo {author} {\bibfnamefont {M.~Z.}\ \bibnamefont
			{Hasan}}\ and\ \bibinfo {author} {\bibfnamefont {C.~L.}\ \bibnamefont
			{Kane}},\ }\bibfield  {title} {\bibinfo {title} {Colloquium: {T}opological
			insulators},\ }\href {https://doi.org/10.1103/RevModPhys.82.3045} {\bibfield
		{journal} {\bibinfo  {journal} {Rev. Mod. Phys.}\ }\textbf {\bibinfo {volume}
			{82}},\ \bibinfo {pages} {3045} (\bibinfo {year} {2010})}\BibitemShut
	{NoStop}%
	 \bibitem [{\citenamefont {Qi}\ and\ \citenamefont {Zhang}(2011)}]{Qi2011}%
	 \BibitemOpen
	 \bibfield  {author} {\bibinfo {author} {\bibfnamefont {X.-L.}\ \bibnamefont
	 		{Qi}}\ and\ \bibinfo {author} {\bibfnamefont {S.-C.}\ \bibnamefont {Zhang}},\
	 }\bibfield  {title} {\bibinfo {title} {Topological insulators and
	 		superconductors},\ }\href {https://doi.org/10.1103/RevModPhys.83.1057}
	 {\bibfield  {journal} {\bibinfo  {journal} {Rev. Mod. Phys.}\ }\textbf
	 	{\bibinfo {volume} {83}},\ \bibinfo {pages} {1057} (\bibinfo {year}
	 	{2011})}\BibitemShut {NoStop}%
	 \bibitem [{\citenamefont {von Klitzing}\ \emph {et~al.}(1980)\citenamefont {von
	 		Klitzing}, \citenamefont {Dorda},\ and\ \citenamefont
	 	{Pepper}}]{Klitzing1980}%
	 \BibitemOpen
	 \bibfield  {author} {\bibinfo {author} {\bibfnamefont {K.}~\bibnamefont {von
	 			Klitzing}}, \bibinfo {author} {\bibfnamefont {G.}~\bibnamefont {Dorda}},\
	 	and\ \bibinfo {author} {\bibfnamefont {M.}~\bibnamefont {Pepper}},\
	 }\bibfield  {title} {\bibinfo {title} {New method for high-accuracy
	 		determination of the fine-structure constant based on quantized {H}all
	 		resistance},\ }\href {https://doi.org/10.1103/PhysRevLett.45.494} {\bibfield
	 	{journal} {\bibinfo  {journal} {Phys. Rev. Lett.}\ }\textbf {\bibinfo
	 		{volume} {45}},\ \bibinfo {pages} {494} (\bibinfo {year} {1980})}\BibitemShut
	 {NoStop}%
	 \bibitem [{\citenamefont {Thouless}\ \emph {et~al.}(1982)\citenamefont
	 	{Thouless}, \citenamefont {Kohmoto}, \citenamefont {Nightingale},\ and\
	 	\citenamefont {den Nijs}}]{Thouless1982}%
	 \BibitemOpen
	 \bibfield  {author} {\bibinfo {author} {\bibfnamefont {D.~J.}\ \bibnamefont
	 		{Thouless}}, \bibinfo {author} {\bibfnamefont {M.}~\bibnamefont {Kohmoto}},
	 	\bibinfo {author} {\bibfnamefont {M.~P.}\ \bibnamefont {Nightingale}},\ and\
	 	\bibinfo {author} {\bibfnamefont {M.}~\bibnamefont {den Nijs}},\ }\bibfield
	 {title} {\bibinfo {title} {Quantized {H}all conductance in a two-dimensional
	 		periodic potential},\ }\href {https://doi.org/10.1103/PhysRevLett.49.405}
	 {\bibfield  {journal} {\bibinfo  {journal} {Phys. Rev. Lett.}\ }\textbf
	 	{\bibinfo {volume} {49}},\ \bibinfo {pages} {405} (\bibinfo {year}
	 	{1982})}\BibitemShut {NoStop}%
	 \bibitem [{\citenamefont {Bansil}\ \emph {et~al.}(2016)\citenamefont {Bansil},
	 	\citenamefont {Lin},\ and\ \citenamefont {Das}}]{Bansil2016}%
	 \BibitemOpen
	 \bibfield  {author} {\bibinfo {author} {\bibfnamefont {A.}~\bibnamefont
	 		{Bansil}}, \bibinfo {author} {\bibfnamefont {H.}~\bibnamefont {Lin}},\ and\
	 	\bibinfo {author} {\bibfnamefont {T.}~\bibnamefont {Das}},\ }\bibfield
	 {title} {\bibinfo {title} {Colloquium: Topological band theory},\ }\href
	 {https://doi.org/10.1103/RevModPhys.88.021004} {\bibfield  {journal}
	 	{\bibinfo  {journal} {Rev. Mod. Phys.}\ }\textbf {\bibinfo {volume} {88}},\
	 	\bibinfo {pages} {021004} (\bibinfo {year} {2016})}\BibitemShut {NoStop}%
	 \bibitem [{\citenamefont {Yuan}\ \emph {et~al.}(2018)\citenamefont {Yuan},
	 	\citenamefont {Lin}, \citenamefont {Xiao},\ and\ \citenamefont
	 	{Fan}}]{Yuan2018}%
	 \BibitemOpen
	 \bibfield  {author} {\bibinfo {author} {\bibfnamefont {L.}~\bibnamefont
	 		{Yuan}}, \bibinfo {author} {\bibfnamefont {Q.}~\bibnamefont {Lin}}, \bibinfo
	 	{author} {\bibfnamefont {M.}~\bibnamefont {Xiao}},\ and\ \bibinfo {author}
	 	{\bibfnamefont {S.}~\bibnamefont {Fan}},\ }\bibfield  {title} {\bibinfo
	 	{title} {Synthetic dimension in photonics},\ }\href
	 {https://doi.org/10.1364/OPTICA.5.001396} {\bibfield  {journal} {\bibinfo
	 		{journal} {Optica}\ }\textbf {\bibinfo {volume} {5}},\ \bibinfo {pages}
	 	{1396} (\bibinfo {year} {2018})}\BibitemShut {NoStop}%
	 \bibitem [{\citenamefont {Ozawa}\ and\ \citenamefont
	 	{Price}(2019)}]{Ozawa2019}%
	 \BibitemOpen
	 \bibfield  {author} {\bibinfo {author} {\bibfnamefont {T.}~\bibnamefont
	 		{Ozawa}}\ and\ \bibinfo {author} {\bibfnamefont {H.~M.}\ \bibnamefont
	 		{Price}},\ }\bibfield  {title} {\bibinfo {title} {Topological quantum matter
	 		in synthetic dimensions},\ }\href {https://doi.org/10.1038/s42254-019-0045-3}
	 {\bibfield  {journal} {\bibinfo  {journal} {Nat. Rev. Phys.}\ }\textbf
	 	{\bibinfo {volume} {1}},\ \bibinfo {pages} {349} (\bibinfo {year}
	 	{2019})}\BibitemShut {NoStop}%
	 \bibitem [{\citenamefont {Dutt}\ \emph {et~al.}(2020)\citenamefont {Dutt},
	 	\citenamefont {Lin}, \citenamefont {Yuan}, \citenamefont {Minkov},
	 	\citenamefont {Xiao},\ and\ \citenamefont {Fan}}]{Dutt2020}%
	 \BibitemOpen
	 \bibfield  {author} {\bibinfo {author} {\bibfnamefont {A.}~\bibnamefont
	 		{Dutt}}, \bibinfo {author} {\bibfnamefont {Q.}~\bibnamefont {Lin}}, \bibinfo
	 	{author} {\bibfnamefont {L.}~\bibnamefont {Yuan}}, \bibinfo {author}
	 	{\bibfnamefont {M.}~\bibnamefont {Minkov}}, \bibinfo {author} {\bibfnamefont
	 		{M.}~\bibnamefont {Xiao}},\ and\ \bibinfo {author} {\bibfnamefont
	 		{S.}~\bibnamefont {Fan}},\ }\bibfield  {title} {\bibinfo {title} {A single
	 		photonic cavity with two independent physical synthetic dimensions},\ }\href
	 {https://doi.org/10.1126/science.aaz3071} {\bibfield  {journal} {\bibinfo
	 		{journal} {Science}\ }\textbf {\bibinfo {volume} {367}},\ \bibinfo {pages}
	 	{59} (\bibinfo {year} {2020})}\BibitemShut {NoStop}%
	 \bibitem [{\citenamefont {Leefmans}\ \emph {et~al.}(2022)\citenamefont
	 	{Leefmans}, \citenamefont {Dutt}, \citenamefont {Williams}, \citenamefont
	 	{Yuan}, \citenamefont {Parto}, \citenamefont {Nori}, \citenamefont {Fan},\
	 	and\ \citenamefont {Marandi}}]{Leefmans2022}%
	 \BibitemOpen
	 \bibfield  {author} {\bibinfo {author} {\bibfnamefont {C.}~\bibnamefont
	 		{Leefmans}}, \bibinfo {author} {\bibfnamefont {A.}~\bibnamefont {Dutt}},
	 	\bibinfo {author} {\bibfnamefont {J.}~\bibnamefont {Williams}}, \bibinfo
	 	{author} {\bibfnamefont {L.}~\bibnamefont {Yuan}}, \bibinfo {author}
	 	{\bibfnamefont {M.}~\bibnamefont {Parto}}, \bibinfo {author} {\bibfnamefont
	 		{F.}~\bibnamefont {Nori}}, \bibinfo {author} {\bibfnamefont {S.}~\bibnamefont
	 		{Fan}},\ and\ \bibinfo {author} {\bibfnamefont {A.}~\bibnamefont {Marandi}},\
	 }\bibfield  {title} {\bibinfo {title} {Topological dissipation in a
	 		time-multiplexed photonic resonator network},\ }\href
	 {https://doi.org/10.1038/s41567-021-01492-w} {\bibfield  {journal} {\bibinfo
	 		{journal} {Nat. Phys.}\ }\textbf {\bibinfo {volume} {18}},\ \bibinfo {pages}
	 	{442} (\bibinfo {year} {2022})}\BibitemShut {NoStop}%
	 \bibitem [{\citenamefont {Thouless}(1983)}]{Thouless1983}%
	 \BibitemOpen
	 \bibfield  {author} {\bibinfo {author} {\bibfnamefont {D.~J.}\ \bibnamefont
	 		{Thouless}},\ }\bibfield  {title} {\bibinfo {title} {Quantization of particle
	 		transport},\ }\href {https://doi.org/10.1103/PhysRevB.27.6083} {\bibfield
	 	{journal} {\bibinfo  {journal} {Phys. Rev. B}\ }\textbf {\bibinfo {volume}
	 		{27}},\ \bibinfo {pages} {6083} (\bibinfo {year} {1983})}\BibitemShut
	 {NoStop}%
	 \bibitem [{\citenamefont {Lohse}\ \emph {et~al.}(2018)\citenamefont {Lohse},
	 	\citenamefont {Schweizer}, \citenamefont {Price}, \citenamefont
	 	{Zilberberg},\ and\ \citenamefont {Bloch}}]{Lohse2018}%
	 \BibitemOpen
	 \bibfield  {author} {\bibinfo {author} {\bibfnamefont {M.}~\bibnamefont
	 		{Lohse}}, \bibinfo {author} {\bibfnamefont {C.}~\bibnamefont {Schweizer}},
	 	\bibinfo {author} {\bibfnamefont {H.~M.}\ \bibnamefont {Price}}, \bibinfo
	 	{author} {\bibfnamefont {O.}~\bibnamefont {Zilberberg}},\ and\ \bibinfo
	 	{author} {\bibfnamefont {I.}~\bibnamefont {Bloch}},\ }\bibfield  {title}
	 {\bibinfo {title} {Exploring 4{D} quantum {H}all physics with a 2{D}
	 		topological charge pump},\ }\href {https://doi.org/10.1038/nature25000}
	 {\bibfield  {journal} {\bibinfo  {journal} {Nature}\ }\textbf {\bibinfo
	 		{volume} {553}},\ \bibinfo {pages} {55} (\bibinfo {year} {2018})}\BibitemShut
	 {NoStop}%
	 \bibitem [{\citenamefont {Zilberberg}\ \emph {et~al.}(2018)\citenamefont
	 	{Zilberberg}, \citenamefont {Huang}, \citenamefont {Guglielmon},
	 	\citenamefont {Wang}, \citenamefont {Chen}, \citenamefont {Kraus},\ and\
	 	\citenamefont {Rechtsman}}]{Zilberberg2018}%
	 \BibitemOpen
	 \bibfield  {author} {\bibinfo {author} {\bibfnamefont {O.}~\bibnamefont
	 		{Zilberberg}}, \bibinfo {author} {\bibfnamefont {S.}~\bibnamefont {Huang}},
	 	\bibinfo {author} {\bibfnamefont {J.}~\bibnamefont {Guglielmon}}, \bibinfo
	 	{author} {\bibfnamefont {M.}~\bibnamefont {Wang}}, \bibinfo {author}
	 	{\bibfnamefont {K.~P.}\ \bibnamefont {Chen}}, \bibinfo {author}
	 	{\bibfnamefont {Y.~E.}\ \bibnamefont {Kraus}},\ and\ \bibinfo {author}
	 	{\bibfnamefont {M.~C.}\ \bibnamefont {Rechtsman}},\ }\bibfield  {title}
	 {\bibinfo {title} {Photonic topological boundary pumping as a probe of 4{D}
	 		quantum {H}all physics},\ }\href {https://doi.org/10.1038/nature25011}
	 {\bibfield  {journal} {\bibinfo  {journal} {Nature}\ }\textbf {\bibinfo
	 		{volume} {553}},\ \bibinfo {pages} {59} (\bibinfo {year} {2018})}\BibitemShut
	 {NoStop}%
	 \bibitem [{\citenamefont {Jotzu}\ \emph {et~al.}(2014)\citenamefont {Jotzu},
	 	\citenamefont {Messer}, \citenamefont {Desbuquois}, \citenamefont {Lebrat},
	 	\citenamefont {Uehlinger}, \citenamefont {Greif},\ and\ \citenamefont
	 	{Esslinger}}]{Jotzu2014}%
	 \BibitemOpen
	 \bibfield  {author} {\bibinfo {author} {\bibfnamefont {G.}~\bibnamefont
	 		{Jotzu}}, \bibinfo {author} {\bibfnamefont {M.}~\bibnamefont {Messer}},
	 	\bibinfo {author} {\bibfnamefont {R.}~\bibnamefont {Desbuquois}}, \bibinfo
	 	{author} {\bibfnamefont {M.}~\bibnamefont {Lebrat}}, \bibinfo {author}
	 	{\bibfnamefont {T.}~\bibnamefont {Uehlinger}}, \bibinfo {author}
	 	{\bibfnamefont {D.}~\bibnamefont {Greif}},\ and\ \bibinfo {author}
	 	{\bibfnamefont {T.}~\bibnamefont {Esslinger}},\ }\bibfield  {title} {\bibinfo
	 	{title} {Experimental realization of the topological {H}aldane model with
	 		ultracold fermions},\ }\href {https://doi.org/10.1038/nature13915} {\bibfield
	 	{journal} {\bibinfo  {journal} {Nature}\ }\textbf {\bibinfo {volume}
	 		{515}},\ \bibinfo {pages} {237} (\bibinfo {year} {2014})}\BibitemShut
	 {NoStop}%
	 \bibitem [{\citenamefont {Aidelsburger}\ \emph {et~al.}(2015)\citenamefont
	 	{Aidelsburger}, \citenamefont {Lohse}, \citenamefont {Schweizer},
	 	\citenamefont {Atala}, \citenamefont {Barreiro}, \citenamefont
	 	{Nascimb{\`e}ne}, \citenamefont {Cooper}, \citenamefont {Bloch},\ and\
	 	\citenamefont {Goldman}}]{Aidelsburger2015}%
	 \BibitemOpen
	 \bibfield  {author} {\bibinfo {author} {\bibfnamefont {M.}~\bibnamefont
	 		{Aidelsburger}}, \bibinfo {author} {\bibfnamefont {M.}~\bibnamefont {Lohse}},
	 	\bibinfo {author} {\bibfnamefont {C.}~\bibnamefont {Schweizer}}, \bibinfo
	 	{author} {\bibfnamefont {M.}~\bibnamefont {Atala}}, \bibinfo {author}
	 	{\bibfnamefont {J.~T.}\ \bibnamefont {Barreiro}}, \bibinfo {author}
	 	{\bibfnamefont {S.}~\bibnamefont {Nascimb{\`e}ne}}, \bibinfo {author}
	 	{\bibfnamefont {N.~R.}\ \bibnamefont {Cooper}}, \bibinfo {author}
	 	{\bibfnamefont {I.}~\bibnamefont {Bloch}},\ and\ \bibinfo {author}
	 	{\bibfnamefont {N.}~\bibnamefont {Goldman}},\ }\bibfield  {title} {\bibinfo
	 	{title} {Measuring the {C}hern number of {H}ofstadter bands with ultracold
	 		bosonic atoms},\ }\href {https://doi.org/10.1038/nphys3171} {\bibfield
	 	{journal} {\bibinfo  {journal} {Nat. Phys.}\ }\textbf {\bibinfo {volume}
	 		{11}},\ \bibinfo {pages} {162} (\bibinfo {year} {2015})}\BibitemShut
	 {NoStop}%
	 \bibitem [{\citenamefont {Chalopin}\ \emph {et~al.}(2020)\citenamefont
	 	{Chalopin}, \citenamefont {Satoor}, \citenamefont {Evrard}, \citenamefont
	 	{Makhalov}, \citenamefont {Dalibard}, \citenamefont {Lopes},\ and\
	 	\citenamefont {Nascimbene}}]{Chalopin2020}%
	 \BibitemOpen
	 \bibfield  {author} {\bibinfo {author} {\bibfnamefont {T.}~\bibnamefont
	 		{Chalopin}}, \bibinfo {author} {\bibfnamefont {T.}~\bibnamefont {Satoor}},
	 	\bibinfo {author} {\bibfnamefont {A.}~\bibnamefont {Evrard}}, \bibinfo
	 	{author} {\bibfnamefont {V.}~\bibnamefont {Makhalov}}, \bibinfo {author}
	 	{\bibfnamefont {J.}~\bibnamefont {Dalibard}}, \bibinfo {author}
	 	{\bibfnamefont {R.}~\bibnamefont {Lopes}},\ and\ \bibinfo {author}
	 	{\bibfnamefont {S.}~\bibnamefont {Nascimbene}},\ }\bibfield  {title}
	 {\bibinfo {title} {Probing chiral edge dynamics and bulk topology of a
	 		synthetic {H}all system},\ }\href {https://doi.org/10.1038/s41567-020-0942-5}
	 {\bibfield  {journal} {\bibinfo  {journal} {Nat. Phys.}\ }\textbf {\bibinfo
	 		{volume} {16}},\ \bibinfo {pages} {1017} (\bibinfo {year}
	 	{2020})}\BibitemShut {NoStop}%
	 \bibitem [{\citenamefont {Wang}\ \emph {et~al.}(2009)\citenamefont {Wang},
	 	\citenamefont {Chong}, \citenamefont {Joannopoulos},\ and\ \citenamefont
	 	{Solja{\v c}i{\'c}}}]{Wang2009}%
	 \BibitemOpen
	 \bibfield  {author} {\bibinfo {author} {\bibfnamefont {Z.}~\bibnamefont
	 		{Wang}}, \bibinfo {author} {\bibfnamefont {Y.}~\bibnamefont {Chong}},
	 	\bibinfo {author} {\bibfnamefont {J.~D.}\ \bibnamefont {Joannopoulos}},\ and\
	 	\bibinfo {author} {\bibfnamefont {M.}~\bibnamefont {Solja{\v c}i{\'c}}},\
	 }\bibfield  {title} {\bibinfo {title} {Observation of unidirectional
	 		backscattering-immune topological electromagnetic states},\ }\href
	 {https://doi.org/10.1038/nature08293} {\bibfield  {journal} {\bibinfo
	 		{journal} {Nature}\ }\textbf {\bibinfo {volume} {461}},\ \bibinfo {pages}
	 	{772} (\bibinfo {year} {2009})}\BibitemShut {NoStop}%
	 \bibitem [{\citenamefont {Hafezi}\ \emph {et~al.}(2013)\citenamefont {Hafezi},
	 	\citenamefont {Mittal}, \citenamefont {Fan}, \citenamefont {Migdall},\ and\
	 	\citenamefont {Taylor}}]{Hafezi2013}%
	 \BibitemOpen
	 \bibfield  {author} {\bibinfo {author} {\bibfnamefont {M.}~\bibnamefont
	 		{Hafezi}}, \bibinfo {author} {\bibfnamefont {S.}~\bibnamefont {Mittal}},
	 	\bibinfo {author} {\bibfnamefont {J.}~\bibnamefont {Fan}}, \bibinfo {author}
	 	{\bibfnamefont {A.}~\bibnamefont {Migdall}},\ and\ \bibinfo {author}
	 	{\bibfnamefont {J.~M.}\ \bibnamefont {Taylor}},\ }\bibfield  {title}
	 {\bibinfo {title} {Imaging topological edge states in silicon photonics},\
	 }\href {https://doi.org/10.1038/nphoton.2013.274} {\bibfield  {journal}
	 	{\bibinfo  {journal} {Nat. Photon.}\ }\textbf {\bibinfo {volume} {7}},\
	 	\bibinfo {pages} {1001} (\bibinfo {year} {2013})}\BibitemShut {NoStop}%
	 \bibitem [{\citenamefont {Khanikaev}\ and\ \citenamefont
	 	{Shvets}(2017)}]{Khanikaev2017}%
	 \BibitemOpen
	 \bibfield  {author} {\bibinfo {author} {\bibfnamefont {A.~B.}\ \bibnamefont
	 		{Khanikaev}}\ and\ \bibinfo {author} {\bibfnamefont {G.}~\bibnamefont
	 		{Shvets}},\ }\bibfield  {title} {\bibinfo {title} {Two-dimensional
	 		topological photonics},\ }\href {https://doi.org/10.1038/s41566-017-0048-5}
	 {\bibfield  {journal} {\bibinfo  {journal} {Nat. Photon.}\ }\textbf {\bibinfo
	 		{volume} {11}},\ \bibinfo {pages} {763} (\bibinfo {year} {2017})}\BibitemShut
	 {NoStop}%
	 \bibitem [{\citenamefont {Ningyuan}\ \emph {et~al.}(2015)\citenamefont
	 	{Ningyuan}, \citenamefont {Owens}, \citenamefont {Sommer}, \citenamefont
	 	{Schuster},\ and\ \citenamefont {Simon}}]{Ningyuan2015}%
	 \BibitemOpen
	 \bibfield  {author} {\bibinfo {author} {\bibfnamefont {J.}~\bibnamefont
	 		{Ningyuan}}, \bibinfo {author} {\bibfnamefont {C.}~\bibnamefont {Owens}},
	 	\bibinfo {author} {\bibfnamefont {A.}~\bibnamefont {Sommer}}, \bibinfo
	 	{author} {\bibfnamefont {D.}~\bibnamefont {Schuster}},\ and\ \bibinfo
	 	{author} {\bibfnamefont {J.}~\bibnamefont {Simon}},\ }\bibfield  {title}
	 {\bibinfo {title} {Time- and site-resolved dynamics in a topological
	 		circuit},\ }\href {https://doi.org/10.1103/PhysRevX.5.021031} {\bibfield
	 	{journal} {\bibinfo  {journal} {Phys. Rev. X}\ }\textbf {\bibinfo {volume}
	 		{5}},\ \bibinfo {pages} {021031} (\bibinfo {year} {2015})}\BibitemShut
	 {NoStop}%
	 \bibitem [{\citenamefont {Klembt}\ \emph {et~al.}(2018)\citenamefont {Klembt},
	 	\citenamefont {Harder}, \citenamefont {Egorov}, \citenamefont {Winkler},
	 	\citenamefont {Ge}, \citenamefont {Bandres}, \citenamefont {Emmerling},
	 	\citenamefont {Worschech}, \citenamefont {Liew}, \citenamefont {Segev},
	 	\citenamefont {Schneider},\ and\ \citenamefont {H{\"o}fling}}]{Klembt2018}%
	 \BibitemOpen
	 \bibfield  {author} {\bibinfo {author} {\bibfnamefont {S.}~\bibnamefont
	 		{Klembt}}, \bibinfo {author} {\bibfnamefont {T.~H.}\ \bibnamefont {Harder}},
	 	\bibinfo {author} {\bibfnamefont {O.~A.}\ \bibnamefont {Egorov}}, \bibinfo
	 	{author} {\bibfnamefont {K.}~\bibnamefont {Winkler}}, \bibinfo {author}
	 	{\bibfnamefont {R.}~\bibnamefont {Ge}}, \bibinfo {author} {\bibfnamefont
	 		{M.~A.}\ \bibnamefont {Bandres}}, \bibinfo {author} {\bibfnamefont
	 		{M.}~\bibnamefont {Emmerling}}, \bibinfo {author} {\bibfnamefont
	 		{L.}~\bibnamefont {Worschech}}, \bibinfo {author} {\bibfnamefont {T.~C.~H.}\
	 		\bibnamefont {Liew}}, \bibinfo {author} {\bibfnamefont {M.}~\bibnamefont
	 		{Segev}}, \bibinfo {author} {\bibfnamefont {C.}~\bibnamefont {Schneider}},\
	 	and\ \bibinfo {author} {\bibfnamefont {S.}~\bibnamefont {H{\"o}fling}},\
	 }\bibfield  {title} {\bibinfo {title} {Exciton-polariton topological
	 		insulator},\ }\href {https://doi.org/10.1038/s41586-018-0601-5} {\bibfield
	 	{journal} {\bibinfo  {journal} {Nature}\ }\textbf {\bibinfo {volume} {562}},\
	 	\bibinfo {pages} {552} (\bibinfo {year} {2018})}\BibitemShut {NoStop}%
	 \bibitem [{\citenamefont {Georgescu}\ \emph {et~al.}(2014)\citenamefont
	 	{Georgescu}, \citenamefont {Ashhab},\ and\ \citenamefont
	 	{Nori}}]{Georgescu2014}%
	 \BibitemOpen
	 \bibfield  {author} {\bibinfo {author} {\bibfnamefont {I.~M.}\ \bibnamefont
	 		{Georgescu}}, \bibinfo {author} {\bibfnamefont {S.}~\bibnamefont {Ashhab}},\
	 	and\ \bibinfo {author} {\bibfnamefont {F.}~\bibnamefont {Nori}},\ }\bibfield
	 {title} {\bibinfo {title} {Quantum simulation},\ }\href
	 {https://doi.org/10.1103/RevModPhys.86.153} {\bibfield  {journal} {\bibinfo
	 		{journal} {Rev. Mod. Phys.}\ }\textbf {\bibinfo {volume} {86}},\ \bibinfo
	 	{pages} {153} (\bibinfo {year} {2014})}\BibitemShut {NoStop}%
	 \bibitem [{\citenamefont {Satzinger}\ \emph {et~al.}(2021)\citenamefont
	 	{Satzinger}, \citenamefont {Liu}, \citenamefont {Smith}, \citenamefont
	 	{Knapp}, \citenamefont {Newman}, \citenamefont {Jones}, \citenamefont {Chen},
	 	\citenamefont {Quintana}, \citenamefont {Mi}, \citenamefont {Dunsworth},
	 	\citenamefont {Gidney}, \citenamefont {Aleiner}, \citenamefont {Arute},
	 	\citenamefont {Arya}, \citenamefont {Atalaya}, \citenamefont {Babbush},
	 	\citenamefont {Bardin}, \citenamefont {Barends}, \citenamefont {Basso},
	 	\citenamefont {Bengtsson}, \citenamefont {Bilmes}, \citenamefont {Broughton},
	 	\citenamefont {Buckley}, \citenamefont {Buell}, \citenamefont {Burkett},
	 	\citenamefont {Bushnell}, \citenamefont {Chiaro}, \citenamefont {Collins},
	 	\citenamefont {Courtney}, \citenamefont {Demura}, \citenamefont {Derk},
	 	\citenamefont {Eppens}, \citenamefont {Erickson}, \citenamefont {Faoro},
	 	\citenamefont {Farhi}, \citenamefont {Fowler}, \citenamefont {Foxen},
	 	\citenamefont {Giustina}, \citenamefont {Greene}, \citenamefont {Gross},
	 	\citenamefont {Harrigan}, \citenamefont {Harrington}, \citenamefont {Hilton},
	 	\citenamefont {Hong}, \citenamefont {Huang}, \citenamefont {Huggins},
	 	\citenamefont {Ioffe}, \citenamefont {Isakov}, \citenamefont {Jeffrey},
	 	\citenamefont {Jiang}, \citenamefont {Kafri}, \citenamefont {Kechedzhi},
	 	\citenamefont {Khattar}, \citenamefont {Kim}, \citenamefont {Klimov},
	 	\citenamefont {Korotkov}, \citenamefont {Kostritsa}, \citenamefont
	 	{Landhuis}, \citenamefont {Laptev}, \citenamefont {Locharla}, \citenamefont
	 	{Lucero}, \citenamefont {Martin}, \citenamefont {R.}, \citenamefont {McEwen},
	 	\citenamefont {Miao}, \citenamefont {Mohseni}, \citenamefont {Montazeri},
	 	\citenamefont {Mruczkiewicz}, \citenamefont {Mutus}, \citenamefont {Naaman},
	 	\citenamefont {Neeley}, \citenamefont {Neill}, \citenamefont {Niu},
	 	\citenamefont {O'Brien}, \citenamefont {Opremcak}, \citenamefont {Pat{\'o}},
	 	\citenamefont {Petukhov}, \citenamefont {Rubin}, \citenamefont {Sank},
	 	\citenamefont {Shvarts}, \citenamefont {Strain}, \citenamefont {Szalay},
	 	\citenamefont {Villalonga}, \citenamefont {White}, \citenamefont {Yao},
	 	\citenamefont {Yeh}, \citenamefont {Yoo}, \citenamefont {Zalcman},
	 	\citenamefont {Neven}, \citenamefont {Boixo}, \citenamefont {Megrant},
	 	\citenamefont {Chen}, \citenamefont {Kelly}, \citenamefont {Smelyanskiy},
	 	\citenamefont {Kitaev}, \citenamefont {Knap}, \citenamefont {Pollmann},\ and\
	 	\citenamefont {Roushan}}]{Satzinger2021}%
	 \BibitemOpen
	 \bibfield  {author} {\bibinfo {author} {\bibfnamefont {K.~J.}\ \bibnamefont
	 		{Satzinger}}, \bibinfo {author} {\bibfnamefont {Y.-J.}\ \bibnamefont {Liu}},
	 	\bibinfo {author} {\bibfnamefont {A.}~\bibnamefont {Smith}}, \bibinfo
	 	{author} {\bibfnamefont {C.}~\bibnamefont {Knapp}}, \bibinfo {author}
	 	{\bibfnamefont {M.}~\bibnamefont {Newman}}, \bibinfo {author} {\bibfnamefont
	 		{C.}~\bibnamefont {Jones}}, \bibinfo {author} {\bibfnamefont
	 		{Z.}~\bibnamefont {Chen}}, \bibinfo {author} {\bibfnamefont {C.}~\bibnamefont
	 		{Quintana}}, \bibinfo {author} {\bibfnamefont {X.}~\bibnamefont {Mi}},
	 	\bibinfo {author} {\bibfnamefont {A.}~\bibnamefont {Dunsworth}}, \bibinfo
	 	{author} {\bibfnamefont {C.}~\bibnamefont {Gidney}}, \bibinfo {author}
	 	{\bibfnamefont {I.}~\bibnamefont {Aleiner}}, \bibinfo {author} {\bibfnamefont
	 		{F.}~\bibnamefont {Arute}}, \bibinfo {author} {\bibfnamefont
	 		{K.}~\bibnamefont {Arya}}, \bibinfo {author} {\bibfnamefont {J.}~\bibnamefont
	 		{Atalaya}}, \bibinfo {author} {\bibfnamefont {R.}~\bibnamefont {Babbush}},
	 	\bibinfo {author} {\bibfnamefont {J.~C.}\ \bibnamefont {Bardin}}, \bibinfo
	 	{author} {\bibfnamefont {R.}~\bibnamefont {Barends}}, \bibinfo {author}
	 	{\bibfnamefont {J.}~\bibnamefont {Basso}}, \bibinfo {author} {\bibfnamefont
	 		{A.}~\bibnamefont {Bengtsson}}, \bibinfo {author} {\bibfnamefont
	 		{A.}~\bibnamefont {Bilmes}}, \bibinfo {author} {\bibfnamefont
	 		{M.}~\bibnamefont {Broughton}}, \bibinfo {author} {\bibfnamefont {B.~B.}\
	 		\bibnamefont {Buckley}}, \bibinfo {author} {\bibfnamefont {D.~A.}\
	 		\bibnamefont {Buell}}, \bibinfo {author} {\bibfnamefont {B.}~\bibnamefont
	 		{Burkett}}, \bibinfo {author} {\bibfnamefont {N.}~\bibnamefont {Bushnell}},
	 	\bibinfo {author} {\bibfnamefont {B.}~\bibnamefont {Chiaro}}, \bibinfo
	 	{author} {\bibfnamefont {R.}~\bibnamefont {Collins}}, \bibinfo {author}
	 	{\bibfnamefont {W.}~\bibnamefont {Courtney}}, \bibinfo {author}
	 	{\bibfnamefont {S.}~\bibnamefont {Demura}}, \bibinfo {author} {\bibfnamefont
	 		{A.~R.}\ \bibnamefont {Derk}}, \bibinfo {author} {\bibfnamefont
	 		{D.}~\bibnamefont {Eppens}}, \bibinfo {author} {\bibfnamefont
	 		{C.}~\bibnamefont {Erickson}}, \bibinfo {author} {\bibfnamefont
	 		{L.}~\bibnamefont {Faoro}}, \bibinfo {author} {\bibfnamefont
	 		{E.}~\bibnamefont {Farhi}}, \bibinfo {author} {\bibfnamefont {A.~G.}\
	 		\bibnamefont {Fowler}}, \bibinfo {author} {\bibfnamefont {B.}~\bibnamefont
	 		{Foxen}}, \bibinfo {author} {\bibfnamefont {M.}~\bibnamefont {Giustina}},
	 	\bibinfo {author} {\bibfnamefont {A.}~\bibnamefont {Greene}}, \bibinfo
	 	{author} {\bibfnamefont {J.~A.}\ \bibnamefont {Gross}}, \bibinfo {author}
	 	{\bibfnamefont {M.~P.}\ \bibnamefont {Harrigan}}, \bibinfo {author}
	 	{\bibfnamefont {S.~D.}\ \bibnamefont {Harrington}}, \bibinfo {author}
	 	{\bibfnamefont {J.}~\bibnamefont {Hilton}}, \bibinfo {author} {\bibfnamefont
	 		{S.}~\bibnamefont {Hong}}, \bibinfo {author} {\bibfnamefont {T.}~\bibnamefont
	 		{Huang}}, \bibinfo {author} {\bibfnamefont {W.~J.}\ \bibnamefont {Huggins}},
	 	\bibinfo {author} {\bibfnamefont {L.~B.}\ \bibnamefont {Ioffe}}, \bibinfo
	 	{author} {\bibfnamefont {S.~V.}\ \bibnamefont {Isakov}}, \bibinfo {author}
	 	{\bibfnamefont {E.}~\bibnamefont {Jeffrey}}, \bibinfo {author} {\bibfnamefont
	 		{Z.}~\bibnamefont {Jiang}}, \bibinfo {author} {\bibfnamefont
	 		{D.}~\bibnamefont {Kafri}}, \bibinfo {author} {\bibfnamefont
	 		{K.}~\bibnamefont {Kechedzhi}}, \bibinfo {author} {\bibfnamefont
	 		{T.}~\bibnamefont {Khattar}}, \bibinfo {author} {\bibfnamefont
	 		{S.}~\bibnamefont {Kim}}, \bibinfo {author} {\bibfnamefont {P.~V.}\
	 		\bibnamefont {Klimov}}, \bibinfo {author} {\bibfnamefont {A.~N.}\
	 		\bibnamefont {Korotkov}}, \bibinfo {author} {\bibfnamefont {F.}~\bibnamefont
	 		{Kostritsa}}, \bibinfo {author} {\bibfnamefont {D.}~\bibnamefont {Landhuis}},
	 	\bibinfo {author} {\bibfnamefont {P.}~\bibnamefont {Laptev}}, \bibinfo
	 	{author} {\bibfnamefont {A.}~\bibnamefont {Locharla}}, \bibinfo {author}
	 	{\bibfnamefont {E.}~\bibnamefont {Lucero}}, \bibinfo {author} {\bibfnamefont
	 		{O.}~\bibnamefont {Martin}}, \bibinfo {author} {\bibfnamefont {M.~J.}\
	 		\bibnamefont {R.}}, \bibinfo {author} {\bibfnamefont {M.}~\bibnamefont
	 		{McEwen}}, \bibinfo {author} {\bibfnamefont {K.~C.}\ \bibnamefont {Miao}},
	 	\bibinfo {author} {\bibfnamefont {M.}~\bibnamefont {Mohseni}}, \bibinfo
	 	{author} {\bibfnamefont {S.}~\bibnamefont {Montazeri}}, \bibinfo {author}
	 	{\bibfnamefont {W.}~\bibnamefont {Mruczkiewicz}}, \bibinfo {author}
	 	{\bibfnamefont {J.}~\bibnamefont {Mutus}}, \bibinfo {author} {\bibfnamefont
	 		{O.}~\bibnamefont {Naaman}}, \bibinfo {author} {\bibfnamefont
	 		{M.}~\bibnamefont {Neeley}}, \bibinfo {author} {\bibfnamefont
	 		{C.}~\bibnamefont {Neill}}, \bibinfo {author} {\bibfnamefont {M.~Y.}\
	 		\bibnamefont {Niu}}, \bibinfo {author} {\bibfnamefont {T.~E.}\ \bibnamefont
	 		{O'Brien}}, \bibinfo {author} {\bibfnamefont {A.}~\bibnamefont {Opremcak}},
	 	\bibinfo {author} {\bibfnamefont {B.}~\bibnamefont {Pat{\'o}}}, \bibinfo
	 	{author} {\bibfnamefont {A.}~\bibnamefont {Petukhov}}, \bibinfo {author}
	 	{\bibfnamefont {N.~C.}\ \bibnamefont {Rubin}}, \bibinfo {author}
	 	{\bibfnamefont {D.}~\bibnamefont {Sank}}, \bibinfo {author} {\bibfnamefont
	 		{V.}~\bibnamefont {Shvarts}}, \bibinfo {author} {\bibfnamefont
	 		{D.}~\bibnamefont {Strain}}, \bibinfo {author} {\bibfnamefont
	 		{M.}~\bibnamefont {Szalay}}, \bibinfo {author} {\bibfnamefont
	 		{B.}~\bibnamefont {Villalonga}}, \bibinfo {author} {\bibfnamefont {T.~C.}\
	 		\bibnamefont {White}}, \bibinfo {author} {\bibfnamefont {Z.}~\bibnamefont
	 		{Yao}}, \bibinfo {author} {\bibfnamefont {P.}~\bibnamefont {Yeh}}, \bibinfo
	 	{author} {\bibfnamefont {J.}~\bibnamefont {Yoo}}, \bibinfo {author}
	 	{\bibfnamefont {A.}~\bibnamefont {Zalcman}}, \bibinfo {author} {\bibfnamefont
	 		{H.}~\bibnamefont {Neven}}, \bibinfo {author} {\bibfnamefont
	 		{S.}~\bibnamefont {Boixo}}, \bibinfo {author} {\bibfnamefont
	 		{A.}~\bibnamefont {Megrant}}, \bibinfo {author} {\bibfnamefont
	 		{Y.}~\bibnamefont {Chen}}, \bibinfo {author} {\bibfnamefont {J.}~\bibnamefont
	 		{Kelly}}, \bibinfo {author} {\bibfnamefont {V.}~\bibnamefont {Smelyanskiy}},
	 	\bibinfo {author} {\bibfnamefont {A.}~\bibnamefont {Kitaev}}, \bibinfo
	 	{author} {\bibfnamefont {M.}~\bibnamefont {Knap}}, \bibinfo {author}
	 	{\bibfnamefont {F.}~\bibnamefont {Pollmann}},\ and\ \bibinfo {author}
	 	{\bibfnamefont {P.}~\bibnamefont {Roushan}},\ }\bibfield  {title} {\bibinfo
	 	{title} {Realizing topologically ordered states on a quantum processor},\
	 }\href {https://doi.org/10.1126/science.abi8378} {\bibfield  {journal}
	 	{\bibinfo  {journal} {Science}\ }\textbf {\bibinfo {volume} {374}},\ \bibinfo
	 	{pages} {1237} (\bibinfo {year} {2021})}\BibitemShut {NoStop}%
	 \bibitem [{\citenamefont {Semeghini}\ \emph {et~al.}(2021)\citenamefont
	 	{Semeghini}, \citenamefont {Levine}, \citenamefont {Keesling}, \citenamefont
	 	{Ebadi}, \citenamefont {Wang~T.}, \citenamefont {Bluvstein}, \citenamefont
	 	{Verresen}, \citenamefont {Pichler}, \citenamefont {Kalinowski},
	 	\citenamefont {Samajdar}, \citenamefont {Omran}, \citenamefont {Sachdev},
	 	\citenamefont {Vishwanath}, \citenamefont {Greiner}, \citenamefont
	 	{Vuleti{\'c}},\ and\ \citenamefont {Lukin}}]{Semeghini2021}%
	 \BibitemOpen
	 \bibfield  {author} {\bibinfo {author} {\bibfnamefont {G.}~\bibnamefont
	 		{Semeghini}}, \bibinfo {author} {\bibfnamefont {H.}~\bibnamefont {Levine}},
	 	\bibinfo {author} {\bibfnamefont {A.}~\bibnamefont {Keesling}}, \bibinfo
	 	{author} {\bibfnamefont {S.}~\bibnamefont {Ebadi}}, \bibinfo {author}
	 	{\bibfnamefont {T.}~\bibnamefont {Wang~T.}}, \bibinfo {author} {\bibfnamefont
	 		{D.}~\bibnamefont {Bluvstein}}, \bibinfo {author} {\bibfnamefont
	 		{R.}~\bibnamefont {Verresen}}, \bibinfo {author} {\bibfnamefont
	 		{H.}~\bibnamefont {Pichler}}, \bibinfo {author} {\bibfnamefont
	 		{M.}~\bibnamefont {Kalinowski}}, \bibinfo {author} {\bibfnamefont
	 		{R.}~\bibnamefont {Samajdar}}, \bibinfo {author} {\bibfnamefont
	 		{A.}~\bibnamefont {Omran}}, \bibinfo {author} {\bibfnamefont
	 		{S.}~\bibnamefont {Sachdev}}, \bibinfo {author} {\bibfnamefont
	 		{A.}~\bibnamefont {Vishwanath}}, \bibinfo {author} {\bibfnamefont
	 		{M.}~\bibnamefont {Greiner}}, \bibinfo {author} {\bibfnamefont
	 		{V.}~\bibnamefont {Vuleti{\'c}}},\ and\ \bibinfo {author} {\bibfnamefont
	 		{M.~D.}\ \bibnamefont {Lukin}},\ }\bibfield  {title} {\bibinfo {title}
	 	{Probing topological spin liquids on a programmable quantum simulator},\
	 }\href {https://doi.org/10.1126/science.abi8794} {\bibfield  {journal}
	 	{\bibinfo  {journal} {Science}\ }\textbf {\bibinfo {volume} {374}},\ \bibinfo
	 	{pages} {1242} (\bibinfo {year} {2021})}\BibitemShut {NoStop}%
	 \bibitem [{\citenamefont {Gu}\ \emph {et~al.}(2017)\citenamefont {Gu},
	 	\citenamefont {Kockum}, \citenamefont {Miranowicz}, \citenamefont {Liu},\
	 	and\ \citenamefont {Nori}}]{Gu2017}%
	 \BibitemOpen
	 \bibfield  {author} {\bibinfo {author} {\bibfnamefont {X.}~\bibnamefont
	 		{Gu}}, \bibinfo {author} {\bibfnamefont {A.~F.}\ \bibnamefont {Kockum}},
	 	\bibinfo {author} {\bibfnamefont {A.}~\bibnamefont {Miranowicz}}, \bibinfo
	 	{author} {\bibfnamefont {Y.-X.}\ \bibnamefont {Liu}},\ and\ \bibinfo {author}
	 	{\bibfnamefont {F.}~\bibnamefont {Nori}},\ }\bibfield  {title} {\bibinfo
	 	{title} {Microwave photonics with superconducting quantum circuits},\ }\href
	 {http://www.sciencedirect.com/science/article/pii/S0370157317303290}
	 {\bibfield  {journal} {\bibinfo  {journal} {Phys. Rep.}\ }\textbf {\bibinfo
	 		{volume} {718-719}},\ \bibinfo {pages} {1} (\bibinfo {year}
	 	{2017})}\BibitemShut {NoStop}%
	 \bibitem [{\citenamefont {You}\ \emph {et~al.}(2007)\citenamefont {You},
	 	\citenamefont {Hu}, \citenamefont {Ashhab},\ and\ \citenamefont
	 	{Nori}}]{You2007}%
	 \BibitemOpen
	 \bibfield  {author} {\bibinfo {author} {\bibfnamefont {J.~Q.}\ \bibnamefont
	 		{You}}, \bibinfo {author} {\bibfnamefont {X.~D.}\ \bibnamefont {Hu}},
	 	\bibinfo {author} {\bibfnamefont {S.}~\bibnamefont {Ashhab}},\ and\ \bibinfo
	 	{author} {\bibfnamefont {F.}~\bibnamefont {Nori}},\ }\bibfield  {title}
	 {\bibinfo {title} {Low-decoherence flux qubit},\ }\href
	 {https://doi.org/10.1103/PhysRevB.75.140515} {\bibfield  {journal} {\bibinfo
	 		{journal} {Phys. Rev. B}\ }\textbf {\bibinfo {volume} {75}},\ \bibinfo
	 	{pages} {140515} (\bibinfo {year} {2007})}\BibitemShut {NoStop}%
	 \bibitem [{\citenamefont {Koch}\ \emph {et~al.}(2007)\citenamefont {Koch},
	 	\citenamefont {Yu}, \citenamefont {Gambetta}, \citenamefont {Houck},
	 	\citenamefont {Schuster}, \citenamefont {Majer}, \citenamefont {Blais},
	 	\citenamefont {Devoret}, \citenamefont {Girvin},\ and\ \citenamefont
	 	{Schoelkopf}}]{Koch2007}%
	 \BibitemOpen
	 \bibfield  {author} {\bibinfo {author} {\bibfnamefont {J.}~\bibnamefont
	 		{Koch}}, \bibinfo {author} {\bibfnamefont {T.~M.}\ \bibnamefont {Yu}},
	 	\bibinfo {author} {\bibfnamefont {J.}~\bibnamefont {Gambetta}}, \bibinfo
	 	{author} {\bibfnamefont {A.~A.}\ \bibnamefont {Houck}}, \bibinfo {author}
	 	{\bibfnamefont {D.~I.}\ \bibnamefont {Schuster}}, \bibinfo {author}
	 	{\bibfnamefont {J.}~\bibnamefont {Majer}}, \bibinfo {author} {\bibfnamefont
	 		{A.}~\bibnamefont {Blais}}, \bibinfo {author} {\bibfnamefont {M.~H.}\
	 		\bibnamefont {Devoret}}, \bibinfo {author} {\bibfnamefont {S.~M.}\
	 		\bibnamefont {Girvin}},\ and\ \bibinfo {author} {\bibfnamefont {R.~J.}\
	 		\bibnamefont {Schoelkopf}},\ }\bibfield  {title} {\bibinfo {title}
	 	{Charge-insensitive qubit design derived from the {C}ooper pair box},\ }\href
	 {https://doi.org/10.1103/PhysRevA.76.042319} {\bibfield  {journal} {\bibinfo
	 		{journal} {Phys. Rev. A}\ }\textbf {\bibinfo {volume} {76}},\ \bibinfo
	 	{pages} {042319} (\bibinfo {year} {2007})}\BibitemShut {NoStop}%
	 \bibitem [{\citenamefont {Dagotto}\ and\ \citenamefont
	 	{Rice}(1996)}]{Elbio1996}%
	 \BibitemOpen
	 \bibfield  {author} {\bibinfo {author} {\bibfnamefont {E.}~\bibnamefont
	 		{Dagotto}}\ and\ \bibinfo {author} {\bibfnamefont {T.~M.}\ \bibnamefont
	 		{Rice}},\ }\bibfield  {title} {\bibinfo {title} {Surprises on the way from
	 		one- to two-dimensional quantum magnets: The ladder materials},\ }\href
	 {https://doi.org/10.1126/science.271.5249.618} {\bibfield  {journal}
	 	{\bibinfo  {journal} {Science}\ }\textbf {\bibinfo {volume} {271}},\ \bibinfo
	 	{pages} {618} (\bibinfo {year} {1996})}\BibitemShut {NoStop}%
	 \bibitem [{\citenamefont {Roushan}\ \emph {et~al.}(2017)\citenamefont
	 	{Roushan}, \citenamefont {Neill}, \citenamefont {Tangpanitanon},
	 	\citenamefont {Bastidas}, \citenamefont {Megrant}, \citenamefont {Barends},
	 	\citenamefont {Chen}, \citenamefont {Chen}, \citenamefont {Chiaro},
	 	\citenamefont {Dunsworth}, \citenamefont {Fowler}, \citenamefont {Foxen},
	 	\citenamefont {Giustina}, \citenamefont {Jeffrey}, \citenamefont {Kelly},
	 	\citenamefont {Lucero}, \citenamefont {Mutus}, \citenamefont {Neeley},
	 	\citenamefont {Quintana}, \citenamefont {Sank}, \citenamefont {Vainsencher},
	 	\citenamefont {Wenner}, \citenamefont {White}, \citenamefont {Neven},
	 	\citenamefont {Angelakis},\ and\ \citenamefont {Martinis}}]{Roushan2017}%
	 \BibitemOpen
	 \bibfield  {author} {\bibinfo {author} {\bibfnamefont {P.}~\bibnamefont
	 		{Roushan}}, \bibinfo {author} {\bibfnamefont {C.}~\bibnamefont {Neill}},
	 	\bibinfo {author} {\bibfnamefont {J.}~\bibnamefont {Tangpanitanon}}, \bibinfo
	 	{author} {\bibfnamefont {V.~M.}\ \bibnamefont {Bastidas}}, \bibinfo {author}
	 	{\bibfnamefont {A.}~\bibnamefont {Megrant}}, \bibinfo {author} {\bibfnamefont
	 		{R.}~\bibnamefont {Barends}}, \bibinfo {author} {\bibfnamefont
	 		{Y.}~\bibnamefont {Chen}}, \bibinfo {author} {\bibfnamefont {Z.}~\bibnamefont
	 		{Chen}}, \bibinfo {author} {\bibfnamefont {B.}~\bibnamefont {Chiaro}},
	 	\bibinfo {author} {\bibfnamefont {A.}~\bibnamefont {Dunsworth}}, \bibinfo
	 	{author} {\bibfnamefont {A.}~\bibnamefont {Fowler}}, \bibinfo {author}
	 	{\bibfnamefont {B.}~\bibnamefont {Foxen}}, \bibinfo {author} {\bibfnamefont
	 		{M.}~\bibnamefont {Giustina}}, \bibinfo {author} {\bibfnamefont
	 		{E.}~\bibnamefont {Jeffrey}}, \bibinfo {author} {\bibfnamefont
	 		{J.}~\bibnamefont {Kelly}}, \bibinfo {author} {\bibfnamefont
	 		{E.}~\bibnamefont {Lucero}}, \bibinfo {author} {\bibfnamefont
	 		{J.}~\bibnamefont {Mutus}}, \bibinfo {author} {\bibfnamefont
	 		{M.}~\bibnamefont {Neeley}}, \bibinfo {author} {\bibfnamefont
	 		{C.}~\bibnamefont {Quintana}}, \bibinfo {author} {\bibfnamefont
	 		{D.}~\bibnamefont {Sank}}, \bibinfo {author} {\bibfnamefont {A.}~\bibnamefont
	 		{Vainsencher}}, \bibinfo {author} {\bibfnamefont {J.}~\bibnamefont {Wenner}},
	 	\bibinfo {author} {\bibfnamefont {T.}~\bibnamefont {White}}, \bibinfo
	 	{author} {\bibfnamefont {H.}~\bibnamefont {Neven}}, \bibinfo {author}
	 	{\bibfnamefont {D.~G.}\ \bibnamefont {Angelakis}},\ and\ \bibinfo {author}
	 	{\bibfnamefont {J.}~\bibnamefont {Martinis}},\ }\bibfield  {title} {\bibinfo
	 	{title} {Spectroscopic signatures of localization with interacting photons in
	 		superconducting qubits},\ }\href {https://doi.org/10.1126/science.aao1401}
	 {\bibfield  {journal} {\bibinfo  {journal} {Science}\ }\textbf {\bibinfo
	 		{volume} {358}},\ \bibinfo {pages} {1175} (\bibinfo {year}
	 	{2017})}\BibitemShut {NoStop}%
	 \bibitem [{\citenamefont {Ye}\ \emph {et~al.}(2019)\citenamefont {Ye},
	 	\citenamefont {Ge}, \citenamefont {Wu}, \citenamefont {Wang}, \citenamefont
	 	{Gong}, \citenamefont {Zhang}, \citenamefont {Zhu}, \citenamefont {Yang},
	 	\citenamefont {Li}, \citenamefont {Liang}, \citenamefont {Lin}, \citenamefont
	 	{Xu}, \citenamefont {Guo}, \citenamefont {Sun}, \citenamefont {Cheng},
	 	\citenamefont {Ma}, \citenamefont {Meng}, \citenamefont {Deng}, \citenamefont
	 	{Rong}, \citenamefont {Lu}, \citenamefont {Peng}, \citenamefont {Fan},
	 	\citenamefont {Zhu},\ and\ \citenamefont {Pan}}]{Ye2019}%
	 \BibitemOpen
	 \bibfield  {author} {\bibinfo {author} {\bibfnamefont {Y.}~\bibnamefont
	 		{Ye}}, \bibinfo {author} {\bibfnamefont {Z.-Y.}\ \bibnamefont {Ge}}, \bibinfo
	 	{author} {\bibfnamefont {Y.}~\bibnamefont {Wu}}, \bibinfo {author}
	 	{\bibfnamefont {S.}~\bibnamefont {Wang}}, \bibinfo {author} {\bibfnamefont
	 		{M.}~\bibnamefont {Gong}}, \bibinfo {author} {\bibfnamefont {Y.-R.}\
	 		\bibnamefont {Zhang}}, \bibinfo {author} {\bibfnamefont {Q.}~\bibnamefont
	 		{Zhu}}, \bibinfo {author} {\bibfnamefont {R.}~\bibnamefont {Yang}}, \bibinfo
	 	{author} {\bibfnamefont {S.}~\bibnamefont {Li}}, \bibinfo {author}
	 	{\bibfnamefont {F.}~\bibnamefont {Liang}}, \bibinfo {author} {\bibfnamefont
	 		{J.}~\bibnamefont {Lin}}, \bibinfo {author} {\bibfnamefont {Y.}~\bibnamefont
	 		{Xu}}, \bibinfo {author} {\bibfnamefont {C.}~\bibnamefont {Guo}}, \bibinfo
	 	{author} {\bibfnamefont {L.}~\bibnamefont {Sun}}, \bibinfo {author}
	 	{\bibfnamefont {C.}~\bibnamefont {Cheng}}, \bibinfo {author} {\bibfnamefont
	 		{N.}~\bibnamefont {Ma}}, \bibinfo {author} {\bibfnamefont {Z.~Y.}\
	 		\bibnamefont {Meng}}, \bibinfo {author} {\bibfnamefont {H.}~\bibnamefont
	 		{Deng}}, \bibinfo {author} {\bibfnamefont {H.}~\bibnamefont {Rong}}, \bibinfo
	 	{author} {\bibfnamefont {C.-Y.}\ \bibnamefont {Lu}}, \bibinfo {author}
	 	{\bibfnamefont {C.-Z.}\ \bibnamefont {Peng}}, \bibinfo {author}
	 	{\bibfnamefont {H.}~\bibnamefont {Fan}}, \bibinfo {author} {\bibfnamefont
	 		{X.}~\bibnamefont {Zhu}},\ and\ \bibinfo {author} {\bibfnamefont {J.-W.}\
	 		\bibnamefont {Pan}},\ }\bibfield  {title} {\bibinfo {title} {Propagation and
	 		localization of collective excitations on a 24-qubit superconducting
	 		processor},\ }\href {https://doi.org/10.1103/PhysRevLett.123.050502}
	 {\bibfield  {journal} {\bibinfo  {journal} {Phys. Rev. Lett.}\ }\textbf
	 	{\bibinfo {volume} {123}},\ \bibinfo {pages} {050502} (\bibinfo {year}
	 	{2019})}\BibitemShut {NoStop}%
	 \bibitem [{\citenamefont {Yan}\ \emph {et~al.}(2019)\citenamefont {Yan},
	 	\citenamefont {Zhang}, \citenamefont {Gong}, \citenamefont {Wu},
	 	\citenamefont {Zheng}, \citenamefont {Li}, \citenamefont {Wang},
	 	\citenamefont {Liang}, \citenamefont {Lin}, \citenamefont {Xu}, \citenamefont
	 	{Guo}, \citenamefont {Sun}, \citenamefont {Peng}, \citenamefont {Xia},
	 	\citenamefont {Deng}, \citenamefont {Rong}, \citenamefont {You},
	 	\citenamefont {Nori}, \citenamefont {Fan}, \citenamefont {Zhu},\ and\
	 	\citenamefont {Pan}}]{Yan2019}%
	 \BibitemOpen
	 \bibfield  {author} {\bibinfo {author} {\bibfnamefont {Z.}~\bibnamefont
	 		{Yan}}, \bibinfo {author} {\bibfnamefont {Y.-R.}\ \bibnamefont {Zhang}},
	 	\bibinfo {author} {\bibfnamefont {M.}~\bibnamefont {Gong}}, \bibinfo {author}
	 	{\bibfnamefont {Y.}~\bibnamefont {Wu}}, \bibinfo {author} {\bibfnamefont
	 		{Y.}~\bibnamefont {Zheng}}, \bibinfo {author} {\bibfnamefont
	 		{S.}~\bibnamefont {Li}}, \bibinfo {author} {\bibfnamefont {C.}~\bibnamefont
	 		{Wang}}, \bibinfo {author} {\bibfnamefont {F.}~\bibnamefont {Liang}},
	 	\bibinfo {author} {\bibfnamefont {J.}~\bibnamefont {Lin}}, \bibinfo {author}
	 	{\bibfnamefont {Y.}~\bibnamefont {Xu}}, \bibinfo {author} {\bibfnamefont
	 		{C.}~\bibnamefont {Guo}}, \bibinfo {author} {\bibfnamefont {L.}~\bibnamefont
	 		{Sun}}, \bibinfo {author} {\bibfnamefont {C.-Z.}\ \bibnamefont {Peng}},
	 	\bibinfo {author} {\bibfnamefont {K.}~\bibnamefont {Xia}}, \bibinfo {author}
	 	{\bibfnamefont {H.}~\bibnamefont {Deng}}, \bibinfo {author} {\bibfnamefont
	 		{H.}~\bibnamefont {Rong}}, \bibinfo {author} {\bibfnamefont {J.~Q.}\
	 		\bibnamefont {You}}, \bibinfo {author} {\bibfnamefont {F.}~\bibnamefont
	 		{Nori}}, \bibinfo {author} {\bibfnamefont {H.}~\bibnamefont {Fan}}, \bibinfo
	 	{author} {\bibfnamefont {X.}~\bibnamefont {Zhu}},\ and\ \bibinfo {author}
	 	{\bibfnamefont {J.-W.}\ \bibnamefont {Pan}},\ }\bibfield  {title} {\bibinfo
	 	{title} {Strongly correlated quantum walks with a 12-qubit superconducting
	 		processor},\ }\href {https://doi.org/10.1126/science.aaw1611} {\bibfield
	 	{journal} {\bibinfo  {journal} {Science}\ }\textbf {\bibinfo {volume}
	 		{364}},\ \bibinfo {pages} {753} (\bibinfo {year} {2019})}\BibitemShut
	 {NoStop}%
	 \bibitem [{\citenamefont {Kraus}\ and\ \citenamefont
	 	{Zilberberg}(2012)}]{Kraus2012a}%
	 \BibitemOpen
	 \bibfield  {author} {\bibinfo {author} {\bibfnamefont {Y.~E.}\ \bibnamefont
	 		{Kraus}}\ and\ \bibinfo {author} {\bibfnamefont {O.}~\bibnamefont
	 		{Zilberberg}},\ }\bibfield  {title} {\bibinfo {title} {Topological
	 		equivalence between the {F}ibonacci quasicrystal and the {H}arper model},\
	 }\href {https://doi.org/10.1103/PhysRevLett.109.116404} {\bibfield  {journal}
	 	{\bibinfo  {journal} {Phys. Rev. Lett.}\ }\textbf {\bibinfo {volume} {109}},\
	 	\bibinfo {pages} {116404} (\bibinfo {year} {2012})}\BibitemShut {NoStop}%
	 \bibitem [{\citenamefont {Harper}(1955)}]{Harper1955}%
	 \BibitemOpen
	 \bibfield  {author} {\bibinfo {author} {\bibfnamefont {P.~G.}\ \bibnamefont
	 		{Harper}},\ }\bibfield  {title} {\bibinfo {title} {The general motion of
	 		conduction electrons in a uniform magnetic field, with application to the
	 		diamagnetism of metals},\ }\href
	 {https://doi.org/10.1088/0370-1298/68/10/305} {\bibfield  {journal} {\bibinfo
	 		{journal} {Proc. Phys. Soc. A}\ }\textbf {\bibinfo {volume} {68}},\ \bibinfo
	 	{pages} {879} (\bibinfo {year} {1955})}\BibitemShut {NoStop}%
	 \bibitem [{\citenamefont {Aubry}\ and\ \citenamefont
	 	{Andr{\'e}}(1980)}]{aubry1980}%
	 \BibitemOpen
	 \bibfield  {author} {\bibinfo {author} {\bibfnamefont {S.}~\bibnamefont
	 		{Aubry}}\ and\ \bibinfo {author} {\bibfnamefont {G.}~\bibnamefont
	 		{Andr{\'e}}},\ }\bibfield  {title} {\bibinfo {title} {Analyticity breaking
	 		and {A}nderson localization in incommensurate lattices},\ }\href@noop {}
	 {\bibfield  {journal} {\bibinfo  {journal} {Ann. Israel Phys. Soc}\ }\textbf
	 	{\bibinfo {volume} {3}},\ \bibinfo {pages} {18} (\bibinfo {year}
	 	{1980})}\BibitemShut {NoStop}%
	 \bibitem [{\citenamefont {Hofstadter}(1976)}]{Hofstadter1976}%
	 \BibitemOpen
	 \bibfield  {author} {\bibinfo {author} {\bibfnamefont {D.~R.}\ \bibnamefont
	 		{Hofstadter}},\ }\bibfield  {title} {\bibinfo {title} {Energy levels and wave
	 		functions of {B}loch electrons in rational and irrational magnetic fields},\
	 }\href {https://doi.org/10.1103/PhysRevB.14.2239} {\bibfield  {journal}
	 	{\bibinfo  {journal} {Phys. Rev. B}\ }\textbf {\bibinfo {volume} {14}},\
	 	\bibinfo {pages} {2239} (\bibinfo {year} {1976})}\BibitemShut {NoStop}%
	 \bibitem [{\citenamefont {Senko}\ \emph {et~al.}(2014)\citenamefont {Senko},
	 	\citenamefont {Smith}, \citenamefont {Richerme}, \citenamefont {Lee},
	 	\citenamefont {Campbell},\ and\ \citenamefont {Monroe}}]{Senko2014}%
	 \BibitemOpen
	 \bibfield  {author} {\bibinfo {author} {\bibfnamefont {C.}~\bibnamefont
	 		{Senko}}, \bibinfo {author} {\bibfnamefont {J.}~\bibnamefont {Smith}},
	 	\bibinfo {author} {\bibfnamefont {P.}~\bibnamefont {Richerme}}, \bibinfo
	 	{author} {\bibfnamefont {A.}~\bibnamefont {Lee}}, \bibinfo {author}
	 	{\bibfnamefont {W.~C.}\ \bibnamefont {Campbell}},\ and\ \bibinfo {author}
	 	{\bibfnamefont {C.}~\bibnamefont {Monroe}},\ }\bibfield  {title} {\bibinfo
	 	{title} {Coherent imaging spectroscopy of a quantum many-body spin system},\
	 }\href {https://doi.org/10.1126/science.1251422} {\bibfield  {journal}
	 	{\bibinfo  {journal} {Science}\ }\textbf {\bibinfo {volume} {345}},\ \bibinfo
	 	{pages} {430} (\bibinfo {year} {2014})}\BibitemShut {NoStop}%
	 \bibitem [{\citenamefont {Jurcevic}\ \emph {et~al.}(2015)\citenamefont
	 	{Jurcevic}, \citenamefont {Hauke}, \citenamefont {Maier}, \citenamefont
	 	{Hempel}, \citenamefont {Lanyon}, \citenamefont {Blatt},\ and\ \citenamefont
	 	{Roos}}]{Jurcevic2015}%
	 \BibitemOpen
	 \bibfield  {author} {\bibinfo {author} {\bibfnamefont {P.}~\bibnamefont
	 		{Jurcevic}}, \bibinfo {author} {\bibfnamefont {P.}~\bibnamefont {Hauke}},
	 	\bibinfo {author} {\bibfnamefont {C.}~\bibnamefont {Maier}}, \bibinfo
	 	{author} {\bibfnamefont {C.}~\bibnamefont {Hempel}}, \bibinfo {author}
	 	{\bibfnamefont {B.~P.}\ \bibnamefont {Lanyon}}, \bibinfo {author}
	 	{\bibfnamefont {R.}~\bibnamefont {Blatt}},\ and\ \bibinfo {author}
	 	{\bibfnamefont {C.~F.}\ \bibnamefont {Roos}},\ }\bibfield  {title} {\bibinfo
	 	{title} {Spectroscopy of interacting quasiparticles in trapped ions},\ }\href
	 {https://doi.org/10.1103/PhysRevLett.115.100501} {\bibfield  {journal}
	 	{\bibinfo  {journal} {Phys. Rev. Lett.}\ }\textbf {\bibinfo {volume} {115}},\
	 	\bibinfo {pages} {100501} (\bibinfo {year} {2015})}\BibitemShut {NoStop}%
	 \bibitem [{\citenamefont {Cai}\ \emph {et~al.}(2019)\citenamefont {Cai},
	 	\citenamefont {Han}, \citenamefont {Mei}, \citenamefont {Xu}, \citenamefont
	 	{Ma}, \citenamefont {Li}, \citenamefont {Wang}, \citenamefont {Song},
	 	\citenamefont {Xue}, \citenamefont {Yin}, \citenamefont {Jia},\ and\
	 	\citenamefont {Sun}}]{Cai2019}%
	 \BibitemOpen
	 \bibfield  {author} {\bibinfo {author} {\bibfnamefont {W.}~\bibnamefont
	 		{Cai}}, \bibinfo {author} {\bibfnamefont {J.}~\bibnamefont {Han}}, \bibinfo
	 	{author} {\bibfnamefont {F.}~\bibnamefont {Mei}}, \bibinfo {author}
	 	{\bibfnamefont {Y.}~\bibnamefont {Xu}}, \bibinfo {author} {\bibfnamefont
	 		{Y.}~\bibnamefont {Ma}}, \bibinfo {author} {\bibfnamefont {X.}~\bibnamefont
	 		{Li}}, \bibinfo {author} {\bibfnamefont {H.}~\bibnamefont {Wang}}, \bibinfo
	 	{author} {\bibfnamefont {Y.~P.}\ \bibnamefont {Song}}, \bibinfo {author}
	 	{\bibfnamefont {Z.-Y.}\ \bibnamefont {Xue}}, \bibinfo {author} {\bibfnamefont
	 		{Z.-q.}\ \bibnamefont {Yin}}, \bibinfo {author} {\bibfnamefont
	 		{S.}~\bibnamefont {Jia}},\ and\ \bibinfo {author} {\bibfnamefont
	 		{L.}~\bibnamefont {Sun}},\ }\bibfield  {title} {\bibinfo {title} {Observation
	 		of topological magnon insulator states in a superconducting circuit},\ }\href
	 {https://doi.org/10.1103/PhysRevLett.123.080501} {\bibfield  {journal}
	 	{\bibinfo  {journal} {Phys. Rev. Lett.}\ }\textbf {\bibinfo {volume} {123}},\
	 	\bibinfo {pages} {080501} (\bibinfo {year} {2019})}\BibitemShut {NoStop}%
	 \bibitem [{\citenamefont {Kraus}\ \emph {et~al.}(2012)\citenamefont {Kraus},
	 	\citenamefont {Lahini}, \citenamefont {Ringel}, \citenamefont {Verbin},\ and\
	 	\citenamefont {Zilberberg}}]{Kraus2012}%
	 \BibitemOpen
	 \bibfield  {author} {\bibinfo {author} {\bibfnamefont {Y.~E.}\ \bibnamefont
	 		{Kraus}}, \bibinfo {author} {\bibfnamefont {Y.}~\bibnamefont {Lahini}},
	 	\bibinfo {author} {\bibfnamefont {Z.}~\bibnamefont {Ringel}}, \bibinfo
	 	{author} {\bibfnamefont {M.}~\bibnamefont {Verbin}},\ and\ \bibinfo {author}
	 	{\bibfnamefont {O.}~\bibnamefont {Zilberberg}},\ }\bibfield  {title}
	 {\bibinfo {title} {Topological states and adiabatic pumping in
	 		quasicrystals},\ }\href {https://doi.org/10.1103/PhysRevLett.109.106402}
	 {\bibfield  {journal} {\bibinfo  {journal} {Phys. Rev. Lett.}\ }\textbf
	 	{\bibinfo {volume} {109}},\ \bibinfo {pages} {106402} (\bibinfo {year}
	 	{2012})}\BibitemShut {NoStop}%
	 \bibitem [{\citenamefont {Niu}\ and\ \citenamefont {Thouless}(1984)}]{Niu1984}%
	 \BibitemOpen
	 \bibfield  {author} {\bibinfo {author} {\bibfnamefont {Q.}~\bibnamefont
	 		{Niu}}\ and\ \bibinfo {author} {\bibfnamefont {D.~J.}\ \bibnamefont
	 		{Thouless}},\ }\bibfield  {title} {\bibinfo {title} {Quantised adiabatic
	 		charge transport in the presence of substrate disorder and many-body
	 		interaction},\ }\href {https://doi.org/10.1088/0305-4470/17/12/016}
	 {\bibfield  {journal} {\bibinfo  {journal} {J. Phys. A}\ }\textbf {\bibinfo
	 		{volume} {17}},\ \bibinfo {pages} {2453} (\bibinfo {year}
	 	{1984})}\BibitemShut {NoStop}%
	 \bibitem [{\citenamefont {Laughlin}(1981)}]{Laughlin1981}%
	 \BibitemOpen
	 \bibfield  {author} {\bibinfo {author} {\bibfnamefont {R.~B.}\ \bibnamefont
	 		{Laughlin}},\ }\bibfield  {title} {\bibinfo {title} {Quantized {H}all
	 		conductivity in two dimensions},\ }\href
	 {https://doi.org/10.1103/PhysRevB.23.5632} {\bibfield  {journal} {\bibinfo
	 		{journal} {Phys. Rev. B}\ }\textbf {\bibinfo {volume} {23}},\ \bibinfo
	 	{pages} {5632} (\bibinfo {year} {1981})}\BibitemShut {NoStop}%
	 \bibitem [{\citenamefont {Fabre}\ \emph {et~al.}(2022)\citenamefont {Fabre},
	 	\citenamefont {Bouhiron}, \citenamefont {Satoor}, \citenamefont {Lopes},\
	 	and\ \citenamefont {Nascimbene}}]{Fabre2022}%
	 \BibitemOpen
	 \bibfield  {author} {\bibinfo {author} {\bibfnamefont {A.}~\bibnamefont
	 		{Fabre}}, \bibinfo {author} {\bibfnamefont {J.-B.}\ \bibnamefont {Bouhiron}},
	 	\bibinfo {author} {\bibfnamefont {T.}~\bibnamefont {Satoor}}, \bibinfo
	 	{author} {\bibfnamefont {R.}~\bibnamefont {Lopes}},\ and\ \bibinfo {author}
	 	{\bibfnamefont {S.}~\bibnamefont {Nascimbene}},\ }\bibfield  {title}
	 {\bibinfo {title} {Laughlin's topological charge pump in an atomic {H}all
	 		cylinder},\ }\href {https://doi.org/10.1103/PhysRevLett.128.173202}
	 {\bibfield  {journal} {\bibinfo  {journal} {Phys. Rev. Lett.}\ }\textbf
	 	{\bibinfo {volume} {128}},\ \bibinfo {pages} {173202} (\bibinfo {year}
	 	{2022})}\BibitemShut {NoStop}%
	 \bibitem [{\citenamefont {Fedorova}\ \emph {et~al.}(2020)\citenamefont
	 	{Fedorova}, \citenamefont {Qiu}, \citenamefont {Linden},\ and\ \citenamefont
	 	{Kroha}}]{Fedorova2020}%
	 \BibitemOpen
	 \bibfield  {author} {\bibinfo {author} {\bibfnamefont {Z.}~\bibnamefont
	 		{Fedorova}}, \bibinfo {author} {\bibfnamefont {H.}~\bibnamefont {Qiu}},
	 	\bibinfo {author} {\bibfnamefont {S.}~\bibnamefont {Linden}},\ and\ \bibinfo
	 	{author} {\bibfnamefont {J.}~\bibnamefont {Kroha}},\ }\bibfield  {title}
	 {\bibinfo {title} {Observation of topological transport quantization by
	 		dissipation in fast {T}houless pumps},\ }\href
	 {https://doi.org/10.1038/s41467-020-17510-z} {\bibfield  {journal} {\bibinfo
	 		{journal} {Nat. Commun.}\ }\textbf {\bibinfo {volume} {11}},\ \bibinfo
	 	{pages} {3758} (\bibinfo {year} {2020})}\BibitemShut {NoStop}%
	 \bibitem [{\citenamefont {Rozhkov}\ \emph {et~al.}(2016)\citenamefont
	 	{Rozhkov}, \citenamefont {Sboychakov}, \citenamefont {Rakhmanov},\ and\
	 	\citenamefont {Nori}}]{Rozhkov2016}%
	 \BibitemOpen
	 \bibfield  {author} {\bibinfo {author} {\bibfnamefont {A.~V.}\ \bibnamefont
	 		{Rozhkov}}, \bibinfo {author} {\bibfnamefont {A.~O.}\ \bibnamefont
	 		{Sboychakov}}, \bibinfo {author} {\bibfnamefont {A.~L.}\ \bibnamefont
	 		{Rakhmanov}},\ and\ \bibinfo {author} {\bibfnamefont {F.}~\bibnamefont
	 		{Nori}},\ }\bibfield  {title} {\bibinfo {title} {Electronic properties of
	 		graphene-based bilayer systems},\ }\href
	 {https://doi.org/10.1016/j.physrep.2016.07.003} {\bibfield  {journal}
	 	{\bibinfo  {journal} {Phys. Rep.}\ }\textbf {\bibinfo {volume} {648}},\
	 	\bibinfo {pages} {1} (\bibinfo {year} {2016})}\BibitemShut {NoStop}%
	 \bibitem [{\citenamefont {Lau}\ \emph {et~al.}(2015)\citenamefont {Lau},
	 	\citenamefont {Ortix},\ and\ \citenamefont {van~den Brink}}]{Lau2015}%
	 \BibitemOpen
	 \bibfield  {author} {\bibinfo {author} {\bibfnamefont {A.}~\bibnamefont
	 		{Lau}}, \bibinfo {author} {\bibfnamefont {C.}~\bibnamefont {Ortix}},\ and\
	 	\bibinfo {author} {\bibfnamefont {J.}~\bibnamefont {van~den Brink}},\
	 }\bibfield  {title} {\bibinfo {title} {Topological edge states with zero
	 		{H}all conductivity in a dimerized {H}ofstadter model},\ }\href
	 {https://doi.org/10.1103/PhysRevLett.115.216805} {\bibfield  {journal}
	 	{\bibinfo  {journal} {Phys. Rev. Lett.}\ }\textbf {\bibinfo {volume} {115}},\
	 	\bibinfo {pages} {216805} (\bibinfo {year} {2015})}\BibitemShut {NoStop}%
	 \bibitem [{\citenamefont {Ge}\ \emph {et~al.}(2020)\citenamefont {Ge},
	 	\citenamefont {Liu}, \citenamefont {Li}, \citenamefont {Li}, \citenamefont
	 	{Luo}, \citenamefont {Wu}, \citenamefont {Xu},\ and\ \citenamefont
	 	{Wang}}]{Ge2020}%
	 \BibitemOpen
	 \bibfield  {author} {\bibinfo {author} {\bibfnamefont {J.}~\bibnamefont
	 		{Ge}}, \bibinfo {author} {\bibfnamefont {Y.}~\bibnamefont {Liu}}, \bibinfo
	 	{author} {\bibfnamefont {J.}~\bibnamefont {Li}}, \bibinfo {author}
	 	{\bibfnamefont {H.}~\bibnamefont {Li}}, \bibinfo {author} {\bibfnamefont
	 		{T.}~\bibnamefont {Luo}}, \bibinfo {author} {\bibfnamefont {Y.}~\bibnamefont
	 		{Wu}}, \bibinfo {author} {\bibfnamefont {Y.}~\bibnamefont {Xu}},\ and\
	 	\bibinfo {author} {\bibfnamefont {J.}~\bibnamefont {Wang}},\ }\bibfield
	 {title} {\bibinfo {title} {High-{C}hern-number and high-temperature quantum
	 		{H}all effect without {L}andau levels},\ }\href
	 {https://doi.org/10.1093/nsr/nwaa089} {\bibfield  {journal} {\bibinfo
	 		{journal} {Natl. Sci. Rev.}\ }\textbf {\bibinfo {volume} {7}},\ \bibinfo
	 	{pages} {1280} (\bibinfo {year} {2020})}\BibitemShut {NoStop}%
	 \bibitem [{\citenamefont {Chen}\ \emph {et~al.}(2020)\citenamefont {Chen},
	 	\citenamefont {Sharpe}, \citenamefont {Fox}, \citenamefont {Zhang},
	 	\citenamefont {Wang}, \citenamefont {Jiang}, \citenamefont {Lyu},
	 	\citenamefont {Li}, \citenamefont {Watanabe}, \citenamefont {Taniguchi},
	 	\citenamefont {Shi}, \citenamefont {Senthil}, \citenamefont
	 	{Goldhaber-Gordon}, \citenamefont {Zhang},\ and\ \citenamefont
	 	{Wang}}]{Chen2020}%
	 \BibitemOpen
	 \bibfield  {author} {\bibinfo {author} {\bibfnamefont {G.}~\bibnamefont
	 		{Chen}}, \bibinfo {author} {\bibfnamefont {A.~L.}\ \bibnamefont {Sharpe}},
	 	\bibinfo {author} {\bibfnamefont {E.~J.}\ \bibnamefont {Fox}}, \bibinfo
	 	{author} {\bibfnamefont {Y.-H.}\ \bibnamefont {Zhang}}, \bibinfo {author}
	 	{\bibfnamefont {S.}~\bibnamefont {Wang}}, \bibinfo {author} {\bibfnamefont
	 		{L.}~\bibnamefont {Jiang}}, \bibinfo {author} {\bibfnamefont
	 		{B.}~\bibnamefont {Lyu}}, \bibinfo {author} {\bibfnamefont {H.}~\bibnamefont
	 		{Li}}, \bibinfo {author} {\bibfnamefont {K.}~\bibnamefont {Watanabe}},
	 	\bibinfo {author} {\bibfnamefont {T.}~\bibnamefont {Taniguchi}}, \bibinfo
	 	{author} {\bibfnamefont {Z.}~\bibnamefont {Shi}}, \bibinfo {author}
	 	{\bibfnamefont {T.}~\bibnamefont {Senthil}}, \bibinfo {author} {\bibfnamefont
	 		{D.}~\bibnamefont {Goldhaber-Gordon}}, \bibinfo {author} {\bibfnamefont
	 		{Y.}~\bibnamefont {Zhang}},\ and\ \bibinfo {author} {\bibfnamefont
	 		{F.}~\bibnamefont {Wang}},\ }\bibfield  {title} {\bibinfo {title} {Tunable
	 		correlated {C}hern insulator and ferromagnetism in a moir{\'e}
	 		superlattice},\ }\href {https://doi.org/10.1038/s41586-020-2049-7} {\bibfield
	 	{journal} {\bibinfo  {journal} {Nature}\ }\textbf {\bibinfo {volume}
	 		{579}},\ \bibinfo {pages} {56} (\bibinfo {year} {2020})}\BibitemShut
	 {NoStop}%
	 \bibitem [{\citenamefont {Zhao}\ \emph {et~al.}(2020)\citenamefont {Zhao},
	 	\citenamefont {Zhang}, \citenamefont {Mei}, \citenamefont {Zhou},
	 	\citenamefont {Yi}, \citenamefont {Zhang}, \citenamefont {Yu}, \citenamefont
	 	{Xiao}, \citenamefont {Wang}, \citenamefont {Samarth}, \citenamefont {Chan},
	 	\citenamefont {Liu},\ and\ \citenamefont {Chang}}]{Zhao2020}%
	 \BibitemOpen
	 \bibfield  {author} {\bibinfo {author} {\bibfnamefont {Y.-F.}\ \bibnamefont
	 		{Zhao}}, \bibinfo {author} {\bibfnamefont {R.}~\bibnamefont {Zhang}},
	 	\bibinfo {author} {\bibfnamefont {R.}~\bibnamefont {Mei}}, \bibinfo {author}
	 	{\bibfnamefont {L.-J.}\ \bibnamefont {Zhou}}, \bibinfo {author}
	 	{\bibfnamefont {H.}~\bibnamefont {Yi}}, \bibinfo {author} {\bibfnamefont
	 		{Y.-Q.}\ \bibnamefont {Zhang}}, \bibinfo {author} {\bibfnamefont
	 		{J.}~\bibnamefont {Yu}}, \bibinfo {author} {\bibfnamefont {R.}~\bibnamefont
	 		{Xiao}}, \bibinfo {author} {\bibfnamefont {K.}~\bibnamefont {Wang}}, \bibinfo
	 	{author} {\bibfnamefont {N.}~\bibnamefont {Samarth}}, \bibinfo {author}
	 	{\bibfnamefont {M.~H.~W.}\ \bibnamefont {Chan}}, \bibinfo {author}
	 	{\bibfnamefont {C.-X.}\ \bibnamefont {Liu}},\ and\ \bibinfo {author}
	 	{\bibfnamefont {C.-Z.}\ \bibnamefont {Chang}},\ }\bibfield  {title} {\bibinfo
	 	{title} {Tuning the {C}hern number in quantum anomalous {H}all insulators},\
	 }\href {https://doi.org/10.1038/s41586-020-3020-3} {\bibfield  {journal}
	 	{\bibinfo  {journal} {Nature}\ }\textbf {\bibinfo {volume} {588}},\ \bibinfo
	 	{pages} {419} (\bibinfo {year} {2020})}\BibitemShut {NoStop}%
	 \bibitem [{\citenamefont {Zhu}\ \emph {et~al.}(2022)\citenamefont {Zhu},
	 	\citenamefont {Sun}, \citenamefont {Gong}, \citenamefont {Chen},
	 	\citenamefont {Zhang}, \citenamefont {Wu}, \citenamefont {Ye}, \citenamefont
	 	{Zha}, \citenamefont {Li}, \citenamefont {Guo}, \citenamefont {Qian},
	 	\citenamefont {Huang}, \citenamefont {Yu}, \citenamefont {Deng},
	 	\citenamefont {Rong}, \citenamefont {Lin}, \citenamefont {Xu}, \citenamefont
	 	{Sun}, \citenamefont {Guo}, \citenamefont {Li}, \citenamefont {Liang},
	 	\citenamefont {Peng}, \citenamefont {Fan}, \citenamefont {Zhu},\ and\
	 	\citenamefont {Pan}}]{Zhu2022}%
	 \BibitemOpen
	 \bibfield  {author} {\bibinfo {author} {\bibfnamefont {Q.}~\bibnamefont
	 		{Zhu}}, \bibinfo {author} {\bibfnamefont {Z.-H.}\ \bibnamefont {Sun}},
	 	\bibinfo {author} {\bibfnamefont {M.}~\bibnamefont {Gong}}, \bibinfo {author}
	 	{\bibfnamefont {F.}~\bibnamefont {Chen}}, \bibinfo {author} {\bibfnamefont
	 		{Y.-R.}\ \bibnamefont {Zhang}}, \bibinfo {author} {\bibfnamefont
	 		{Y.}~\bibnamefont {Wu}}, \bibinfo {author} {\bibfnamefont {Y.}~\bibnamefont
	 		{Ye}}, \bibinfo {author} {\bibfnamefont {C.}~\bibnamefont {Zha}}, \bibinfo
	 	{author} {\bibfnamefont {S.}~\bibnamefont {Li}}, \bibinfo {author}
	 	{\bibfnamefont {S.}~\bibnamefont {Guo}}, \bibinfo {author} {\bibfnamefont
	 		{H.}~\bibnamefont {Qian}}, \bibinfo {author} {\bibfnamefont {H.-L.}\
	 		\bibnamefont {Huang}}, \bibinfo {author} {\bibfnamefont {J.}~\bibnamefont
	 		{Yu}}, \bibinfo {author} {\bibfnamefont {H.}~\bibnamefont {Deng}}, \bibinfo
	 	{author} {\bibfnamefont {H.}~\bibnamefont {Rong}}, \bibinfo {author}
	 	{\bibfnamefont {J.}~\bibnamefont {Lin}}, \bibinfo {author} {\bibfnamefont
	 		{Y.}~\bibnamefont {Xu}}, \bibinfo {author} {\bibfnamefont {L.}~\bibnamefont
	 		{Sun}}, \bibinfo {author} {\bibfnamefont {C.}~\bibnamefont {Guo}}, \bibinfo
	 	{author} {\bibfnamefont {N.}~\bibnamefont {Li}}, \bibinfo {author}
	 	{\bibfnamefont {F.}~\bibnamefont {Liang}}, \bibinfo {author} {\bibfnamefont
	 		{C.-Z.}\ \bibnamefont {Peng}}, \bibinfo {author} {\bibfnamefont
	 		{H.}~\bibnamefont {Fan}}, \bibinfo {author} {\bibfnamefont {X.}~\bibnamefont
	 		{Zhu}},\ and\ \bibinfo {author} {\bibfnamefont {J.-W.}\ \bibnamefont {Pan}},\
	 }\bibfield  {title} {\bibinfo {title} {Observation of thermalization and
	 		information scrambling in a superconducting quantum processor},\ }\href
	 {https://doi.org/10.1103/PhysRevLett.128.160502} {\bibfield  {journal}
	 	{\bibinfo  {journal} {Phys. Rev. Lett.}\ }\textbf {\bibinfo {volume} {128}},\
	 	\bibinfo {pages} {160502} (\bibinfo {year} {2022})}\BibitemShut {NoStop}%
	 \bibitem [{\citenamefont {Kane}\ and\ \citenamefont {Mele}(2005)}]{Kane2005}%
	 \BibitemOpen
	 \bibfield  {author} {\bibinfo {author} {\bibfnamefont {C.~L.}\ \bibnamefont
	 		{Kane}}\ and\ \bibinfo {author} {\bibfnamefont {E.~J.}\ \bibnamefont
	 		{Mele}},\ }\bibfield  {title} {\bibinfo {title} {Quantum spin {H}all effect
	 		in graphene},\ }\href {https://doi.org/10.1103/PhysRevLett.95.226801}
	 {\bibfield  {journal} {\bibinfo  {journal} {Phys. Rev. Lett.}\ }\textbf
	 	{\bibinfo {volume} {95}},\ \bibinfo {pages} {226801} (\bibinfo {year}
	 	{2005})}\BibitemShut {NoStop}%
	 \bibitem [{\citenamefont {Li}\ \emph {et~al.}(2020)\citenamefont {Li},
	 	\citenamefont {Ge},\ and\ \citenamefont {Fan}}]{Li2020}%
	 \BibitemOpen
	 \bibfield  {author} {\bibinfo {author} {\bibfnamefont {S.-S.}\ \bibnamefont
	 		{Li}}, \bibinfo {author} {\bibfnamefont {Z.-Y.}\ \bibnamefont {Ge}},\ and\
	 	\bibinfo {author} {\bibfnamefont {H.}~\bibnamefont {Fan}},\ }\bibfield
	 {title} {\bibinfo {title} {Localization of rung pairs in a hard-core
	 		{B}ose-{H}ubbard ladder},\ }\href
	 {https://doi.org/10.1103/PhysRevA.102.062409} {\bibfield  {journal} {\bibinfo
	 		{journal} {Phys. Rev. A}\ }\textbf {\bibinfo {volume} {102}},\ \bibinfo
	 	{pages} {062409} (\bibinfo {year} {2020})}\BibitemShut {NoStop}%
\end{thebibliography}

\newpage

\begin{figure*}[t]
	\centering
	\includegraphics[width=0.97\textwidth]{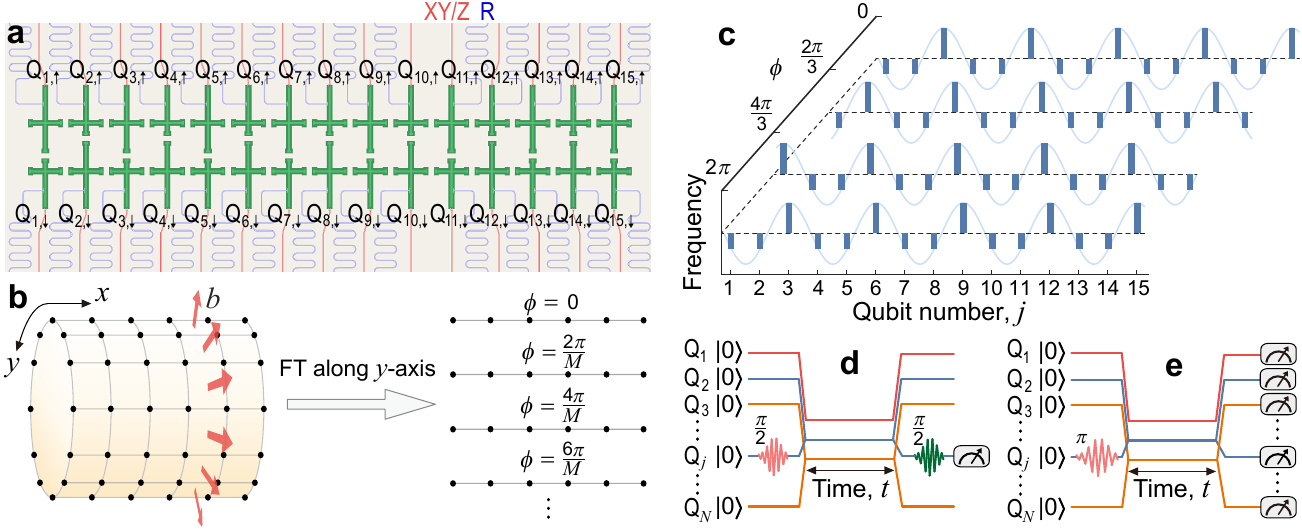}\\
	\caption{\textbf{Quantum simulation of Chern insulators on a 30-qubit-ladder superconducting
			processor.} \textbf{a}, Schematic of the superconducting quantum processor, where 30 superconducting
		qubits constitute a ladder. Each qubit, coupled to an independent readout resonator (R),
		has an independent microwave line for XY and Z controls.
		\textbf{b}, Mapping the 2D Hofstadter model to various configurations of Aubry-Andr\'{e}-Harper (AAH) chains
		with a Fourier transformation (FT) along the $y$-axis with $M$ sites. \textbf{c}, Qubits' frequencies for the
		synthesis of a series of AAH chains with different values of $\phi$ and $b=\frac{1}{3}$.
		\textbf{d}, \textbf{e}, Experimental waveform sequences for the dynamic band structure spectroscopy (\textbf{d})
		and the single-particle quantum walks (QWs) (\textbf{e}).}\label{fig:1}
\end{figure*}

%\clearpage

\begin{figure*}[t]
	\centering
	\includegraphics[width=0.92\textwidth]{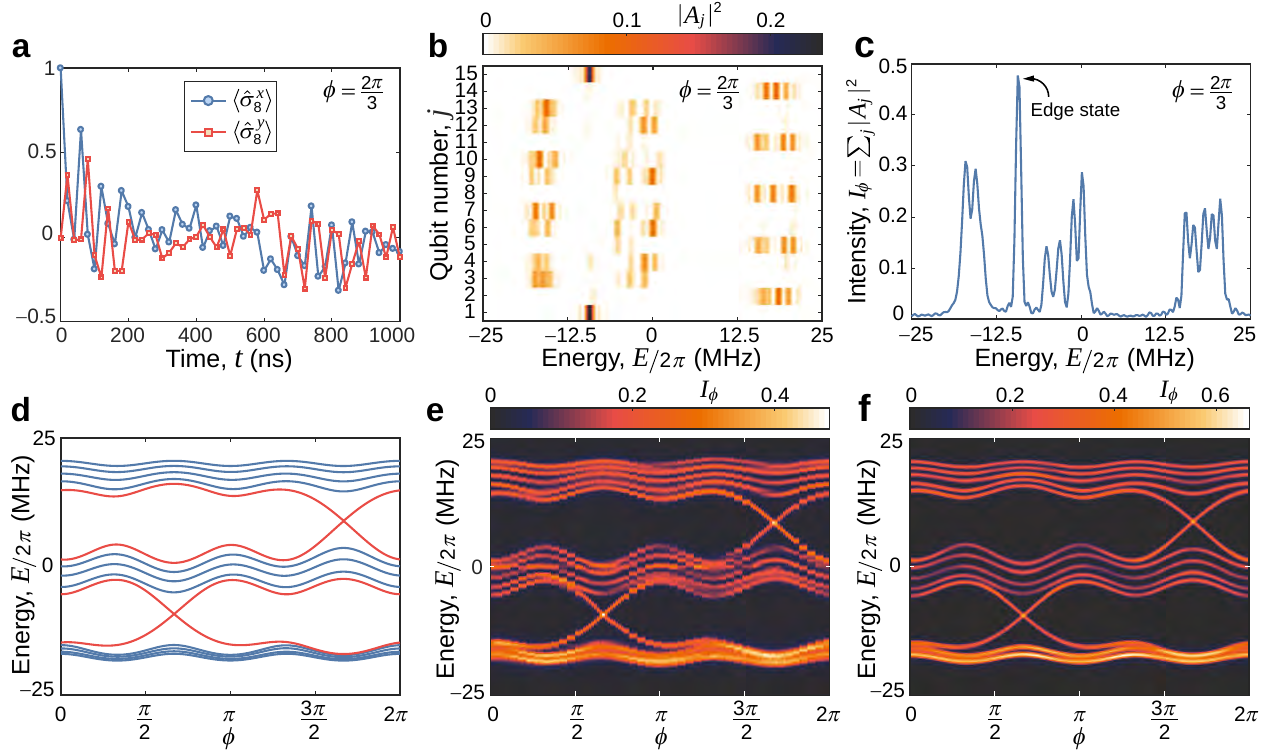}\\
	\caption{\textbf{Band structure spectroscopy of the 2D Chern insulator with a synthetic dimension.}
		\textbf{a}, Typical data of $\langle\hat{\sigma}^x\rangle$ and $\langle\hat{\sigma}^y\rangle$ versus time $t$ when choosing
		Q$_8$ as the target qubit for $b=\frac{1}{3}$, $\Delta/2\pi=12$~MHz and $\phi=\frac{2\pi}{3}$.
		\textbf{b}, Squared FT magnitudes $|A_j|^2$ of the response functions
		$\chi_j(t)\equiv\langle\hat{\sigma}_j^x(t)\rangle+i\langle\hat{\sigma}_j^y(t)\rangle$
		for all fifteen qubits.
		Note that the edge states at the edge qubits: Q$_1$ and Q$_{15}$.
		\textbf{c}, Summation of the squared FT magnitudes $I_\phi\equiv\sum_{j}|A_j|^2$.
		\textbf{d}, Band structure of the 2D Chern insulator for $b=\frac{1}{3}$ and $\Delta=12$~MHz
		with 15 sites along the $x$-direction and periodic along the $y$ direction. Red curves show the edge states.
		\textbf{e}, \textbf{f}, Experimentally measured data for $I_\phi$ (\textbf{e}) for different values of $\phi$ varied from 0 to $2\pi$ are compared with numerically calculated data of $I_\phi$ (\textbf{f})
		obtained by numerically simulating the dynamics of the
		15-qubit system without decoherence.
	}\label{fig:2}
\end{figure*}

\begin{figure*}[t]
	\centering
	\includegraphics[width=0.92\textwidth]{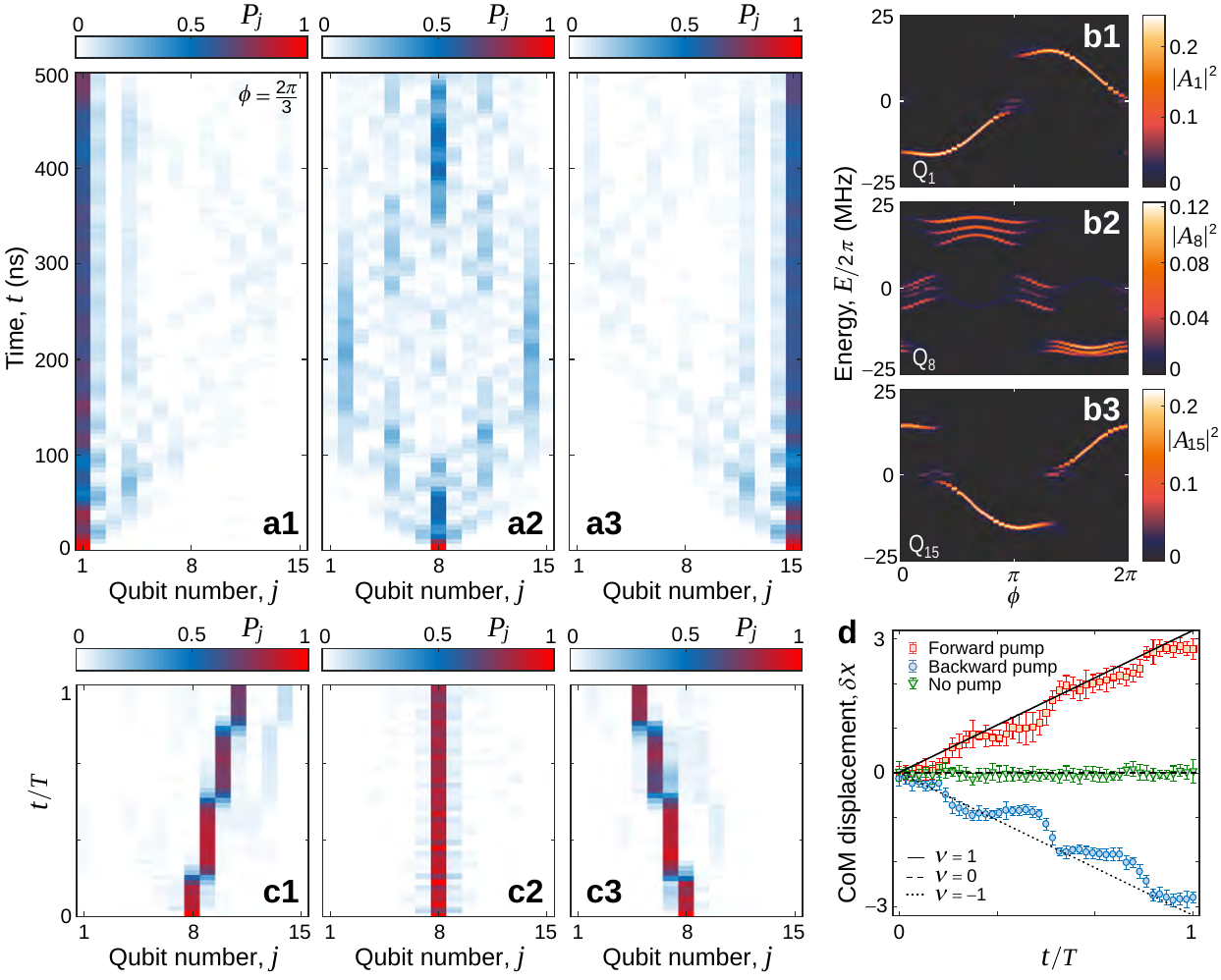}\\
	\caption{\textbf{Dynamical signatures of topological edge states and the topological charge pump.}
		\textbf{a1}--\textbf{a3}, Time evolution of the excitation probability $P_j$ with $b=\frac{1}{3}$,
		$\Delta/2\pi=12$~MHz and
		$\phi=2\pi/3$ after initially exciting the leftmost qubit Q$_1$ (\textbf{a1}), the central qubit Q$_8$ (\textbf{a2}),
		and the rightmost qubit Q$_{15}$ (\textbf{a3}).
		\textbf{b1}--\textbf{b3}, Experimental data of the squared FT magnitudes $|A_{1}|^2$, $|A_{8}|^2$, and $|A_{15}|^2$, when choosing Q$_1$ (\textbf{b1}), Q$_8$ (\textbf{b2}), and Q$_{15}$ (\textbf{b3}) as the target qubits.
		{\textbf{c1}--\textbf{c3}}, Time evolution of an excitation initially prepared at the central qubit Q$_8$ when it is forward pumped (\textbf{c1}),
		not pumped (\textbf{c2}) and backward pumped (\textbf{c3}), respectively,  with $\Delta/2\pi=36$~MHz for an initial $\phi_0=5\pi/3$.
		{\textbf{d}}, Displacement of the centre of mass (CoM) $\delta x$ versus time $t$ in one pumping cycle with period $T$ for the cases in (\textbf{c1}--\textbf{c3}). The error bars are 1 SD, calculated from all 10 groups of experimental results.
	}\label{fig:3}
\end{figure*}

\begin{figure*}[t]
	\centering
	\includegraphics[width=0.97\textwidth]{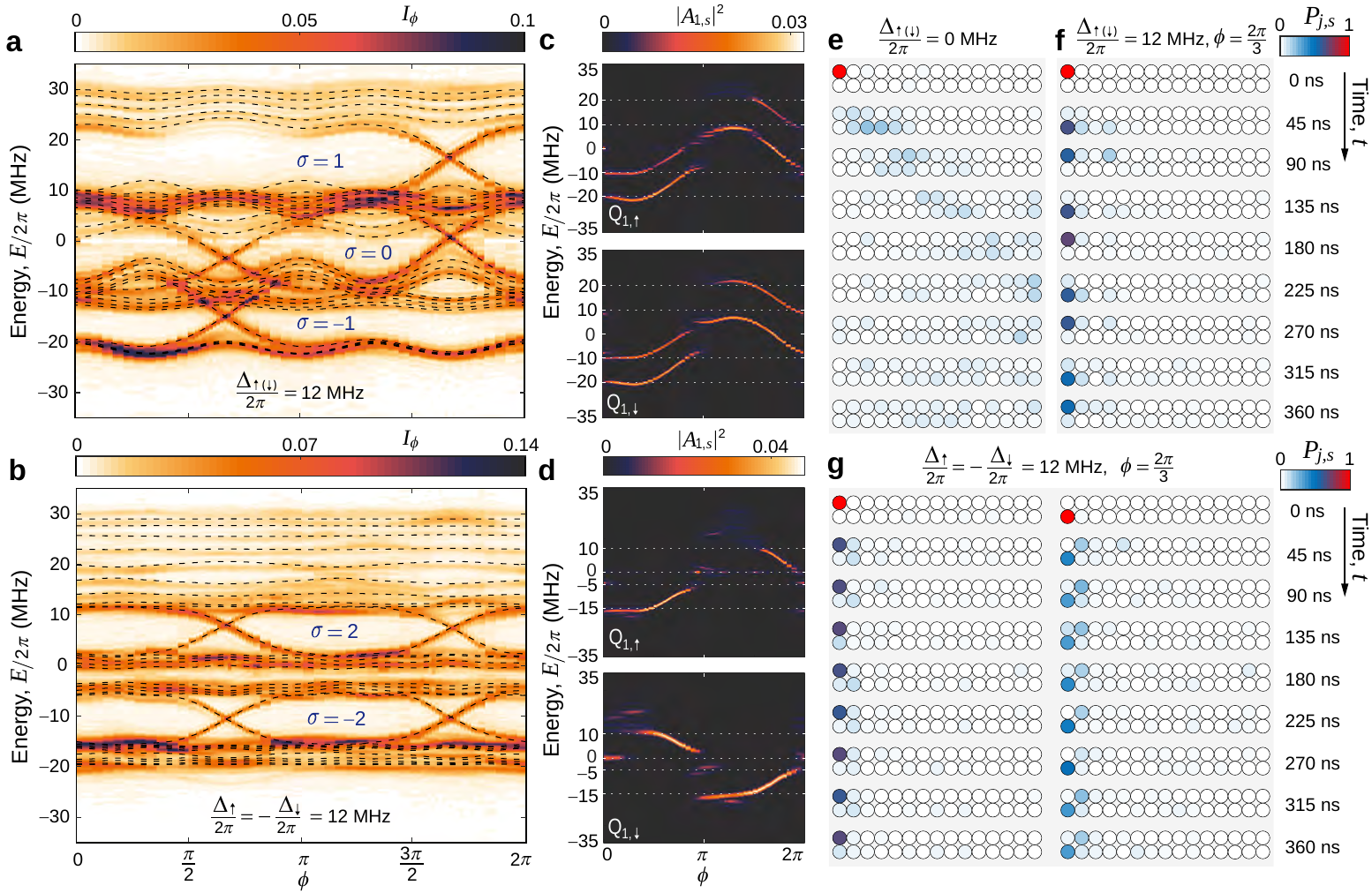}\\
	\caption{\textbf{Quantum simulation of bilayer Chern insulators  %quantum topological systems
			using all thirty qubits on a ladder-type quantum processor.}
		\textbf{a},\textbf{b}, Experimentally measured $I_\phi$ with $b=\frac{1}{3}$  versus $\phi$ varied from 0 to $2\pi$
		for two AAH chains with the same on-site potentials $\Delta_{\uparrow(\downarrow)}/2\pi=12$~MHz (\textbf{a})
		and opposite on-site potentials $\Delta_{\uparrow}/2\pi=-\Delta_{\downarrow}/2\pi=12$~MHz (\textbf{b}), respectively,
		which are compared with the theoretical projected band structures (dashed curves).
		The  Hall conductivity is defined as the summation of the Chern number $\mathcal{C}_n$
		over the occupied bands: $\sigma=\sum_{n} \mathcal{C}_n$ by letting $e^2/h=1$.
		\textbf{c}, \textbf{d}, Experimental data of the squared FT magnitudes $|A_{1,\uparrow}|^2$ and $|A_{1,\downarrow}|^2$ when choosing Q$_{1,\uparrow}$ and Q$_{1,\downarrow}$ as the target qubits for $\Delta_{\uparrow(\downarrow)}/2\pi=12$~MHz (\textbf{c})
		and $\Delta_{\uparrow}/2\pi=-\Delta_{\downarrow}/2\pi=12$~MHz (\textbf{d}), respectively.
		\textbf{e}, \textbf{g}, Time evolutions of the excitation probability $P_{j,s}$, with
		$\phi=\frac{2\pi}{3}$, after initially exciting a corner qubit (Q$_{1,\uparrow}$ or Q$_{1,\downarrow}$)
		for $\Delta_{\uparrow(\downarrow)}/2\pi=0$~MHz (\textbf{e}),
		$\Delta_{\uparrow(\downarrow)}/2\pi=12$~MHz (\textbf{f}), and $\Delta_{\uparrow}/2\pi=-\Delta_{\downarrow}/2\pi=12$~MHz (\textbf{g}), respectively.
		Animations of the time evolutions are available in Supplementary Movie~1.}\label{fig:4}
\end{figure*}

\begin{figure*}[t]
	\centering
	\includegraphics[width=0.97\textwidth]{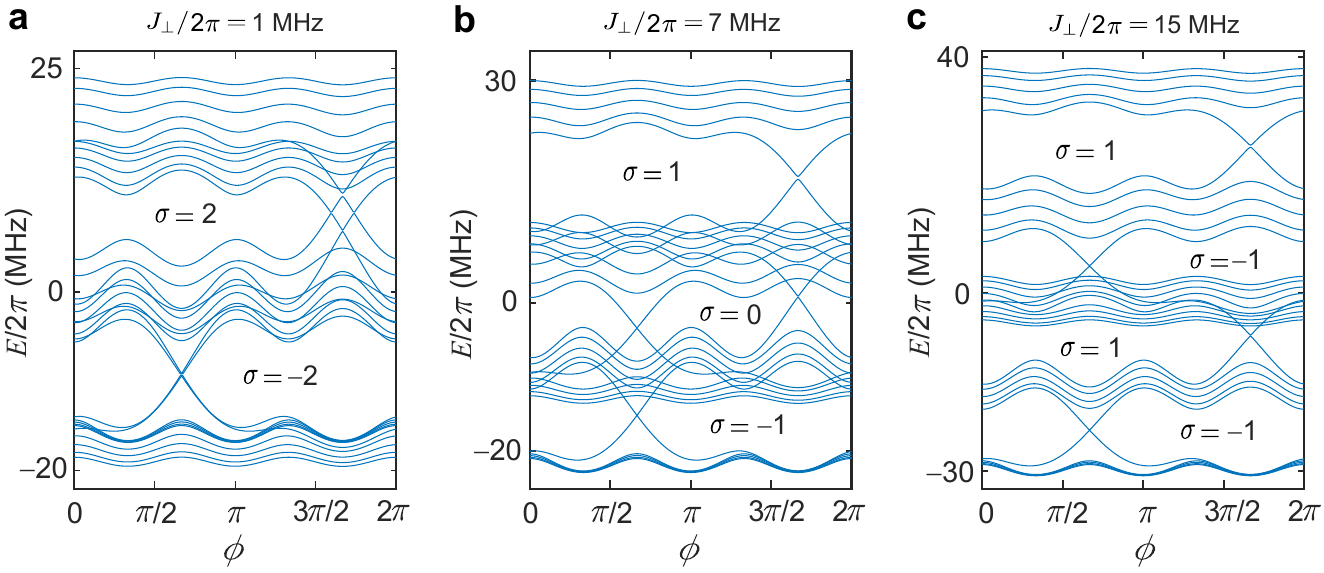}
	\caption{\textbf{Energy spectra for the bilayer topological system with the same periodically modulated on-site potentials.} \textbf{a}--\textbf{c}, Energy spectra versus $\phi$ of a finite ladder with $30$ sites for the same modulated amplitudes $\Delta_\uparrow/2\pi=\Delta_\downarrow/2\pi = 12$~MHz with $J_\perp/2\pi=1$~MHz (\textbf{a}),
		$J_\perp/2\pi=7$~MHz (\textbf{b}), and $J_\perp/2\pi=15$~MHz (\textbf{c}).
		Other parameters used are $J_{\parallel}/2\pi=8$~MHz, $J'_{\parallel}=0.1 J_{\parallel}$, and
		$J_{\times}=0.2 J_{\parallel}$.
		\vspace{15cm}}\label{fig1chain}
\end{figure*}

\begin{figure}[!h]
	\centering
	\includegraphics[width=0.5\textwidth]{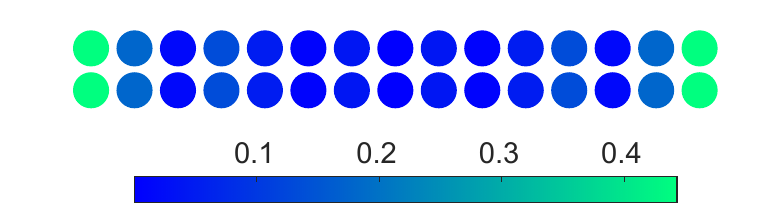}
	\caption{\textbf{Mid-gap state for the bilayer topological system with the same periodically modulated on-site potentials.}  Density distribution of a mid-gap state for $\Delta_\uparrow/2\pi=\Delta_\downarrow/2\pi = 12$~MHz and $\phi=\frac{2\pi}{3}$ with $J_\perp/2\pi=7$MHz, $J_{\parallel}/2\pi=8$~MHz, $J'_{\parallel}=0.1 J_{\parallel}$, and
		$J_{\times}=0.2 J_{\parallel}$.
		\vspace{15cm} }\label{intensity}
\end{figure}

\begin{figure*}[t]
	\centering
	\includegraphics[width=0.7\textwidth]{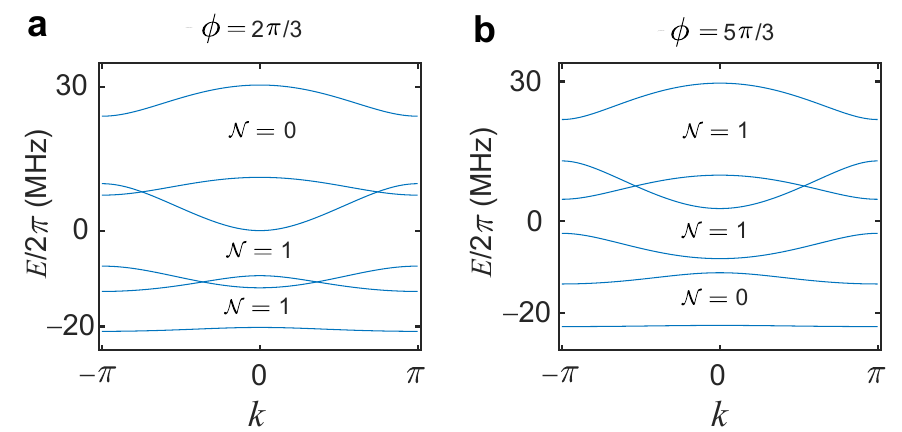}
	\caption{\textbf{Bloch energy bands for the bilayer topological system with the same periodically modulated on-site potentials.}
		\textbf{a}, \textbf{b}, Bloch energy bands versus $k$ for identical chains in a qubit ladder, with modulated amplitudes $\Delta_\uparrow/2\pi=\Delta_\downarrow/2\pi = 12$~MHz with $\phi=\frac{2\pi}{3}$ (\textbf{a}),
		and $\phi=\frac{5\pi}{3}$ (\textbf{b}). 	The topological invariant $\mathcal{N}$ at each band gap is indicated, and $\mathcal{N}=1$ indicates a pair of edge states.
		Other parameters used are $J_\perp/2\pi=7$~MHz, $J_{\parallel}/2\pi=8$~MHz, $J'_{\parallel}=0.1 J_{\parallel}$, and
		$J_{\times}=0.2 J_{\parallel}$.
		\vspace{15cm}}\label{bloch}
\end{figure*}

\begin{figure*}[t]
	\centering
	\includegraphics[width=0.97\textwidth]{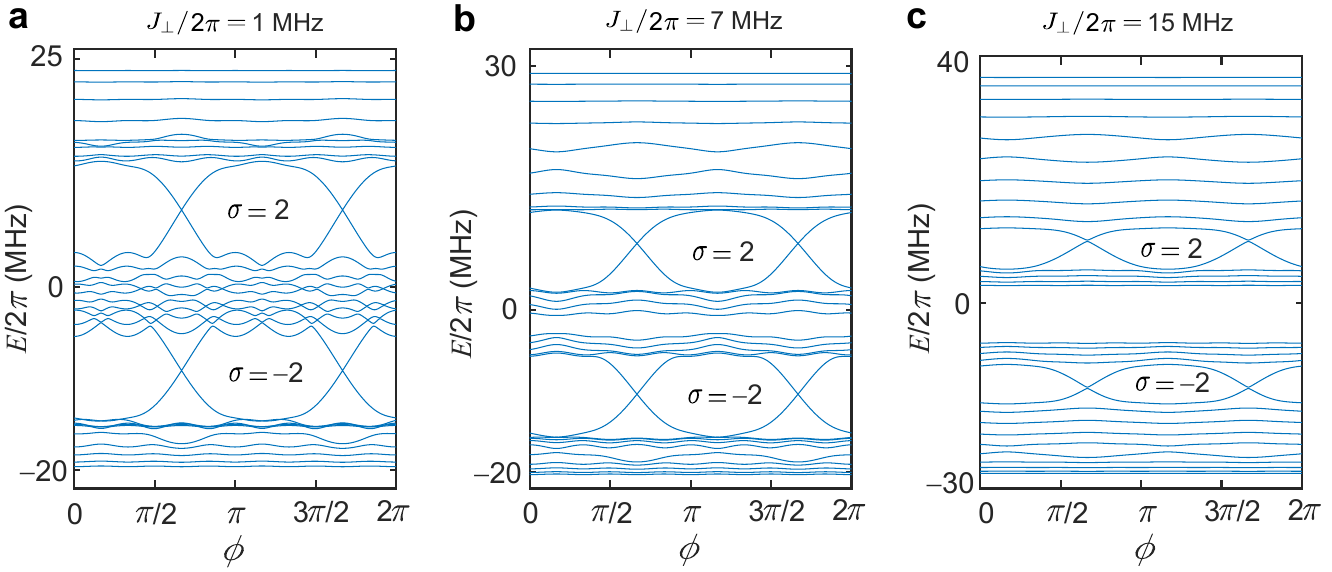}
	\caption{\textbf{Energy spectra for the bilayer topological system with opposite periodically modulated on-site potentials.} \textbf{a}--\textbf{c},
		Energy spectra versus $\phi$ of a finite ladder with $30$ sites for opposite modulated amplitudes $\Delta_\uparrow/2\pi=-\Delta_\downarrow/2\pi = 12$~MHz with $J_\perp/2\pi=1$~MHz (\textbf{a}), $J_\perp/2\pi=7$~MHz (\textbf{b}), and $J_\perp/2\pi=15$~MHz (\textbf{c}).
		Other parameters used are $J_{\parallel}/2\pi=8$~MHz, $J'_{\parallel}=0.1 J_{\parallel}$, and
		$J_{\times}=0.2 J_{\parallel}$.
		\vspace{15cm}}\label{fig2chain}
\end{figure*}

\begin{figure*}[t]
	\centering
	\includegraphics[width=0.97\textwidth]{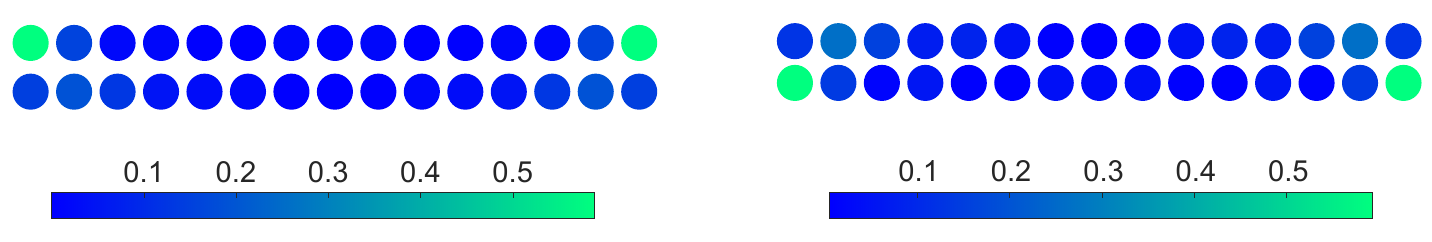}
	\caption{\textbf{Mid-gap states for the bilayer topological system with opposite periodically modulated on-site potentials.}
		Density distribution of mid-gap states for $\Delta_\uparrow/2\pi=-\Delta_\downarrow/2\pi = 12$~MHz and $\phi=\frac{2\pi}{3}$ with $J_\perp/2\pi=7$~MHz, $J_{\parallel}/2\pi=8$~MHz, $J'_{\parallel}=0.1 J_{\parallel}$, and
		$J_{\times}=0.2 J_{\parallel}$. The left panel is for a mid-gap state in the second gap; and
		the right panel is for a mid-gap state in the fourth gap. Note that the first and  fifth gaps in Fig.~\ref{fig2chain}\textbf{b} are not clearly shown due to finite-size effects.}\label{intensity2}
\end{figure*}

\end{document}

\title{Supplementary Information for:\\
Simulating Chern insulators on a superconducting quantum processor}

%\affiliation{Institute of Physics, Chinese Academy of Sciences, Beijing 100190, China}
%\affiliation{School of Physics and Optoelectronics, South China University of Technology, Guangzhou 510640, China}
%\affiliation{Theoretical Quantum Physics Laboratory, RIKEN Cluster for Pioneering Research,
%Wako-shi, Saitama 351-0198, Japan}
%\affiliation{RIKEN Centre for Quantum Computing (RQC),  Wako-shi, Saitama 351-0198, Japan}
%\affiliation{Beijing Academy of Quantum Information Sciences, Beijing 100193, China}
%\affiliation{Key Laboratory of Weak Light Nonlinear Photonics, Ministry of Education, Teda Applied Physics Institute and School of Physics, Nankai University, 300457 Tianjin, China}
%\affiliation{CAS Centre for Excellence in Topological Quantum Computation,
%UCAS, Beijing 100190, China}
%\affiliation{Songshan Lake Materials Laboratory, Dongguan 523808, China}
%\affiliation{Physics Department, University of Michigan, Ann Arbor, Michigan 48109-1040, USA}

\author{Zhong-Cheng Xiang}
\thanks{These authors contributed equally to this work.}
\affiliation{Institute of Physics, Chinese Academy of Sciences, Beijing 100190, China}

\author{Kaixuan Huang}
\thanks{These authors contributed equally to this work.}
\affiliation{Beijing Academy of Quantum Information Sciences, Beijing 100193, China}
\affiliation{Institute of Physics, Chinese Academy of Sciences, Beijing 100190, China}
\affiliation{Key Laboratory of Weak Light Nonlinear Photonics, Ministry of Education, Teda Applied Physics Institute and School of Physics, Nankai University, Tianjin 300457, China}

\author{Yu-Ran Zhang}
\thanks{These authors contributed equally to this work.}
\affiliation{School of Physics and Optoelectronics, South China University of Technology, Guangzhou 510640, China}
\affiliation{Theoretical Quantum Physics Laboratory, Cluster for Pioneering Research, RIKEN, Wako-shi, Saitama 351-0198, Japan}
\affiliation{Center for Quantum Computing, RIKEN, Wako-shi, Saitama 351-0198, Japan}

\author{Tao Liu}
\affiliation{School of Physics and Optoelectronics, South China University of Technology, Guangzhou 510640, China}

  \author{Yun-Hao Shi}
  \affiliation{Institute of Physics, Chinese Academy of Sciences, Beijing 100190, China}

  \author{Cheng-Lin Deng}
  \affiliation{Institute of Physics, Chinese Academy of Sciences, Beijing 100190, China}
 
  \author{Tong Liu}
    \affiliation{Institute of Physics, Chinese Academy of Sciences, Beijing 100190, China}

    \author{Hao Li}
  \affiliation{Institute of Physics, Chinese Academy of Sciences, Beijing 100190, China}

  \author{Gui-Han Liang}
  \affiliation{Institute of Physics, Chinese Academy of Sciences, Beijing 100190, China}

    \author{Zheng-Yang Mei}
  \affiliation{Institute of Physics, Chinese Academy of Sciences, Beijing 100190, China}

  \author{Haifeng Yu}
  \affiliation{Beijing Academy of Quantum Information Sciences, Beijing 100193, China}

  \author{Guangming Xue}
  \affiliation{Beijing Academy of Quantum Information Sciences, Beijing 100193, China}

   \author{Ye Tian}
  \affiliation{Institute of Physics, Chinese Academy of Sciences, Beijing 100190, China}

  \author{Xiaohui Song}
  \affiliation{Institute of Physics, Chinese Academy of Sciences, Beijing 100190, China}

  \author{Zhi-Bo Liu}
\affiliation{Key Laboratory of Weak Light Nonlinear Photonics, Ministry of Education, Teda Applied Physics Institute and School of Physics, Nankai University, Tianjin 300457, China}

    \author{Kai Xu}
        \email{kaixu@iphy.ac.cn}
    \affiliation{Institute of Physics, Chinese Academy of Sciences, Beijing 100190, China}
    \affiliation{Beijing Academy of Quantum Information Sciences, Beijing 100193, China}
    \affiliation{CAS Centre for Excellence in Topological Quantum Computation,
UCAS, Beijing 100190, China}
    \affiliation{Songshan Lake Materials Laboratory, Dongguan 523808, China}

  \author{Dongning Zheng}
  \affiliation{Institute of Physics, Chinese Academy of Sciences, Beijing 100190, China}
  \affiliation{CAS Centre for Excellence in Topological Quantum Computation,
UCAS, Beijing 100190, China}
  \affiliation{Songshan Lake Materials Laboratory, Dongguan 523808, China}

\author{Franco Nori}
\email{fnori@riken.jp}
\affiliation{Theoretical Quantum Physics Laboratory, Cluster for Pioneering Research, RIKEN, Wako-shi,
Saitama 351-0198, Japan}
\affiliation{Center for Quantum Computing, RIKEN, Wako-shi, Saitama 351-0198, Japan}
\affiliation{Physics Department, University of Michigan, Ann Arbor, Michigan 48109-1040, USA}

\author{Heng Fan}
\email{hfan@iphy.ac.cn}
\affiliation{Institute of Physics, Chinese Academy of Sciences, Beijing 100190, China}
\affiliation{Beijing Academy of Quantum Information Sciences, Beijing 100193, China}
\affiliation{CAS Centre for Excellence in Topological Quantum Computation, UCAS, Beijing 100190, China}
\affiliation{Songshan Lake Materials Laboratory, Dongguan 523808, China}

\maketitle

\setcounter{section}{0}
\renewcommand{\thesection}{Supplementary Note \arabic{section}}
\renewcommand{\figurename}{Supplementary Fig.}
\setcounter{figure}{0}  % reset counter

\renewcommand{\tablename}{Supplementary Table}
\setcounter{table}{0}
\renewcommand {\thetable} {\arabic{table}}%

\tableofcontents

\section{Experimental setup}

\begin{figure*}[t]
	\begin{center}
		\includegraphics[width=0.85\textwidth,clip=True]{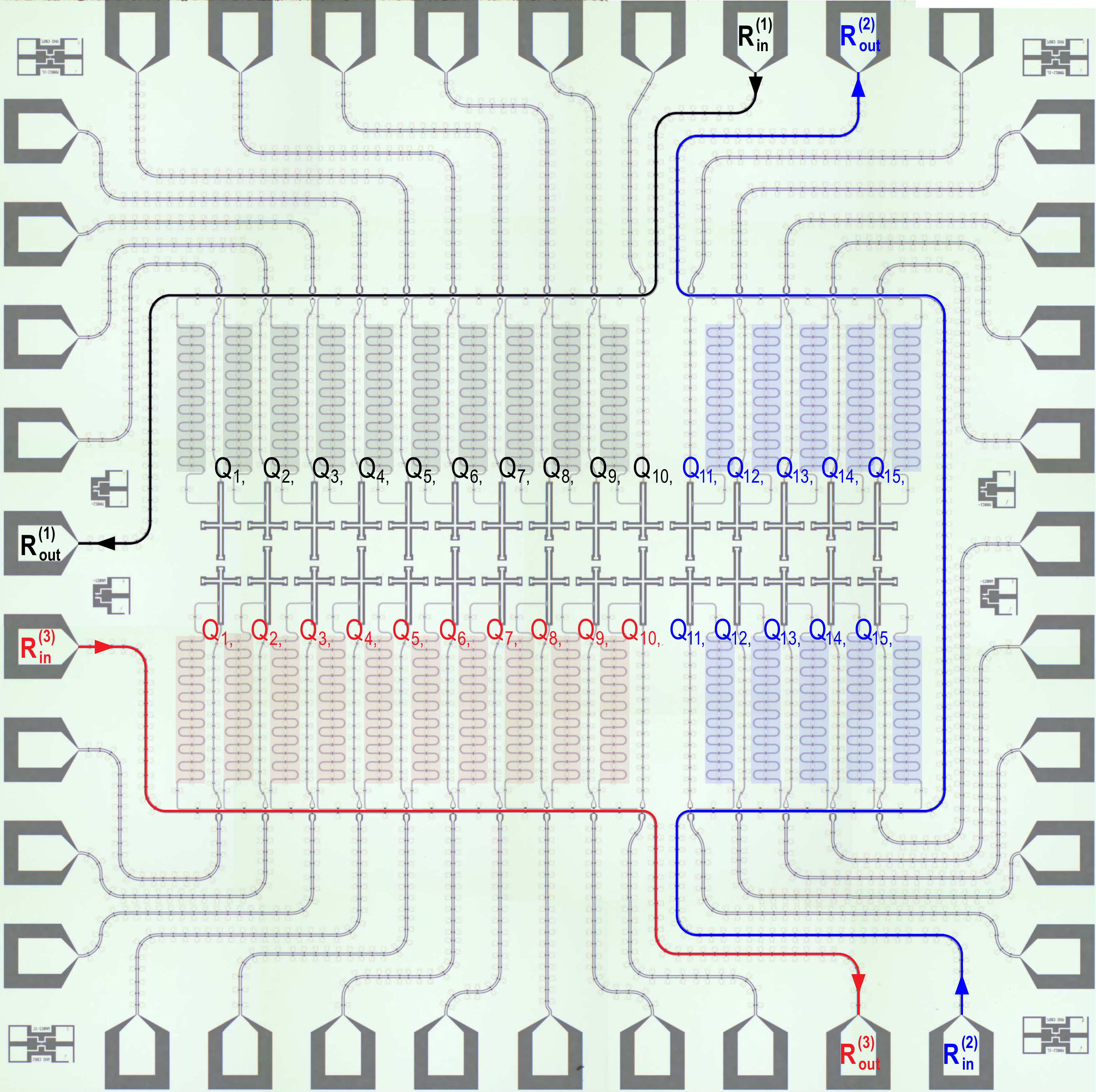}
		\caption{Optical micrograph of the thirty-qubit sample.
Thirty superconducting qubits constitute a ladder. Each qubit has an independent
readout resonator and an independent microwave line for XY and Z controls.
Thirty readout resonators are separated into three groups,
and ten resonators in each group share
a microwave transmission line (highlighted in black, blue or red) for measurements.
}
		\label{fig:device}
	\end{center}
\end{figure*}

\begin{figure*}[t]
	\begin{center}
		\includegraphics[width=0.8\textwidth,clip=True]{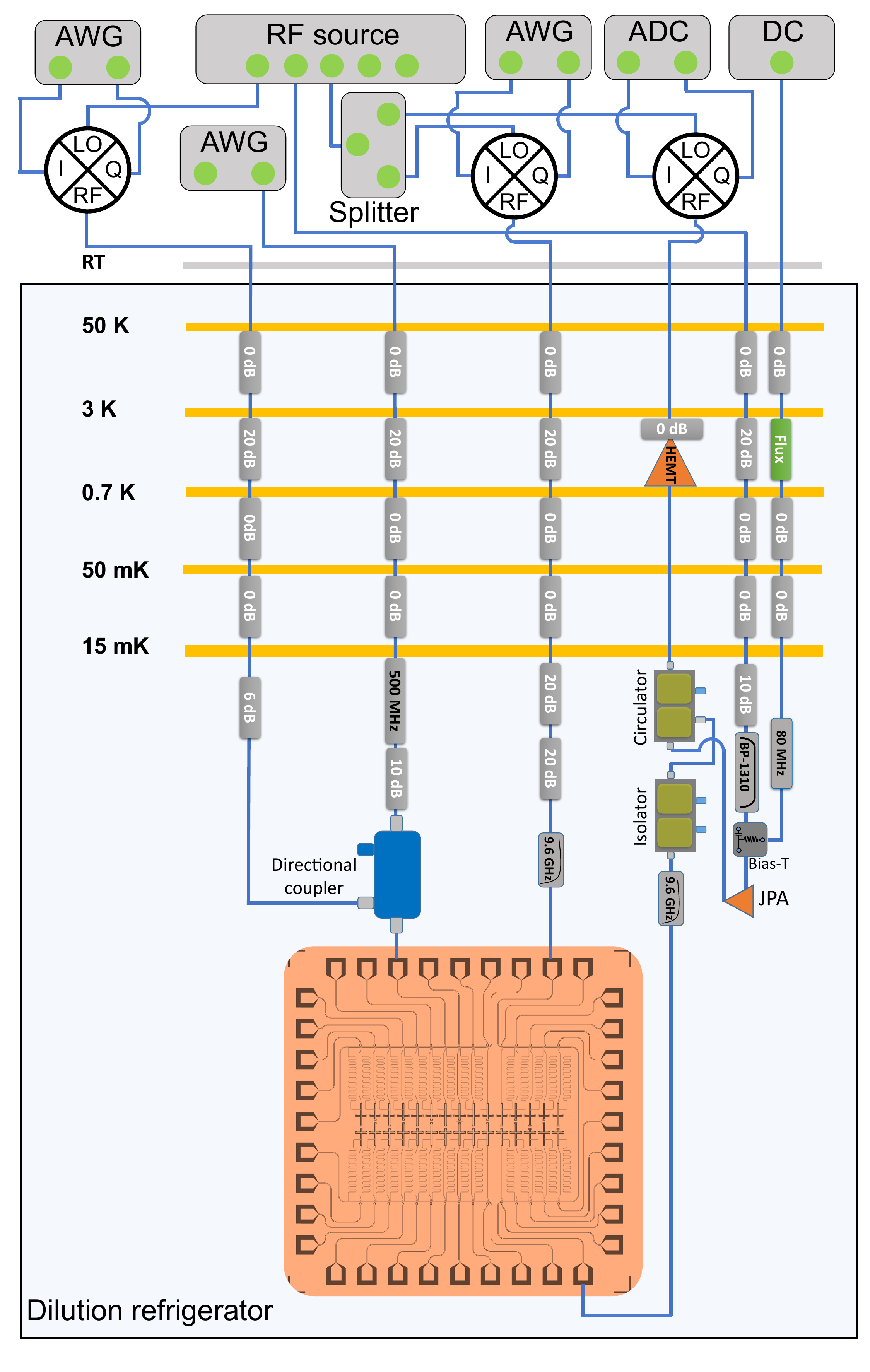}
		\caption{Experimental setup and low-temperature wiring.}
		\label{fig:wiring}
	\end{center}
\end{figure*}

\subsection{Fabrication}
Our experiments are performed on a superconducting circuit consisting of 30 transmon qubits
 (Q$_{j,s}$, with $j$ varied from 1 to 15 and pseudo-spin $s=\{\uparrow,\downarrow\}$),
 which constitute a two-legged qubit ladder,
as shown in Supplementary Fig.~\ref{fig:device}.
Each qubit, coupled to an independent readout resonator (R), has an independent microwave line for XY and Z controls.
This ladder-type 30-superconducting-qubit device is fabricated on a 430~$\mu$m thick sapphire
chip with standard wafer cleaning.

First, a 100~nm thick layer of  Al is deposed on a $10 \times 10$~mm$^2$ sapphire substrate, which is patterned
with optical lithography using a 0.70~$\mu$m thick layer of positive SPR955 resist.
Then, we apply the wet etching
method to produce the large components of the superconducting chip, such as the microwave coplanar waveguide resonators,
the transmission lines, the control lines, and the capacitors of the transmon qubits.

The second step is the preparation process for Josephson junctions. After patterning a bilayer of MMA and PMMA
resists with electron beam lithography,  the Josephson junctions are made using double-angle
evaporations, which include a 65~nm thick Al layer at $+60^\circ$ followed by an oxidation process in pure oxygen for several minutes and then a 100~nm thick second layer of Al at $0^\circ$. Finally, in order to suppress
the parasitic modes, several airbridges \cite{Chen2014} are constructed on the chip.
\begin{table*}[t]
	\begin{tabular}
    {c|ccccccccc}
		\hline
		\hline
		& ~~~$\omega_{j,\uparrow}^{\textrm{idle}}/2\pi$~~~ &  ~~~$\omega_{j,\uparrow}^{\textrm{max}}/2\pi$~~~
    & ~~~~$\bar{T}^1_{j,\uparrow}$~~~~  %&  %~~~~$T_{j,\uparrow}^{2*}$~~~~
    &  ~~~$\omega^{\textrm{R}}_j/2\pi$~~~
    &  ~~~$g_{{j}}/2\pi$~~~  & ~~~$\eta_j/2\pi$~~~ &  ~~~~$F^0_{j,\uparrow}$~~~~  & ~~~~$F^1_{j,\uparrow}$~~~~ \\
    &(GHz)  & (GHz)  &($\mu s$)%&($\mu$s)
    &(GHz)& (MHz)& (GHz)&&\\
		\hline
      ~~~Q$_{1,\uparrow}$~~~ & 4.410 & 4.879 & $\sim$30 %& $\sim$2.3
      & 6.623 & 35.82 & 0.203 & 0.975 & 0.937\\
      Q$_{2,\uparrow}$ & 5.290 & 5.384 & $\sim$27 %& $\sim$11.6
      & 6.640 & 37.57 & 0.252 & 0.977 & 0.916\\
        Q$_{3,\uparrow}$ & 4.517 & 4.921 & $\sim$32 %& $\sim$14.5
        & 6.659 & 33.85 & 0.202 & 0.975 & 0.920\\
      Q$_{4,\uparrow}$ & 5.370 & 5.387 & $\sim$30 %& $\sim$7.7
      & 6.677 & 36.25 & 0.246 & 0.978 & 0.921\\
        Q$_{5,\uparrow}$ & 4.360 & 4.910 & $\sim$41 %& $\sim$10.8
        & 6.703 & 35.41 & 0.202 & 0.977 & 0.927\\
      Q$_{6,\uparrow}$ & 5.230 & 5.458 & $\sim$30  %& $\sim$7.0
      & 6.720 & 41.83 & 0.250 & 0.974 & 0.932\\
        Q$_{7,\uparrow}$ & 4.499 & 4.821 & $\sim$40 %& $\sim$7.7
        & 6.739 & 36.30 & 0.204 & 0.961 & 0.897\\
      Q$_{8,\uparrow}$ & 5.320 & 5.541 & $\sim$33 %& $\sim$5.8
      & 6.758 & 34.07 & 0.247 & 0.963 & 0.916\\
        Q$_{9,\uparrow}$ & 4.430 & 5.027 & $\sim$39 %& $\sim$3.6
        & 6.777 & 33.82 & 0.200 & 0.955 & 0.912\\
      Q$_{10,\uparrow}$ & 5.150 & 5.447 & $\sim$37 %& $\sim$13.3
      & 6.803 & 38.06 & 0.246 & 0.983 & 0.885\\
        Q$_{11,\uparrow}$ & 4.480 & 4.989 & $\sim$38 %& $\sim$11.3
        & 6.618 & 35.80 & 0.200 & 0.967 & 0.901\\
      Q$_{12,\uparrow}$ & 5.350 & 5.420 & $\sim$30 %& $\sim$3.1
      & 6.635 & 39.11 & 0.251 & 0.969 & 0.910\\
        Q$_{13,\uparrow}$ & 4.335 & 5.046 & $\sim$39 %& $\sim$3.5
        & 6.656 & 31.94 & 0.202 & 0.971 & 0.883\\
        Q$_{14,\uparrow}$ & 5.260 & 5.462 & $\sim$30 %& $\sim$5.3
        & 6.671 & 35.10 & 0.244 & 0.977 & 0.895\\
        Q$_{15,\uparrow}$ & 4.450 & 5.070 & $\sim$31 %& $\sim$12.5
        & 6.695 & 33.90 & 0.202 & 0.979 & 0.886\\
		\hline
		\hline
	\end{tabular}
  \centering
  	\caption{Qubit characteristics for the 15-qubit experiment using 15 qubits
    (from Q$_{1,\uparrow}$ to Q$_{15,\uparrow}$) on the same leg.
    $\omega_{j,\uparrow}^{\textrm{idle}}/2\pi$
    is the idle frequency of the qubit, at which the single-qubit gates are performed.
    $\bar{T}^1_{j,\uparrow}$ denotes the average energy relaxation time from the frequency
    about 4.0~GHz to the maximum frequency $\omega_{j,\uparrow}^{\textrm{max}}/2\pi$ of Q$_j$.
    Also, $\omega^{\textrm{R}}_{j,\uparrow}/2\pi$ is
    the readout resonator's frequency of Q$_{j,\uparrow}$ with $g_{{j,\uparrow}}/2\pi$ being
    the coupling strength between the qubit and the readout resonator. The anharmonicity
    $\eta_{j,\uparrow}/2\pi$ of Q$_{j,\uparrow}$  is defined by
    $\eta_{j,\uparrow}\equiv\omega^{10}_{j,\uparrow} - \omega^{21}_{j,\uparrow}$, with
    $\omega^{10}_{j,\uparrow}/2\pi$ ($\omega^{21}_{j,\uparrow}/2\pi$) being the energy difference
    between the $|1\rangle$ and $|0\rangle$ states ($|2\rangle$ and $|1\rangle$ states).
    $F_{j,\uparrow}^{0}$ ($F_{j,\uparrow}^{1}$) is the measurement
    probability of $|0\rangle$ ($|1\rangle$) when the qubit is prepared at $|0\rangle$ ($|1\rangle$),
    which is used to mitigate the readout errors.}\label{tab1}
\end{table*}

\begin{table*}[t]
	\begin{tabular}
    {c|ccccccccc}
		\hline
		\hline
    & ~~~$\omega_{j,s}^{\textrm{idle}}/2\pi$~~~ &  ~~~$\omega_{j,s}^{\textrm{max}}/2\pi$~~~
    & ~~~~$\bar{T}^1_{{j,s}}$~~~~  &  ~~~~$T_{{j,s}}^{2*}$~~~~   &  ~~~$\omega^{\textrm{R}}_{j,s}/2\pi$~~~
    &  ~~~$g_{{j,s}}/2\pi$~~~  & ~~~$\eta_{j,s}/2\pi$~~~ &  ~~~~$F^0_{j,s}$~~~~  & ~~~~$F^1_{j,s}$~~~~ \\
    &(GHz)  & (GHz)  &($\mu s$)&($\mu$s)&(GHz)& (MHz)& (GHz)&&\\
		\hline
      ~~~Q$_{1,\uparrow}$~~~& 4.050 & 4.879 & $\sim$30 & $\sim$1.2 & 6.623 & 35.82 & 0.203 & 0.891 & 0.902\\
  		Q$_{2,\uparrow}$ & 5.350 & 5.384 & $\sim$27 & $\sim$3.7 & 6.640 & 37.57 & 0.252 & 0.973 & 0.886\\
  	    Q$_{3,\uparrow}$ & 4.090 & 4.921 & $\sim$32 & $\sim$0.9 & 6.659 & 33.85 & 0.202 & 0.924 & 0.883\\
  		Q$_{4,\uparrow}$ & 5.385 & 5.387 & $\sim$30 & $\sim$4.0 & 6.677 & 36.25 & 0.246 & 0.965 & 0.891\\
  	    Q$_{5,\uparrow}$ & 4.460 & 4.910 & $\sim$41 & $\sim$4.6 & 6.703 & 35.41 & 0.202 & 0.947 & 0.870\\
  		Q$_{6,\uparrow}$ & 5.450 & 5.458 & $\sim$30 & $\sim$4.6 & 6.720 & 41.83 & 0.250 & 0.967 & 0.894\\
  	    Q$_{7,\uparrow}$ & 4.065 & 4.821 & $\sim$40 & $\sim$3.1 & 6.739 & 36.30 & 0.204 & 0.911 & 0.877\\
  		Q$_{8,\uparrow}$ & 5.410 & 5.541 & $\sim$33 & $\sim$3.3 & 6.758 & 34.07 & 0.247 & 0.973 & 0.884\\
  	    Q$_{9,\uparrow}$ & 4.175 & 5.027 & $\sim$39 & $\sim$2.7 & 6.777 & 33.82 & 0.200 & 0.931 & 0.842\\
  	Q$_{10,\uparrow}$ & 5.140 & 5.447 & $\sim$37 & $\sim$4.4 & 6.803 & 38.06 & 0.246 & 0.966 & 0.900\\
  	    Q$_{11,\uparrow}$ & 4.405 & 4.989 & $\sim$38 & $\sim$1.3 & 6.618 & 35.80 & 0.200 & 0.908 & 0.853\\
  		Q$_{12,\uparrow}$ & 5.310 & 5.420 & $\sim$30 & $\sim$3.3 & 6.635 & 39.11 & 0.251 & 0.981 & 0.893\\
  	    Q$_{13,\uparrow}$ & 4.100 & 5.046 & $\sim$39 & $\sim$2.4 & 6.656 & 31.94 & 0.202 & 0.925 & 0.880\\
  	    Q$_{14,\uparrow}$ & 5.180 & 5.462 & $\sim$30 & $\sim$2.9 & 6.671 & 35.10 & 0.244 & 0.974 & 0.894\\
  	    Q$_{15,\uparrow}$ & 4.530 & 5.070 & $\sim$31 & $\sim$4.1 & 6.695 & 33.90 & 0.202 & 0.946 & 0.880\\
  	  Q$_{15,\downarrow}$ & 5.370 & 5.413 & $\sim$28 & $\sim$6.2 & 6.725 & 35.80 & 0.246 & 0.984 & 0.905\\
  	   Q$_{14,\downarrow}$ & 4.390 & 4.926 & $\sim$38 & $\sim$1.6 & 6.746 & 36.64 & 0.198 & 0.949 & 0.902\\
  	Q$_{13,\downarrow}$ & 5.240 & 5.326 & $\sim$26 & $\sim$4.0 & 6.769 & 40.11 & 0.246 & 0.978 & 0.859\\
  	    Q$_{12,\downarrow}$ & 4.350 & 4.905 & $\sim$40 & $\sim$1.9 & 6.791 & 35.97 & 0.198 & 0.950 & 0.896\\
  	   Q$_{11,\downarrow}$ & 5.438 & 5.486 & $\sim$29 & $\sim$3.0 & 6.812 & 38.98 & 0.244 & 0.972 & 0.914\\
  	   Q$_{10,\downarrow}$ & 4.485 & 5.087 & $\sim$38 & $\sim$1.2 & 6.628 & 32.70 & 0.202 & 0.963 & 0.875\\
  	    Q$_{9,\downarrow}$ & 4.985 & 5.421 & $\sim$36 & $\sim$1.6 & 6.640 & 32.02 & 0.245 & 0.977 & 0.902\\
  	   Q$_{8,\downarrow}$ & 4.445 & 4.930 & $\sim$40 & $\sim$1.9 & 6.665 & 35.24 & 0.203 & 0.958 & 0.872\\
  	Q$_{7,\downarrow}$ & 5.305 & 5.371 & $\sim$39 & $\sim$1.9 & 6.684 & 35.47 & 0.250 & 0.947 & 0.897\\
  	    Q$_{6,\downarrow}$ & 4.503 & 4.968 & $\sim$28 & $\sim$1.7 & 6.710 & 36.88 & 0.200 & 0.925 & 0.904\\
  	    Q$_{5,\downarrow}$ & 5.075 & 5.518 & $\sim$34 & $\sim$1.6 & 6.731 & 41.09 & 0.242 & 0.952 & 0.871\\
  	    Q$_{4,\downarrow}$ & 4.365 & 4.976 & $\sim$41 & $\sim$1.2 & 6.751 & 37.09 & 0.201 & 0.951 & 0.864\\
  	    Q$_{3,\downarrow}$ & 4.870 & 5.289 & $\sim$35 & $\sim$1.7 & 6.769 & 38.91 & 0.248 & 0.982 & 0.904\\
  	    Q$_{2,\downarrow}$ & 4.323 & 4.922 & $\sim$31 & $\sim$1.5 & 6.791 & 34.72 & 0.203 & 0.939 & 0.891\\
  	    Q$_{1,\downarrow}$ & 4.925 & 5.390 & $\sim$28 & $\sim$1.5 & 6.811 & 42.20 & 0.251 & 0.962 & 0.927\\
		\hline
		\hline
	\end{tabular}
  \centering
    \caption{Qubit characteristics for the 30-qubit experiment using all 30 qubits
    (Q$_{j,s}$ with $j$  from 1 to 15 and $s\in\{\uparrow,\downarrow\}$).
    $\omega_{j,s}^{\textrm{idle}}/2\pi$
    is the idle frequency of the qubit, where single-qubit gates are performed.
    $\bar{T}^1_{j,s}$ denotes the average energy relaxation time from the frequency
    about 4.0~GHz to the maximum frequency $\omega_{j,s}^{\textrm{max}}/2\pi$ of Q$_j$.
    The dephasing time $T_{2}^{\ast}$ is measured at the idle frequency
    $\omega_{j,s}^{\textrm{idle}}/2\pi$. Also, $\omega^{\textrm{R}}_{j,s}/2\pi$ is
    the readout resonator's frequency of Q$_{j,s}$ with $g_{{j,s}}/2\pi$ being
    the coupling strength between the qubit and the readout resonator. The anharmonicity
    $\eta_{j,s}/2\pi$ of Q$_{j,s}$  is defined by
    $\eta_{j,s}\equiv\omega^{10}_{j,s} - \omega^{21}_{j,s}$, with
    $\omega^{10}_{j,s}/2\pi$ ($\omega^{21}_{j,s}/2\pi$) being the energy difference
    between the $|1\rangle$ and $|0\rangle$ states ($|2\rangle$ and $|1\rangle$ states).
    $F_{j,s}^{0}$ ($F_{j,s}^{1}$) is the measurement
    probability of $|0\rangle$ ($|1\rangle$) when the qubit is prepared at $|0\rangle$ ($|1\rangle$),
    which is used to mitigate the readout errors.}\label{tab2}
\end{table*}

\subsection{Wiring}
A schematic diagram of our experimental setup and wiring is shown in Supplementary Fig.~\ref{fig:wiring}.
In our experiments, the superconducting quantum processor is cooled in a
dilution refrigerator (BlueFors, XLD1000sl) with a base temperature of mixing chamber
(MC) about 15~mK. This device has three readout lines, and the readout signals are amplified
by the Josephson-based parametric amplifiers (JPAs) and the high-electron-mobility
transistors (HEMTs). The specific arrangement of the attenuators and filters are used for
noise suppression. The up-conversion principle is used to generate the XY and read-in
signals, while the down-conversion principle is for readout signal. The Z signals are
directly generated by the arbitrary waveform generator (AWG). The XY and Z signals are
coupled together at the MC plate of the dilution refrigerator by using a directional coupler.

\subsection{Device information}
For the 15-qubit experiment,  fifteen superconducting qubits (Q$_{j,\uparrow}$ with $j$
from 1 to 15) on one leg of the qubit-ladder are used, which constitutes a one-dimensional (1D) qubit-chain.
The idle frequencies of 15 qubits, denoted by $\omega^{\textrm{idle}}_{j,\uparrow}/2\pi$,
are carefully arranged to minimise the crosstalk error among qubits during the single-qubit operations,
while the 15 unused qubits (Q$_{j,\downarrow}$ with $j$  from 1 to 15) on the other leg
are detuned far from 3.5 GHz to avoid any unwanted interaction.
The qubits characteristics for the 15-qubit experiment are shown in Supplementary Table~\ref{tab1}.

For the 30-qubit experiment, all thirty superconducting qubits (Q$_{j,s}$ with $j$
from 1 to 15 and $s\in\{\uparrow,\downarrow\}$) are used.
The qubit's characteristics for the 30-qubit experiment are shown in Supplementary  Table~\ref{tab2}.
As shown in %Table~\ref{tab1} and
Supplementary Table~\ref{tab2}, the decoherence times $T^1$ and $T^{2*}$
are much longer than the system's evolution time within 1~$\mu$s.

\subsection{System Hamiltonian}
Our superconducting quantum processor can be described as a Bose-Hubbard ladder
 with a Hamiltonian ($\hbar=1$) \cite{Roushan2017,Ye2019}
 \begin{align}
   H_{\textrm{BH}}=&J_{\parallel}\sum_{j,s}(\hat{a}_{j,s}^\dag\hat{a}_{j+1,s}+\textrm{H.c.})
  +\sum_{j,s}\frac{\eta_{j,s}}{2}\hat{n}_{j,s}(\hat{n}_{j,s}-1)\nonumber\\
   & +J_{\perp}\sum_{j}(\hat{a}_{j,\uparrow}^\dag\hat{a}_{j,\downarrow}+\textrm{H.c.})
   +\sum_{j,s}V_{j,s}\hat{n}_{j,s},\label{BHladder}
 \end{align}
 where $\hat{a}^{\dag}$ ($\hat{a}$) is the bosonic creation
 (annihilation) operator, and $\hat{n}\equiv \hat{a}^{\dag}\hat{a}$ is the number operator.
Here, $J_{\parallel}/2\pi\simeq 8$~MHz and $J_{\perp}/2\pi\simeq 7$~MHz denote
the nearest-neighbour (NN) hopping between nearby qubits on the same leg and on the same rung,
respectively. Also, $\eta$ is the on-site nonlinear interaction,
and $V_{j,s}$ is the tuneable on-site potential.

Our device is designed to fulfil the hard-core limit $|\eta/J|\gg1$, and thus,
the highly occupied states of transmon qubits are blockaded, which represents the fermionisation of strongly
interacting bosons \cite{Yan2019}. The system Hamiltonian can then be simplified as
\begin{align}
     H=&J_{\parallel}\sum_{j,s}(\hat{c}_{j,s}^\dag\hat{c}_{j+1,s}+\textrm{H.c.})
     +\sum_{j,s}V_{j,s}\hat{c}_{j,s}^\dag\hat{c}_{j,s}\nonumber\\
     &+J_{\perp}\sum_{j}(\hat{c}_{j,\uparrow}^\dag\hat{c}_{j,\downarrow}+\textrm{H.c.}),
\end{align}
where $\hat{c}^{\dag}$  ($\hat{c}$) is the hard-core bosonic creation (annihilation) operator
with $(\hat{c}^{\dag})^2=\hat{c}^{2}=0$,
and $[\hat{c}^{\dag}_{j,s},\hat{c}_{i,r}]=\delta_{ji}\delta_{sr}$.

Note that in addition to the hopping between nearest-neighbour (NN) qubits, there also exist
the hopping between next-nearest-neighbour (NNN) qubits on  different legs:
\begin{align}
J_\times\sum_j(\hat{c}^\dag_{j,\uparrow}\hat{c}_{j+1,\downarrow}+\hat{c}^\dag_{j,\downarrow}\hat{c}_{j+1,\uparrow}+\textrm{H.c.}),
\end{align}
and the hopping between third-nearest-neighbour (TNN) qubits on the same leg:
\begin{align}
J'_\parallel\sum_{j,s}(\hat{c}^\dag_{j,s}\hat{c}_{j+2,s}+\textrm{H.c.}).
\end{align}

\subsection{Qubits coupling strengths}
The coupling strengths between the nearest-neighbour (NN), next-nearest-neighbour (NNN), and
third-nearest-neighbour (TNN) qubits are shown in Supplementary Fig.~\ref{fig:Coupling}a and \ref{fig:Coupling}b
for the 15-qubit and 30-qubit experiments, respectively.
The coupling strengths between the selected qubits are measured at 4.7 and 4.5~GHz for
the 15-qubit and 30-qubit experiments, respectively.
For the 15-qubit experiment, we only measured the nearest and next-nearest
coupling strengths, which are considered in the numerical simulation, since the TNN coupling in the 15-qubit chain
is too weak and can be neglected. For the 30-qubit experiment, we measured the NN, NNN
and TNN  coupling strengths, which are used in the numerical simulation.

\begin{figure*}[t]
	\begin{center}
		\includegraphics[width=0.97\textwidth,clip=True]{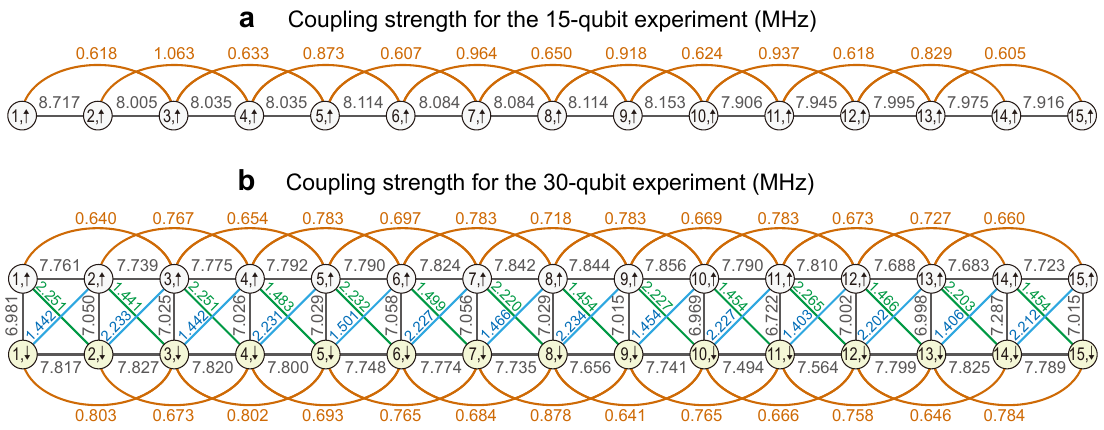}
		\caption{Experimental measured coupling strengths between nearest-neighbour (NN),
    next-nearest-neighbour (NNN), and third-nearest-neighbour (TNN) qubits,
    which are used for the numerical simulations of the 15-qubit experiment (a), and the 30-qubit experiment (b). }
		\label{fig:Coupling}
	\end{center}
\end{figure*}

\subsection{Single-qubit gates}
In our experiments, we initialise a qubit from the $|0\rangle$ state  to the $|1\rangle$ and
$(|0\rangle + |1\rangle)/\sqrt{2}$ states by applying a X$_{\pi}$ and Y$_{\frac{\pi}{2}}$ gates on
this qubit for quantum walks and topological band structure measurements, respectively.
Single-qubit operations, including the Y$_{\frac{\pi}{2}}$ %$X/2$ (=$R_x(\pi/2)$)
and X$_{\pi}$ %$X$ (=$R_x(\pi)$)
gates, have a duration of 120~ns and a full width at half maximum (FWHM) of 60~ns.
By using the derivative removal by adiabatic gate (DRAG) theory,
the quadrature correction terms with a DRAG coefficient %$\alpha$
are optimised
to minimise the leakage to higher energy levels \cite{Motzoi2009}.
The readout pulses for all qubits have a duration of 2.0~$\mu$s, and the readout pulses
 powers and frequencies are optimised to realise high-visibility and low-error readouts.
The readout fidelities, denoted by $F_{0}$ and $F_{1}$, are shown in
Supplementary Table~\ref{tab1} and Supplementary Table~\ref{tab2} for the 15-qubit and 30-qubit experiments, respectively.

Here, to identify the accuracy of the initial state preparation, we calculate the fidelity  of the prepared
states compared with the perfect $|1\rangle$ and $(|0\rangle + |1\rangle)/\sqrt{2}$ states, respectively, by using
the quantum state tomography (QST) technique.  The QST measurements of the prepared states
require individually measuring the qubit in bases formed by the eigenvectors of
$\hat{\sigma}^x$, $\hat{\sigma}^y$, and $\hat{\sigma}^z$,
respectively. In our experiments, the measurement along the $z$-axis of the Bloch sphere can be
directly performed, and those along the $x$- and $y$-axes  are realised by  inserting the
Y$_{-\frac{\pi}{2}}$ and X$_{-\frac{\pi}{2}}$ gates on the qubit before the measurements, respectively.
In total, we have three tomographic operations \{I,~X$_{-\frac{\pi}{2}}$,~Y$_{-\frac{\pi}{2}}$\} and obtain
two occupation probabilities $\{P_0,P_1\}$ for each operation, which allow us to reconstruct the density
matrix $\rho_{\textrm{exp}}$ of the initially prepared state.

The fidelity between the
experimentally measured density matrix $\rho_{\textrm{exp}}$ and the ideal one $|\psi\rangle$
 can be quantified as
 $F(\rho_{\textrm{exp}}, |\psi\rangle) =\langle\psi|\rho_{\textrm{exp}}|\psi\rangle$.
 We maintain a fixed sample of 3,000 repetitions of readouts and repeat the measurement
 procedure ten times for estimating the average value and the standard deviation of the state
 fidelity for each qubit.
 Experimental results of the fidelity for the $(|0\rangle + |1\rangle)/\sqrt{2}$ and $|1\rangle$ states are
 both above 0.990,  which are shown in Supplementary Fig.~\ref{fig:BasicTomo}a and \ref{fig:BasicTomo}b, respectively.

\begin{figure*}[t]
	\begin{center}
		\includegraphics[width=0.97\textwidth,clip=True]{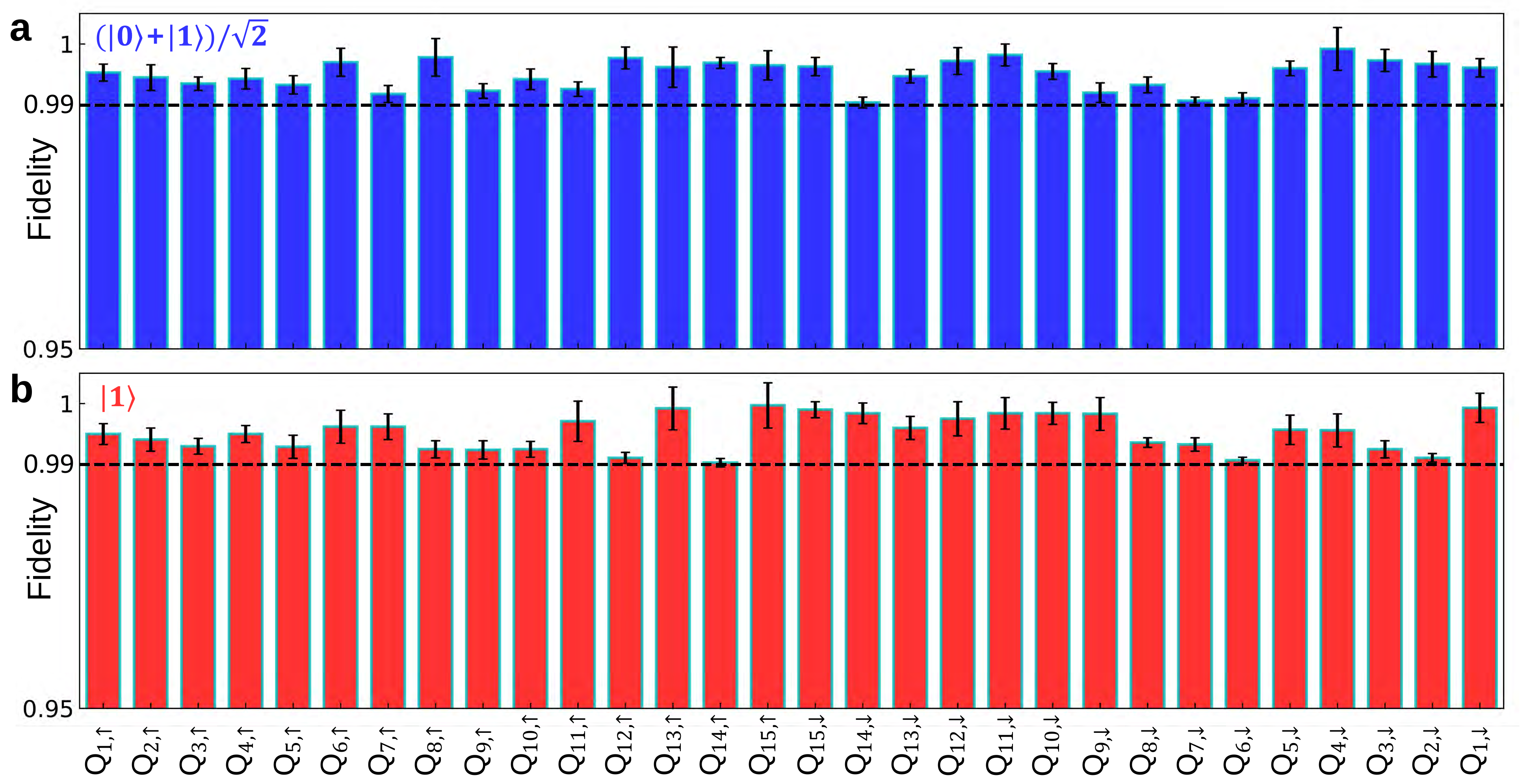}
		\caption{Single-qubit quantum state tomography (QST) for the $(|0\rangle + |1\rangle)/\sqrt{2}$ state (a) and the $|1\rangle$ state (b), which are prepared by applying Y$_{\frac{\pi}{2}}$ and X$_{\pi}$ gates, respectively. }
		\label{fig:BasicTomo}
	\end{center}
\end{figure*}

\section{Calibration}
\subsection{Pulse amplitudes for Z and XY controls}
In our experiments, we periodically calibrate the amplitude of the Z pulse (about 10~$\mu$s) at the
idle frequency and the pulse amplitude of the single-qubit gate on each qubit to avoid
the variations of the performances of the Z and XY controls with time.

\subsection{Readout probability}
We  monitor the time evolution of the readout fidelity during the experiment in order to perform the
real-time correction of the readout errors \cite{Zheng2017}, by applying the inverse of the tensor
product of the matrices
\begin{align}
  \left(\begin{array}{cc}
F^0_{j,s} & 1-F^1_{j,s} \\
1-F^0_{j,s} & F^1_{j,s}  \\
\end{array}\right)
\end{align}
for the selected qubit Q$_{j,s}$ to the measured occupation probability vector $(P_{j,s}^{0},P_{j,s}^{1})^T$.

\subsection{Z pulse distortion}
In the waveform of the Z pulse to adjust the frequency of a qubit, the rising edge, the falling edge,
and the flatness of the Z pulse are very significant for realising a high-fidelity quantum gate and
the long-time quantum state evolution. The Z pulse distortion would cause an unwanted frequency drift
before the readout process. This issue causes errors when we measure in the
$\hat{\sigma}^x$ and $\hat{\sigma}^y$ bases, because the frequency drift would generate
an unwanted phase shift that greatly affects the measured result \cite{Neill2018}.
Therefore, it is of fundamental importance to correct the unwanted Z pulse distortion.
In our experiments, we develop two types of pulse sequences, which aim to correct the Z pulse
distortions occurring in the short-time and long-time scales, respectively.

\begin{figure*}[t]
	\begin{center}
		\includegraphics[width=0.85\textwidth,clip=True]{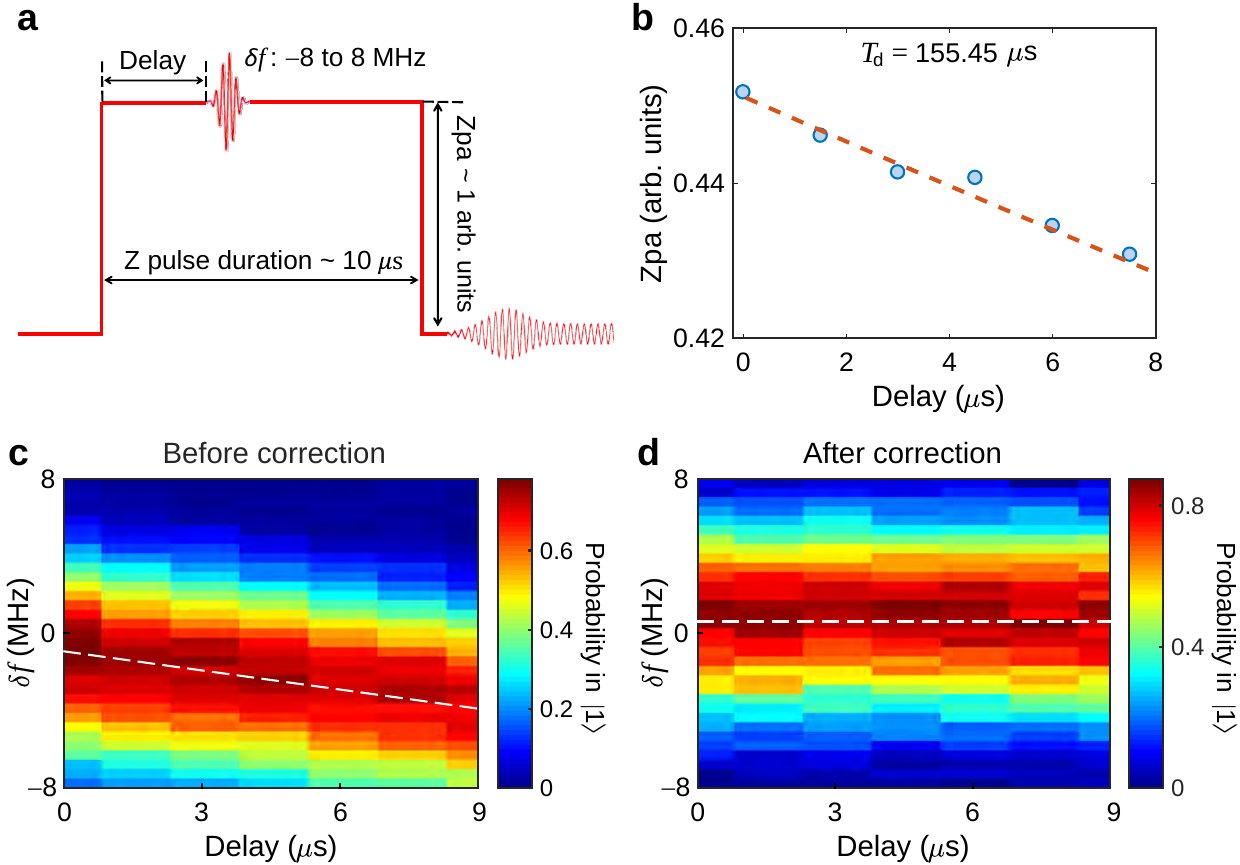}
		\caption{Calibration of the Z pulse distortion for the long-time scale caused by the DC block.
	(a) Schematic of the calibration pulse sequence. We apply a Z pulse with a fixed duration
  ($\sim 10$~$\mu$s) and a fixed amplitude ($\sim 1$~arb. units) and  excite the qubit
  around the target frequency by a X$_\pi$ pulse. Then, we measure the population of the $|1\rangle$ state
  for different time delays between the X$_\pi$ pulse and the Z pulse, given a difference of the
  frequency $\delta\!f$.
  (b) An exponential decay function
  is used to estimate the characteristic decay factor as $T_{\textrm{d}} = 155.45$~$\mu$s.
  (c  and  d) Z pulse shapes obtained by measuring population of $|1\rangle$ before (c) and after (d) the
  Z correction, respectively.}
		\label{fig:DC_Zcorrect}
	\end{center}
\end{figure*}

\begin{figure*}[t]
	\begin{center}
		\includegraphics[width=0.85\textwidth,clip=True]{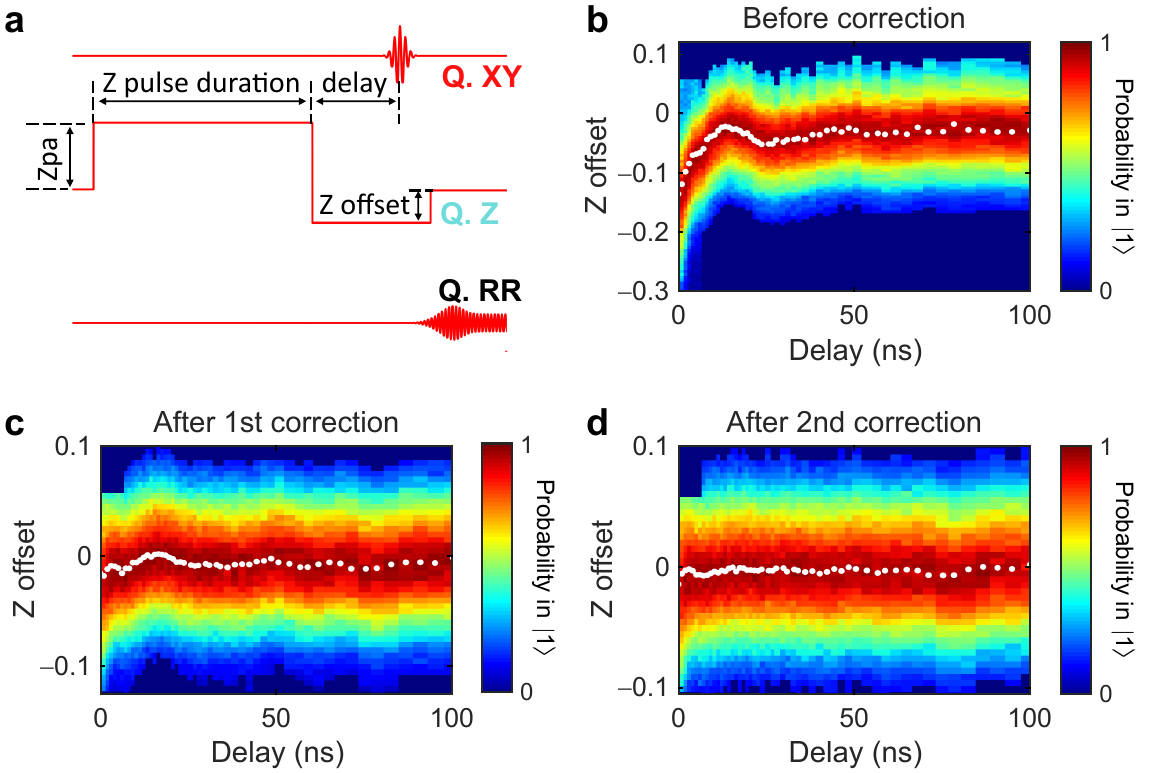}
		\caption{Calibration of the Z pulse distortion for short-time scales.
	(a) Schematic of the experimental pulse sequence. We apply a Z pulse with a fixed duration
   ($\sim1$~$\mu$s) and a fixed amplitude ($\sim1$~arb. units) to the qubit for obtaining the
   shape of the rising (falling) edge by exciting the qubit around the its idle frequency
   (by varying the Z offset parameter). (b) Pulse shape before the correction.
    (c  and  d) Z pulse shapes after the first (c) and second (d) correction procedures, respectively.}
		\label{fig:pulseShape}
	\end{center}
\end{figure*}

The existence of the DC component in the waveform of the Z pulse will shorten the
dephasing time $T^{2*}$ of the qubit. In our experiments, a DC-block element (with a capacitance
 $\sim1$--$2\mu$F) is employed for blocking the DC component. However, the use of the DC-block
element inevitably distorts the shape of the Z pulse with a long time  (over 500~ns), which decreases
 the modulation precision of the qubits frequencies.

 Thus, we design a pulse sequence for correcting the Z
 pulse distortion as shown in  Supplementary Fig.~\ref{fig:DC_Zcorrect}a. First, a X$_\pi$ pulse around the target
 frequency and a Z pulse with a fixed duration and a fixed amplitude are applied to the qubit.
 Then, we vary the starting time of the X$_\pi$ pulse and measure the population of the $|1\rangle$
 state for different time delays. The qubit can be excited to $|1\rangle$ when the
 frequency of the qubit matches that of the applied X$_\pi$ pulse. As shown in  Supplementary Fig.~\ref{fig:DC_Zcorrect}c,
 the measured population of $|1\rangle$  before the Z correction shows the shape of the
 Z pulse that is sensed by the qubit. After obtaining the maximum value of the population of
 $|1\rangle$ for each delay, we can fit the data of the Z pulse amplitude (Zpa) with an exponential
 decay function:
\begin{equation}\label{equ:correctZ_equ}
P_1 (t) = \alpha[1 + \mathrm{exp}(-t/T_{\textrm{d}}) ] + \beta,
\end{equation}
with $\alpha$ and $\beta$ being the fitting parameters and
$T_0$ being the estimated input of the characteristic decay factor $T_{\textrm{d}}$.
As shown in  Supplementary Fig.~\ref{fig:DC_Zcorrect}b, $T_{\textrm{d}} = 155.45$~$\mu$s is fitted for the DC-block
element with a capacitance 2~$\mu$F.
With the obtained characteristic decay factor and the deconvolution method, we can send a
corrected waveform to the arbitrary waveform generator (AWG). Finally, the qubit can sense a
relatively flat Z pulse as shown in  Supplementary Fig.~\ref{fig:DC_Zcorrect}d.

The above procedure can correct the distorted Z pulse for the long-time scale. To calibrate the Z pulse
distortion within a short-time scale (e.g., $<100$~ns), the pulse sequence as shown in
Supplementary Fig.~\ref{fig:pulseShape}a is designed to record the shape of the tail of the Z pulse.
Note that the measurement is performed at the idle frequency of the  qubit.
Therefore, it is better to bias the qubit to a low frequency (1~GHz below the sweet point
$\omega^{\textrm{max}}_{j,s}/2\pi$) for obtaining a Zpa-sensitive region. The experimental result of the
pulse-shape measurement is shown in  Supplementary Fig.~\ref{fig:pulseShape}b, and an obvious distortion is
observed before the correction. In our experiments, a finite impulse response (FIR) filter and several
first-order infinite impulse response (IIR) filters are employed for the correction \cite{Rol2020}.
The FIR filter can be described by a polynomial function with $16$--$24$ parameters, while the IIR filters are
an integration of several exponential functions. The experimental result after the first round of the
correction is shown in  Supplementary Fig.~\ref{fig:pulseShape}c. Generally, in order to obtain a better performance,
we repeat the correction procedure twice and finally obtain a stationary step response behaviour
as shown in  Supplementary Fig.~\ref{fig:pulseShape}d.

\subsection{Z pulse crosstalk}
The Z pulse crosstalks between different qubit pairs is calibrated by experimentally measuring
the Z-crosstalk matrix $\tilde{M}_{Z}$ using a similar method as shown in Supplementary Ref.~\cite{Song2017}.
Each element of $\tilde{M}_{Z}$ denotes the Z pulse amplitude (Zpa) that is sensed by Q$_{j,s}$, when a
unity Zpa (1~arb. units) is applied to Q$_{i,r}$. Note that all elements $(\tilde{M}_{Z})_{i,r;j,s}$ are
less than 4.5$\%$, as shown in Supplementary Fig.~\ref{fig:crosstalk}.
The Zpa applied to each qubit is calibrated based on the mapping relationship
\begin{align}
  {Z}_{\textrm{sensed}} = \tilde{M}_{Z}\cdot{Z}_{\textrm{applied}},
\end{align}
where ${Z}_{\textrm{sensed}}$ and ${Z}_{\textrm{applied}}$
are for  the Zpas that are sensed by the qubit and applied to the qubit, respectively.

\begin{figure*}[t]
	\begin{center}
		\includegraphics[width=0.97\textwidth]{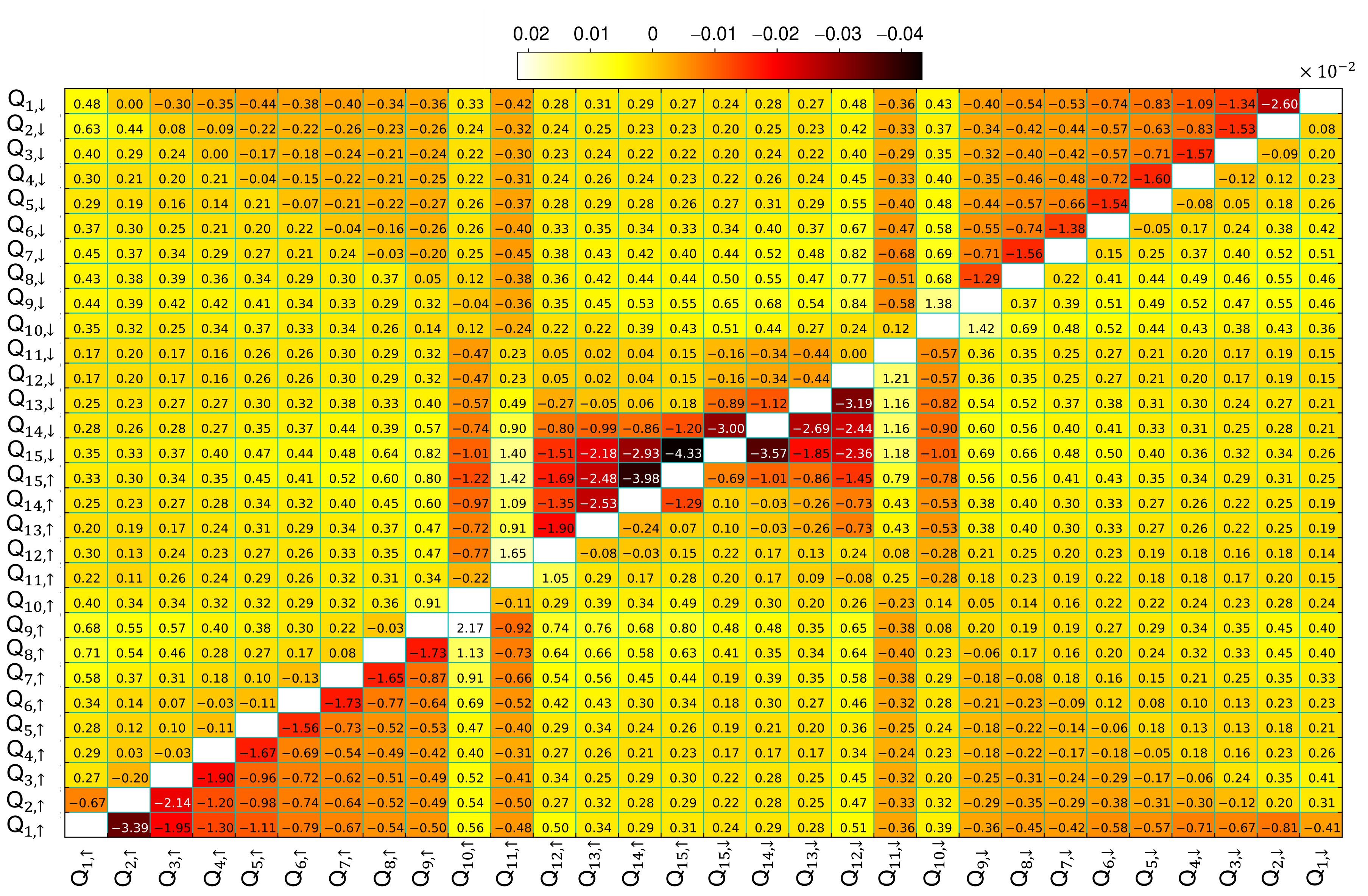}
		\caption{Z crosstalk.
		Z crosstalk matrix $\tilde{M}_{Z}$, of which each element $(\tilde{M}_{Z})_{i,r;j,s}$
    denotes the sensed Zpa by Q$_{j,s}$ when a unity Zpa (1~arb. units) is applied to Q$_{i,r}$.}
		\label{fig:crosstalk}
	\end{center}
\end{figure*}

\subsection{Qubits frequencies}
In the 15-qubit experiment, we simulate the two-dimensional (2D) integer quantum Hall effect (QHE)
using 15 qubits on one leg of the qubit ladder by tuning the qubits frequencies
as $\omega_{j}(\phi) = \omega_0 + \Delta\sum_{j=1}^{15}\cos(2 \pi bj + \phi)$, with the reference
frequency being fixed at $\omega_0/2\pi = 4.7$ GHz.
Here, $\phi$ is varied from 0 to 2$\pi$ in order to obtain various instances of the 1D
Aubry-Andr\'{e}-Harper (AAH) chains with different on-site potentials
$V_j(\phi) = \omega_{j}(\phi) - \omega_0$, in the rotating frame with respect to $\omega_0$.
Then, we experimentally simulate the 1D topological charge pump, which provides an alternative way to
explore the 2D integer QHE, by adiabatically varying $\phi$ initially from $\phi_0 = \frac{5\pi}{3}$ with
$\Delta/2\pi = 36$ MHz. In the 30-qubit experiment, we tune the qubits frequencies
as $\omega'_{j,s}(\phi) = \omega'_0 + \Delta_s\sum_{j=1}^{15}\cos(2 \pi bj + \phi)$, with the reference
frequency being fixed at $\omega'_0/2\pi = 4.5$ GHz.

Thus, it is very significant to precisely adjust the qubits frequencies in our experiments.
For the single-qubit experiment, it is simple to tune the qubit to the target frequency with a
determined relationship between the Zpa and the qubit frequency, as shown in Supplementary Fig.~\ref{fig:rabilong}a.
 However, for the multi-qubit case, the target qubits  frequencies will shift slightly when other qubits Z
 signals are applied simultaneously, even though the Z crosstalks have been calibrated by using
 the Z-crosstalk  matrix $\tilde{M}_{Z}$. This indicates the existence of residual Z crosstalks.
Although we can obtain a larger Z-crosstalk matrix $\tilde{M}'_{Z}$ when other qubits Z pulses are
applied, it is not very worthwhile, because to calibrate all the matrix elements are very time consuming.

\begin{figure*}[t]
	\begin{center}
		\includegraphics[width=0.95\textwidth,clip=True]{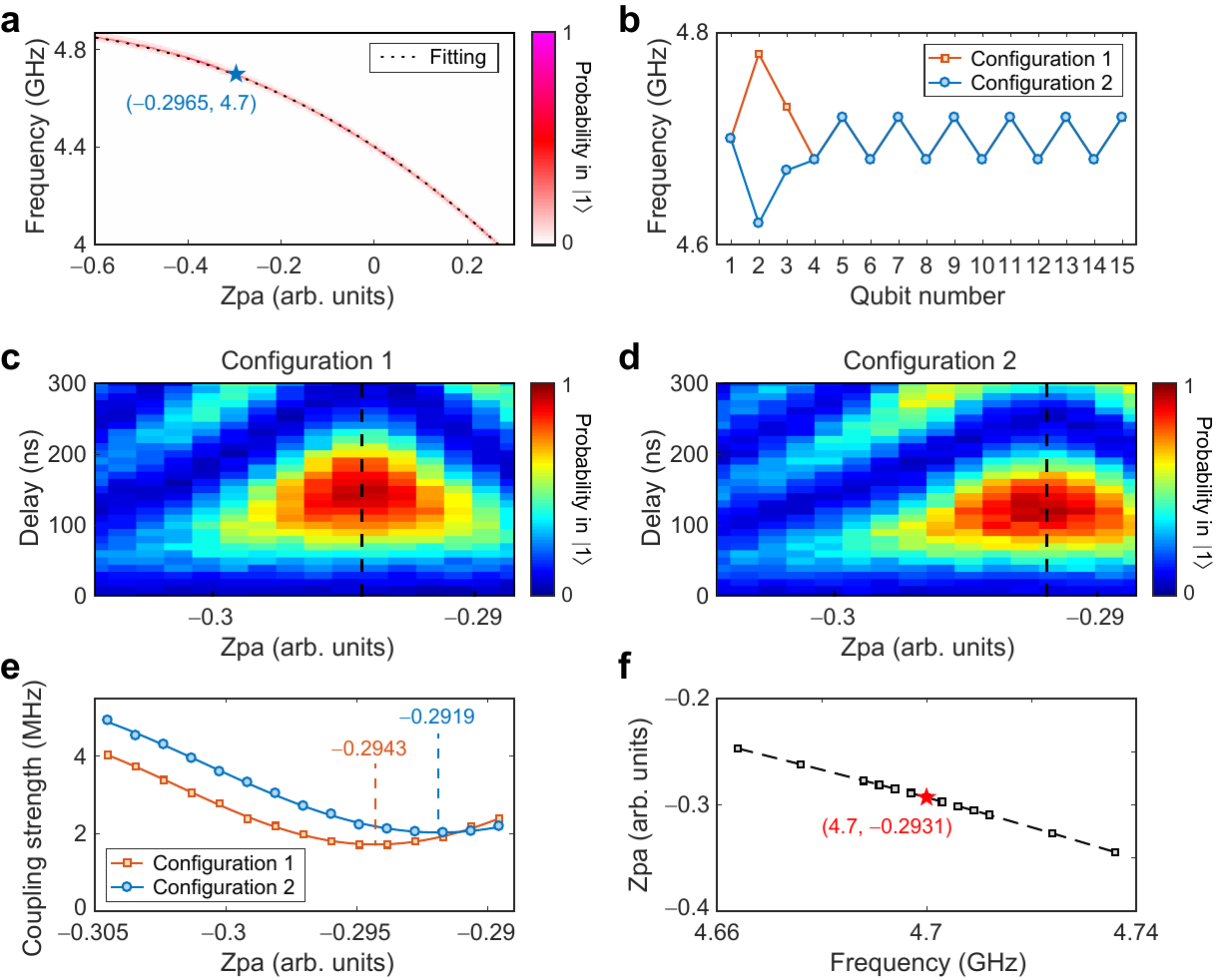}
		\caption{Calibration of qubits frequencies by using the Rabi oscillation measurements.
	(a) Relationship between the Zpa and the qubits frequencies.
  The experimental data are extracted from the automatic spectroscopy measurement.
  (b) Two Zpa configurations for adjusting Q$_{1,\uparrow}$ to 4.7~GHz by using the Rabi oscillation measurements.
  (c and d) Experimental Rabi oscillation results under two Zpa configurations, respectively.
  (e) The calculated driving strengths for two Zpa configurations.
  The smallest driving strength  corresponds to the optimal Zpa.
  (f) The updated  relationship between the Zpa and the qubits frequencies in the region studied.
  The red star is obtained by  averaging the two optimal Zpas when using different Zpa configurations.}
		\label{fig:rabilong}
	\end{center}
\end{figure*}

Here, we perform  direct Rabi oscillation measurements on each stand-alone qubit around the
reference frequency [from ($\omega_0/2\pi - 36$~MHz) to ($\omega_0/2\pi + 36$~MHz)].
As shown in Supplementary Fig.~\ref{fig:rabilong}b, to calibrate the Zpa  for detuning Q$_{1,\uparrow}$ to 4.7~GHz,
we design two Zpa configurations for other qubits to cancel the effect of the residual Z crosstalks.
Given the experimental on-site potentials and the coupling strengths between qubits
as shown in  Supplementary Fig.~\ref{fig:Coupling}a, the nearest-neighbour (NN) qubit (Q$_{2,\uparrow}$), the
next-nearest-neighbour (NNN) qubit (Q$_{3,\uparrow}$) and other qubits are detuned $\pm 80$~MHz,
$\pm 30$~MHz and $\pm 20$~MHz with respect to $\omega_0/2\pi$, respectively.
Note that there are also other Zpa configurations in addition to those as shown in  Supplementary Fig.~\ref{fig:rabilong}b,
which hardly affect the final result.

We fix the driving magnitude of the local transverse field
with a resonant frequency at 4.7~GHz and scan the Zpa around the guessed one (marked with a blue star in
Supplementary Fig.~\ref{fig:rabilong}a). For the two Zpa configurations, the 2D Rabi oscillation results are plotted in
Supplementary Fig.~\ref{fig:rabilong}c and \ref{fig:rabilong}d, respectively. The slowest Rabi oscillation
characterises the optimised  Zpa, corresponding to the smallest driving strength as shown in
Supplementary Fig.~\ref{fig:rabilong}e.

Finally, the Zpa used in our experiments (marked with a red star in
Supplementary Fig.~\ref{fig:rabilong}f) is fixed at the average value of the two optimised Zpa configurations.
Note that the final Zpa indicates a frequency shift of about 2.5~MHz with respect to that shown in
Supplementary Fig.~\ref{fig:rabilong}a. For simplicity, we can select some typical frequencies and repeat the above
optimisation procedure for obtaining a new relationship between the Zpa and qubits frequencies in the
 region studied (see Supplementary Fig.~\ref{fig:rabilong}f).

\subsection{Phase calibration}
As shown in Fig.~1d in the main text for the pulse sequences for the band structure spectroscopy,
a Y$_{\frac{\pi}{2}}$ rotation pulse is applied on the target qubit for measuring the time
evolutions of $\langle\hat{\sigma}^x(t)\rangle$ and $\langle\hat{\sigma}^y(t)\rangle$.
Before the readout of Q$_{j,s}$ at its idle point in the bases of $\hat{\sigma}^x$ and
$\hat{\sigma}^y$, we detune all qubits to their corresponding frequencies for the quench evolution
with a time $t$ by applying a Z pulse to each qubit. This operation will accumulate a dynamical phase that
needs to be compensated when applying the rotation pulse after the time evolution.
For each qubit Q$_{j,s}$, the dynamical phase accumulated during the evolution can be calculated
by $\delta \omega_{j,s} t$, where $\delta \omega_{j,s} \equiv \omega_{j,s} - \omega_0$, and $\omega_0/2\pi$ is the target
frequency of Q$_{j,s}$ for the quench dynamics. Although the Z pulse distortion has been carefully calibrated
before the experiment, there are still some imperfect factors, leading to an additional phase shift, as
compared with the theoretical calculations.

Therefore, we design a  pulse sequence for the phase calibration. First, a Y$_{\frac{\pi}{2}}$
pulse is applied to Q$_{j,s}$, and then, all qubits frequencies are arranged with two configurations, similar
as the case in  Supplementary Fig.~\ref{fig:rabilong}b for the calibration of the qubits frequencies.
Then, we apply another Y$_{\frac{\pi}{2}}$ pulse to Q$_{j,s}$ and vary its microwave phase for obtaining
the probability when Q$_j$ is in $|1\rangle$ as a function of $t$ and the phase difference
$(\varphi - \delta \omega_{j,s}  t)$. For these two Zpa configurations, we obtain the experimental results
shown in  Supplementary Fig.~\ref{fig:PhaseCorr}a and \ref{fig:PhaseCorr}b, respectively.
To calculate the phase compensation for the second rotation pulse, we perform a standard cosine fitting
to the probability of a qubit Q$_{j,s}$ in the $|1\rangle$ state, $P_{j,s}$, as a function of $(\varphi - \delta \omega_{j,s}  t)$ for each evolution time $t$. The results are shown in
Supplementary Fig.~\ref{fig:PhaseCorr}c, and the final phase compensation can be obtained as $(\varphi_1 + \varphi_2)/2$,
where $\varphi_1$ and $\varphi_2$ are the phase compensations under two Zpa configurations, respectively.
After the above phase calibration process and adding the phase compensation on each qubit, we
obtain the experimental results shown in Supplementary  Fig.~\ref{fig:PhaseCorr}d, which indicate that the
phase shift is compensated for all evolution times.

\section{Supplementary experimental data}
\subsection{Single-excitation quantum walks}
After initialising the system with a state $|0\rangle^{\otimes N}$, we excite one qubit Q$_{j,s}$ with
a X$_\pi$ pulse, tune all qubits to their corresponding frequencies and measure them at a time $t$ after
their free evolutions.
The total repeat count for the measurement at each time on each qubit is 3,000.

In the 15-qubit experiment, for $b=\frac{1}{3}$ and different values of $\Delta_\uparrow/2\pi$ and $\phi$, the experimental results
of the time evolution of the excitation probability $P_{j,\uparrow}$
are shown in Supplementary Fig.~\ref{sfig:5}, \ref{sfig:6}, \ref{sfig:7} and \ref{sfig:8} for:
\begin{itemize}
\item[($i$)] $\Delta_\uparrow/2\pi=0$~MHz,
\item[($ii$)] $\Delta_\uparrow/2\pi=12$~MHz with $\phi=\frac{2\pi}{3}$,
\item[($iii$)] $\Delta_\uparrow/2\pi=12$~MHz with $\phi=\frac{\pi}{10}$,
\item[($iv$)] $\Delta_\uparrow/2\pi=12$~MHz with $\phi=\pi$,
\end{itemize}
respectively, which are compared with the theoretical predictions using the parameters in Supplementary Table~\ref{tab1}.

In the 30-qubit experiment, for $b=\frac{1}{3}$ and different values of $\Delta_s/2\pi$ and $\phi$,
the experimental results are shown in Supplementary Fig.~\ref{sfig:09}, \ref{sfig:10}, \ref{sfig:11}, \ref{sfig:12}, and \ref{sfig:13} for:
\begin{itemize}
\item[($i$)] $\Delta_{\uparrow}/2\pi=\Delta_{\downarrow}/2\pi=0$~MHz,
\item[($ii$)] $\Delta_{\uparrow}/2\pi=\Delta_{\downarrow}/2\pi=12$~MHz with $\phi=\frac{2\pi}{3}$,
\item[($iii$)] $\Delta_{\uparrow}/2\pi=-\Delta_{\downarrow}/2\pi=12$~MHz with $\phi=\frac{2\pi}{3}$,
\item[($iv$)] $\Delta_{\uparrow}/2\pi=\Delta_{\downarrow}/2\pi=12$~MHz  with $\phi=\frac{\pi}{10}$,
\item[($v$)] $\Delta_{\uparrow}/2\pi=-\Delta_{\downarrow}/2\pi=12$~MHz  with $\phi=\frac{\pi}{10}$,
\end{itemize}
respectively,
which are compared with the theoretical predictions using the parameters in Supplementary Table~\ref{tab2}.

\begin{figure*}[t]
	\begin{center}
		\includegraphics[width=0.85\textwidth,clip=True]{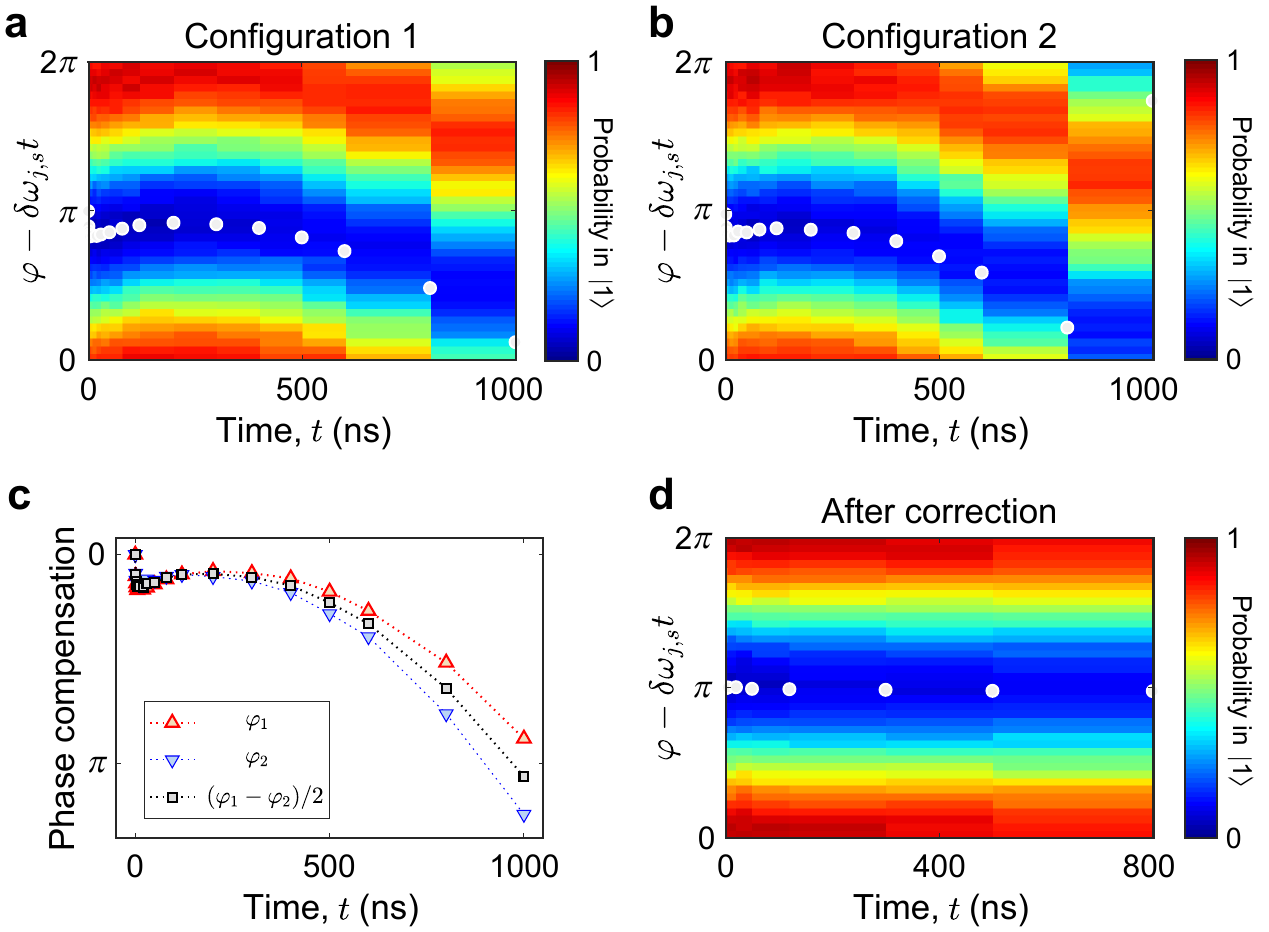}
		\caption{Phase calibration.
 (a and b) Time evolutions of the  probability of  a qubit Q$_{j,s}$ in the
$|1\rangle$ state for two Zpa configurations, respectively.
(c) The phase shift obtained by using the standard cosine fitting for two Zpa configurations.
The final phase compensation can be obtained as $(\varphi_1 + \varphi_2)/2$, where $\varphi_1$ and $\varphi_2$
 are the phase compensations under two Zpa configurations, respectively.
(d) Time evolution of the  probability of  Q$_{j,s}$ in the $|1\rangle$ state after the phase calibration.}
		\label{fig:PhaseCorr}
	\end{center}
\end{figure*}

To evaluate the performance of the experimental results of the quantum walks, we calculate the fidelity of
the evolved state using
\begin{align}
F(t) = \sum_{j,s}\sqrt{p_{j,s}(t)q_{j,s}(t)},
\end{align}
where $p_{j,s}(t)$ and $q_{j,s}(t)$ are the experimental and theoretical probability distributions of
the measurements on Q$_{j,s}$ in the $|1\rangle$ state.
We consider the mean fidelity that is averaged for different initial states with the excitation on different qubits.
As shown in Supplementary Fig.~\ref{fig:fidelity}, the relatively high fidelity (above around 0.9 and 0.85 within 500~ns for the 15-qubit
and 30-qubit experiments, respectively) for all on-site potentials configurations imply that our
experimental results are well consistent with the theoretical simulations.

\subsection{Band structure spectroscopy}
After initialising the selected qubits to their idle points, we place one target qubit Q$_{j,s}$ in the
superposed state $|+_{j,s}\rangle=(|0_{j,s}\rangle+|1_{j,s}\rangle)/\sqrt{2}$,
using a Y$_{\frac{\pi}{2}}$ pulse. Then, we detune all selected qubits  to their corresponding
frequencies for the quench dynamics with a time $t$. Then, we measure Q$_{j,s}$ at its idle point
 in the $\hat{\sigma}^x$ and  $\hat{\sigma}^y$ bases, and
 the total repeat count for each measurement on each qubit is 2,400.
 For each $\phi$, the time evolutions of $\hat{\sigma}^x_{j,s}$ and $\hat{\sigma}^y_{j,s}$ are
recorded. We then calculate the
squared Fourier transformation (FT) magnitude $|A_{j,s}|^2$  of the response function
$\chi_{j,s} (t)\equiv \langle\hat{\sigma}^x_{j,s}(t)\rangle + i\langle\hat{\sigma}^y_{j,s}(t)\rangle$ for
each qubit.

In the 15-qubit experiment, for $b=\frac{1}{3}$ and different values of $\Delta_\uparrow/2\pi$
and $\phi$, the experimental results of the normalised squared FT magnitude
$|A_{j,\uparrow}|_{\textrm{n.m.}}^2$ for each selected target qubit
are shown in Supplementary Fig.~\ref{sfig:1}  and \ref{sfig:2} for $\Delta_\uparrow/2\pi=0$~MHz and
$\Delta_\uparrow/2\pi=12$~MHz, respectively.

In the 30-qubit experiment, for $b=\frac{1}{3}$ and different values of  $\Delta_s/2\pi$
and $\phi$, the experimental results of the normalised squared FT magnitude
$|A_{j,s}|_{\textrm{n.m.}}^2$ for each selected target qubit
are shown in Supplementary Fig.~\ref{sfig:3}  and \ref{sfig:4} for $\Delta_{\uparrow}/2\pi=\Delta_{\downarrow}/2\pi=12$~MHz and
$\Delta_{\uparrow}/2\pi=-\Delta_{\downarrow}/2\pi=12$~MHz, respectively.

For the case $\Delta_{\uparrow}/2\pi=\Delta_{\downarrow}/2\pi=12$~MHz, we choose 15 qubits for
the band structure measurements, which include all four corner qubits, i.e., Q$_{1,\uparrow}$, Q$_{1,\downarrow}$, Q$_{15,\uparrow}$, and Q$_{15,\downarrow}$, and
11 bulk qubits, i.e., Q$_{2,\uparrow}$, Q$_{4,\uparrow}$, Q$_{5,\uparrow}$, Q$_{6,\uparrow}$, Q$_{7,\uparrow}$, Q$_{8,\uparrow}$, Q$_{10,\uparrow}$, Q$_{13,\uparrow}$, Q$_{14,\uparrow}$,  Q$_{12,\downarrow}$,
and  Q$_{3,\downarrow}$.

For the case $\Delta_{\uparrow}/2\pi=-\Delta_{\downarrow}/2\pi=12$~MHz, we choose 20 qubits for
the  band structure measurements, which include all four corner qubits, i.e., Q$_{1,\uparrow}$, Q$_{1,\downarrow}$, Q$_{15,\uparrow}$, and Q$_{15,\downarrow}$, and
16 bulk qubits, i.e., Q$_{2,\uparrow}$, Q$_{4,\uparrow}$, Q$_{5,\uparrow}$, Q$_{6,\uparrow}$, Q$_{8,\uparrow}$, Q$_{10,\uparrow}$, Q$_{12,\uparrow}$, Q$_{14,\downarrow}$, Q$_{12,\downarrow}$,  Q$_{10,\downarrow}$,
Q$_{9,\downarrow}$, Q$_{8,\downarrow}$, Q$_{6,\downarrow}$, Q$_{5,\downarrow}$,
Q$_{4,\downarrow}$, and  Q$_{3,\downarrow}$.

Then, we compare in Supplementary Fig.~\ref{sfig:14} the experimentally measured band structures by summing
the FT magnitudes of all selected qubits with the theoretical predictions using the experimental
parameters in Supplementary Fig.~\ref{fig:Coupling}a and \ref{fig:Coupling}b for the 15-qubit and
30-qubit experiments, respectively.
The average deviations of the peaks of the normalised FT signals
$I_\phi^{\textrm{n.m.}}$ (red crosses) from the theoretical band structures (black dashed curves)
are 0.247~MHz, 0.264~MHz, 0.568~MHz, and 0.512~MHz for:
\begin{itemize}
\item[($i$)]  15-qubit experiment with
$\Delta_\uparrow/2\pi=0$~MHz,
\item[($ii$)]  15-qubit experiment with $\Delta_\uparrow/2\pi=12$~MHz and $b=\frac{1}{3}$,
\item[($iii$)]  30-qubit experiment with  $\Delta_{\uparrow}/2\pi=\Delta_{\downarrow}/2\pi=12$~MHz and $b =\frac{1}{3}$,
 \item[($iv$)]  30-qubit experiment with  $\Delta_{\uparrow}/2\pi=-\Delta_{\downarrow}/2\pi=12$~MHz and $b=\frac{1}{3}$,
 \end{itemize}
 respectively.
 Compared with the ranges of the eigenenergies $-20$ to $20$~MHz and $-30$ to $30$~MHz
  for the 15-qubit  and
 30-qubit systems, respectively, the small average deviations imply that our band
 structure spectroscopic measurements are very accurate.

\subsection{Quantum charge pumping and fast pumping }
In addition to the topological charge pumping data as shown in Fig.~3c1--3c3 and 3d in the main text, 
we have also performed experiments with a faster
pumping rate, which clearly demonstrate fast pumping.
In details, we have decreased the period $T$ from 1,100~ns to 410~ns, corresponding to a faster ramping speed of $\phi$. 
The experimental results for the displacement of the centre of mass (CoM) 
$\delta x$ versus the time $t/T$ are compared with numerical simulations in Supplementary Fig.~\ref{fig:rf1}a.  
It is obvious that the pumping process with a period $T=410$~ns shows fast pumping with a large derivation from the Chern number $+1$.
The experimental data of the evolutions of distributions  $P_j(t)$  during the topological pumping and the displacement of 
the CoM with a period $T=410$~ns are compared with the data with $T=1$,100~ns in Supplementary Fig.~\ref{fig:rf2}.
For fast pumping with $T=410$~ns, we maintained a fixed sample of 3,000
 single-shot readouts and repeated the measurement procedure
 8 times (16 times for the no pump case) for estimating the
 mean values and standard deviations at each evolution time $t$.
Furthermore, Supplementary Fig.~\ref{fig:rf1}b shows the numerical results of the displacement 
of the CoM in one pumping cycle with different periods, which indicates a fast pumping region in our system when $T\lesssim610$~ns.

\subsection{Numerical details}
Numerical simulations are performed using the \textsc{QuTiP} \cite{Johansson:2012aa,Johansson:2013aa} (the quantum toolbox in \textsc{Python}) and \textsc{NumPy}.
The time evolutions of the systems  are numerically simulated using \textsc{QuTiP}’s
master equation solver mesolve, where the parameters in Supplementary Fig.~\ref{fig:Coupling} are used. Because the evolution time is much shorter
than the qubits' decoherence times $t\ll \bar{T}^1, T^{2*}$, we do not consider the effect of decoherence in simulations.

%\bibliography{Supp.bib}


\begin{figure*}[t]
	\centering
	\includegraphics[width=0.97\textwidth]{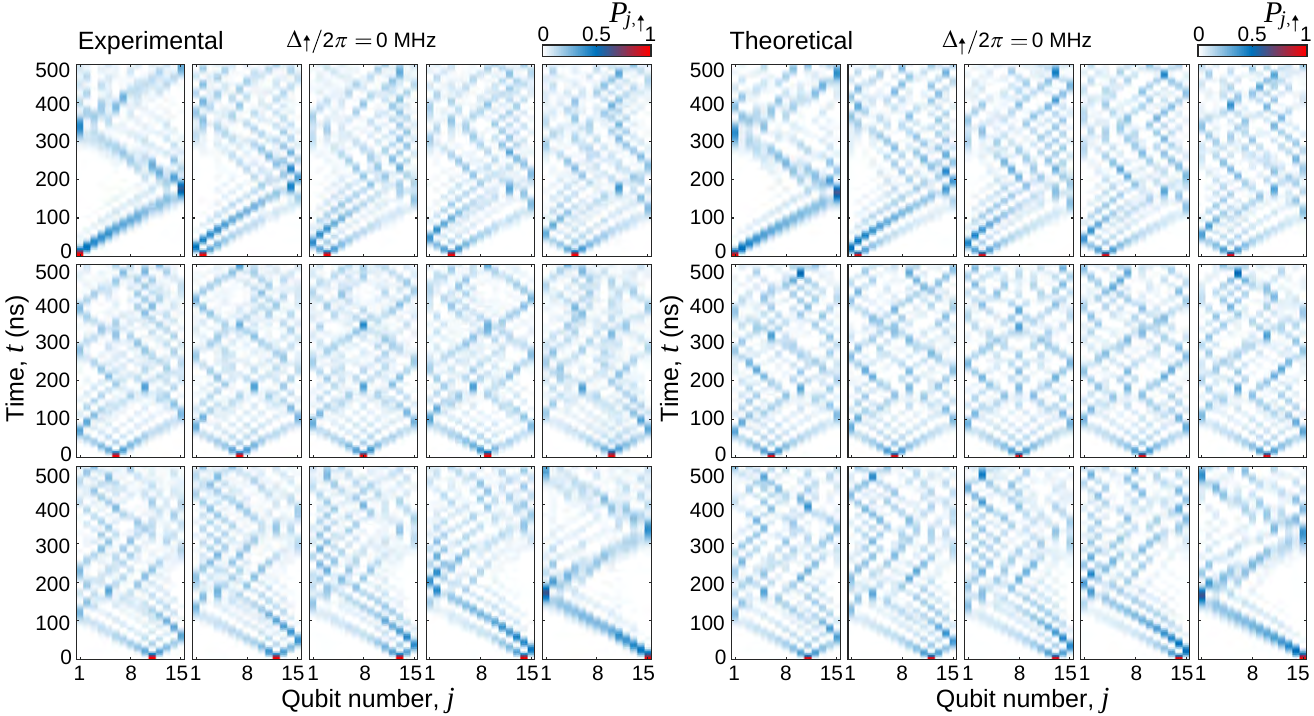}\\
	\caption{Time evolution of the excitation probability
$P_{j,\uparrow}$ with $\Delta_{\uparrow}/2\pi = 0$~MHz  after initially exciting a qubit on the
15-qubit chain. }\label{sfig:5}
\end{figure*}

\begin{figure*}[!ht]
	\centering
	\includegraphics[width=0.97\textwidth]{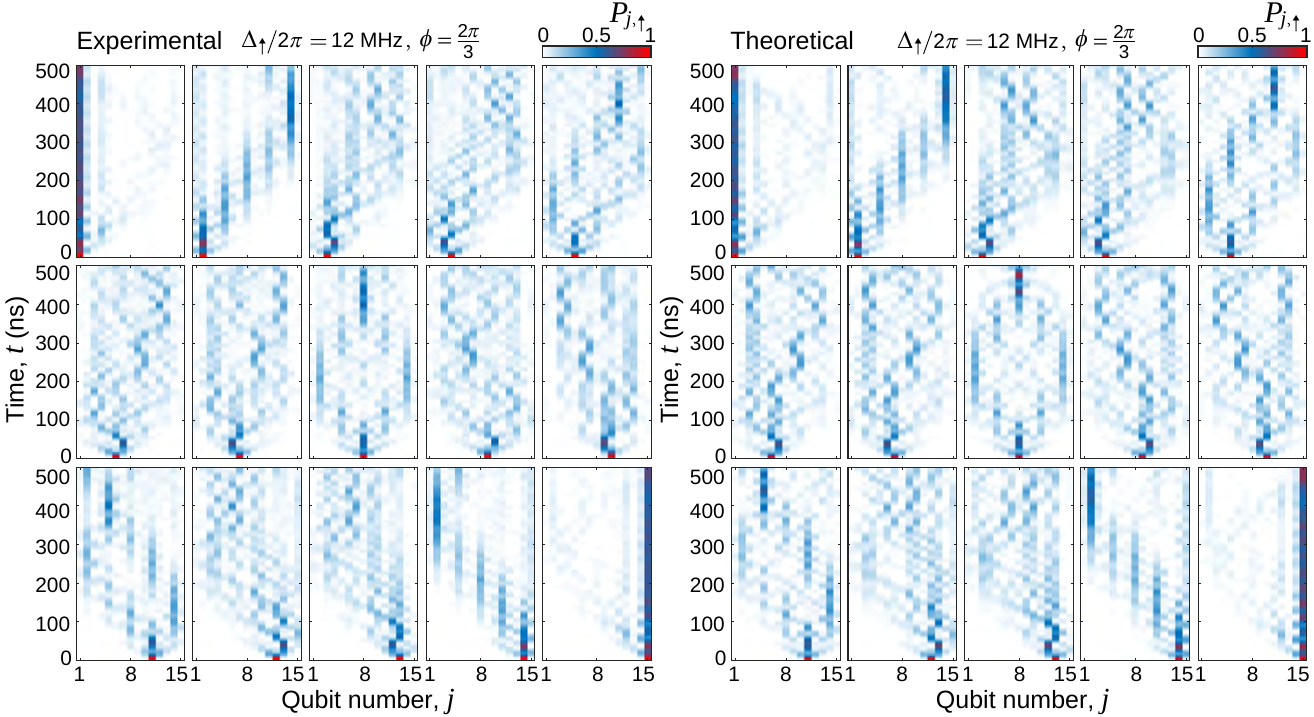}\\
	\caption{Time evolution of the excitation probability
$P_{j,\uparrow}$ with $\Delta_{\uparrow}/2\pi = 12$~MHz, $b=\frac{1}{3}$, and $\phi = \frac{2\pi}{3}$ after initially exciting a qubit on the
15-qubit chain.}\label{sfig:6}
\end{figure*}

\begin{figure*}[!ht]
	\centering
	\includegraphics[width=0.97\textwidth]{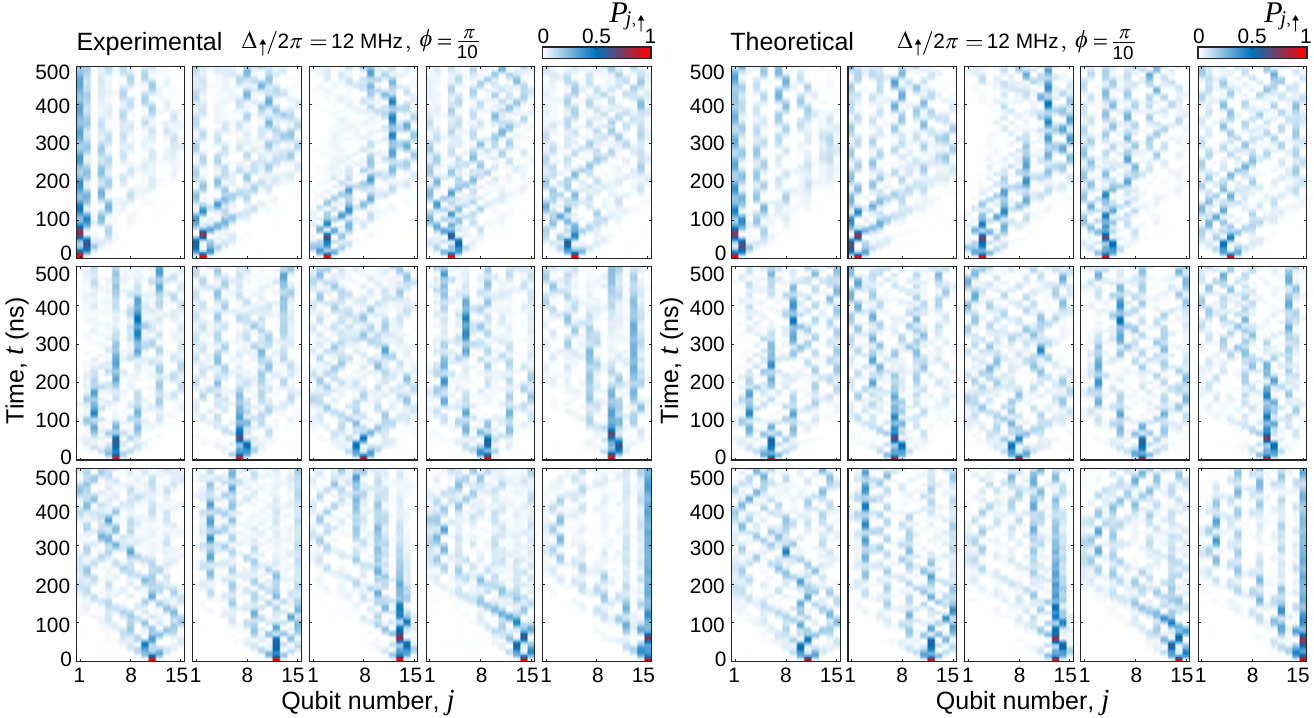}\\
	\caption{Time evolution of the excitation probability
$P_{j,\uparrow}$ with $\Delta_{\uparrow}/2\pi = 12$~MHz, $b=\frac{1}{3}$, and $\phi = \frac{\pi}{10}$ after initially exciting a qubit on the
15-qubit chain. }\label{sfig:7}
\end{figure*}

\begin{figure*}[!ht]
	\centering
	\includegraphics[width=0.97\textwidth]{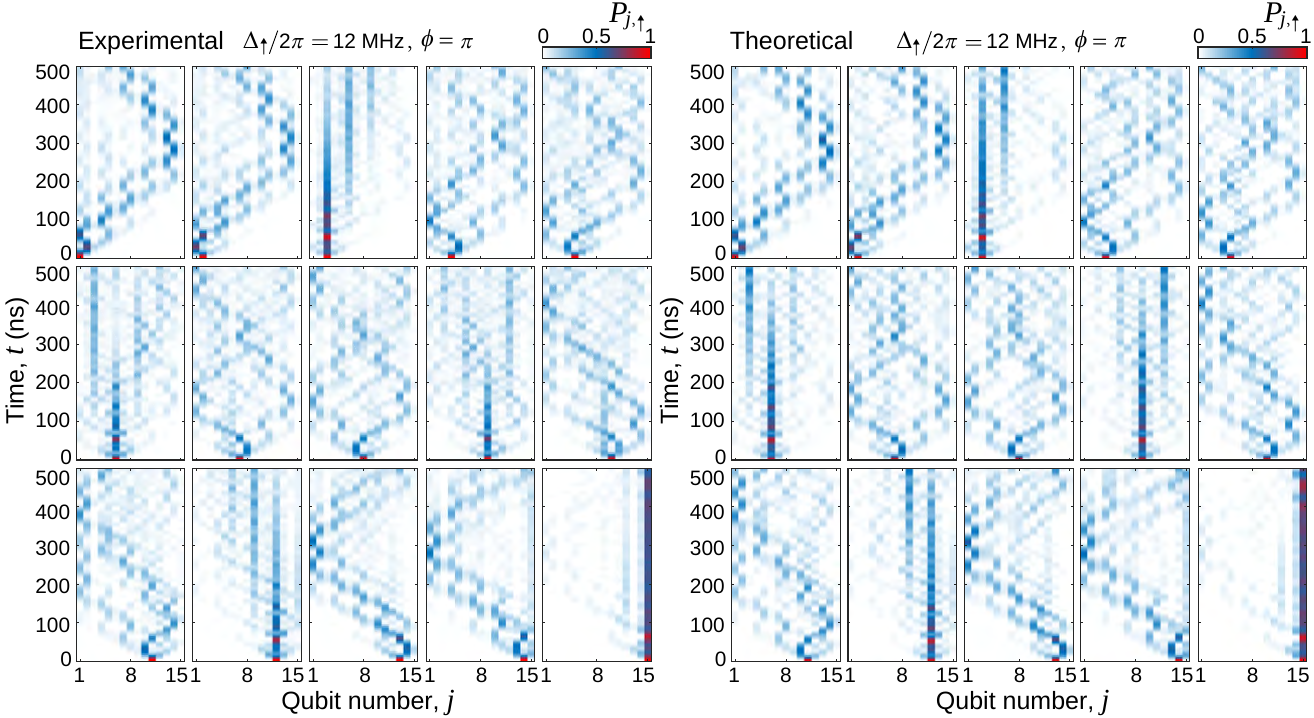}\\
	\caption{ Time evolution of the excitation probability
$P_{j,\uparrow}$ with $\Delta_{\uparrow}/2\pi = 12$~MHz, $b=\frac{1}{3}$, and $\phi = \pi$ after initially exciting each qubit on the
15-qubit chain. }\label{sfig:8}
\end{figure*}

  \begin{figure*}[!ht]
  	\centering
  	\includegraphics[width=0.97\textwidth]{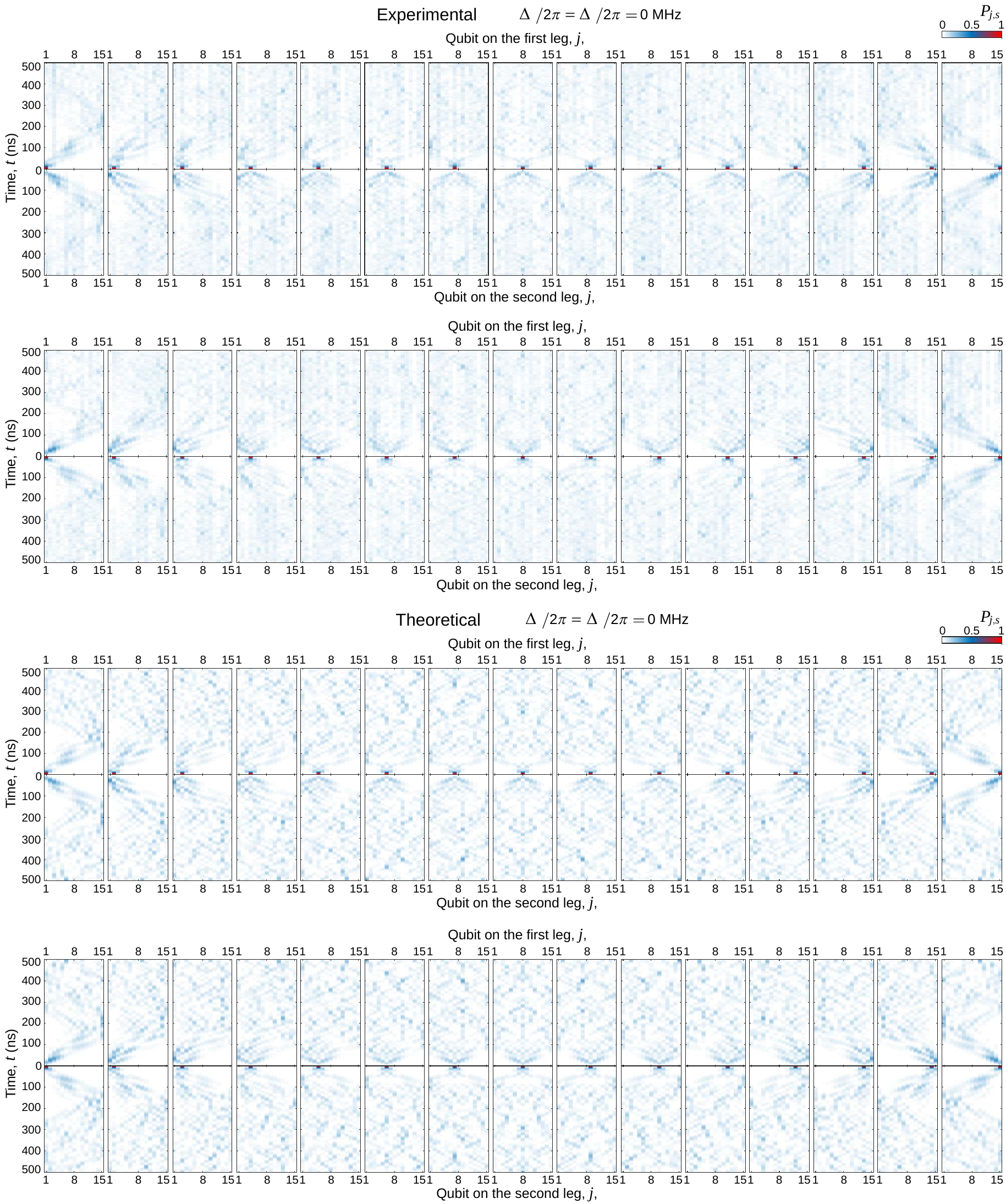}\\
  	\caption{ Time evolution of the excitation probability
  $P_{j,s}$ with $\Delta_{\uparrow}/2\pi =\Delta_{\downarrow}/2\pi = 0$~MHz after initially exciting each qubit on the
  30-qubit ladder. }\label{sfig:09}
  \end{figure*}

  \begin{figure*}[!ht]
    \centering
    \includegraphics[width=0.97\textwidth]{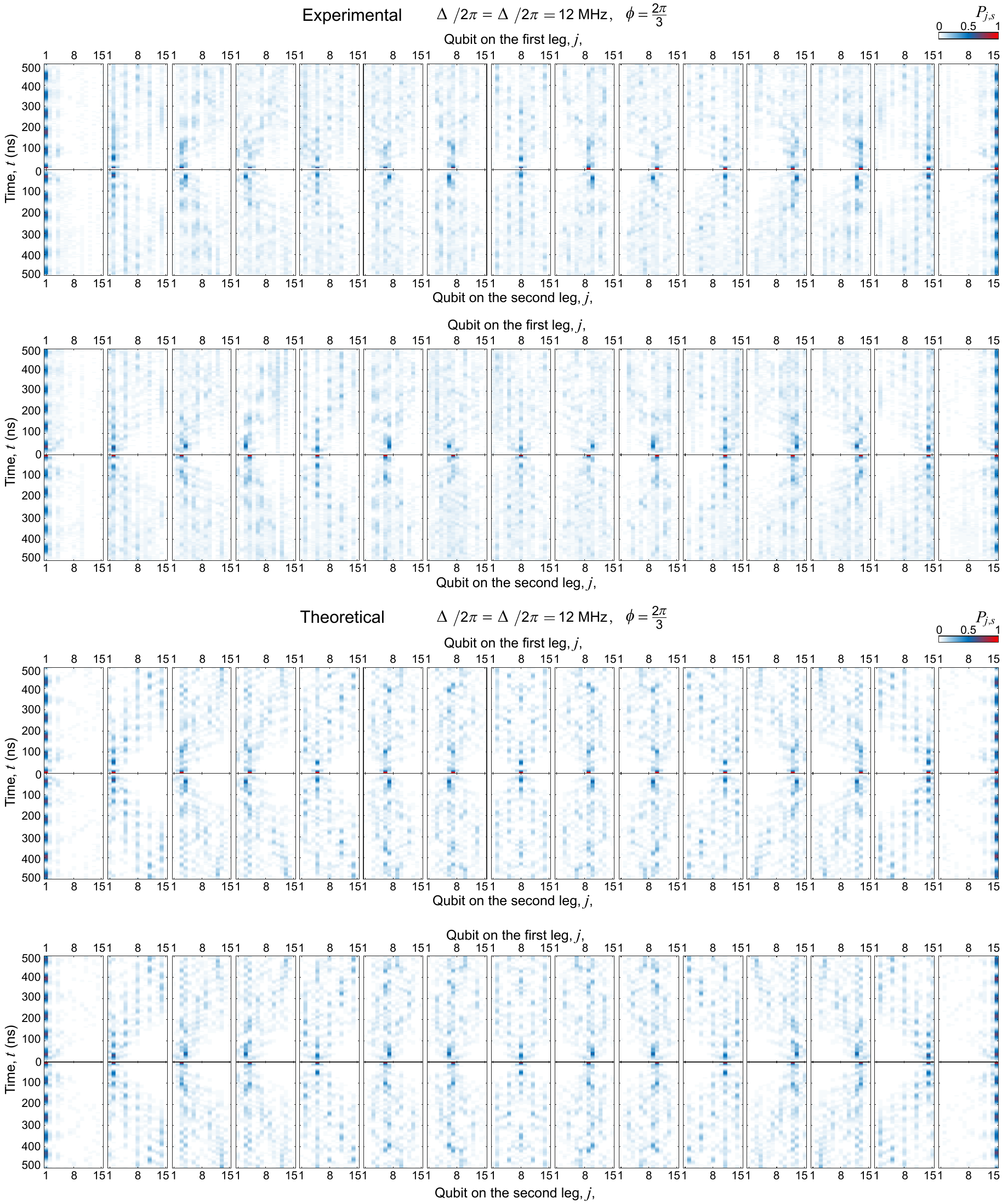}\\
    \caption{ Time evolution of the excitation probability
  $P_{j,s}$ with $\Delta_{\uparrow}/2\pi =\Delta_{\downarrow}/2\pi = 12$~MHz, $b=\frac{1}{3}$, and $\phi = \frac{2\pi}{3}$ after initially exciting each qubit on the
  30-qubit ladder. }\label{sfig:10}
  \end{figure*}

  \begin{figure*}[t]
    \centering
    \includegraphics[width=0.97\textwidth]{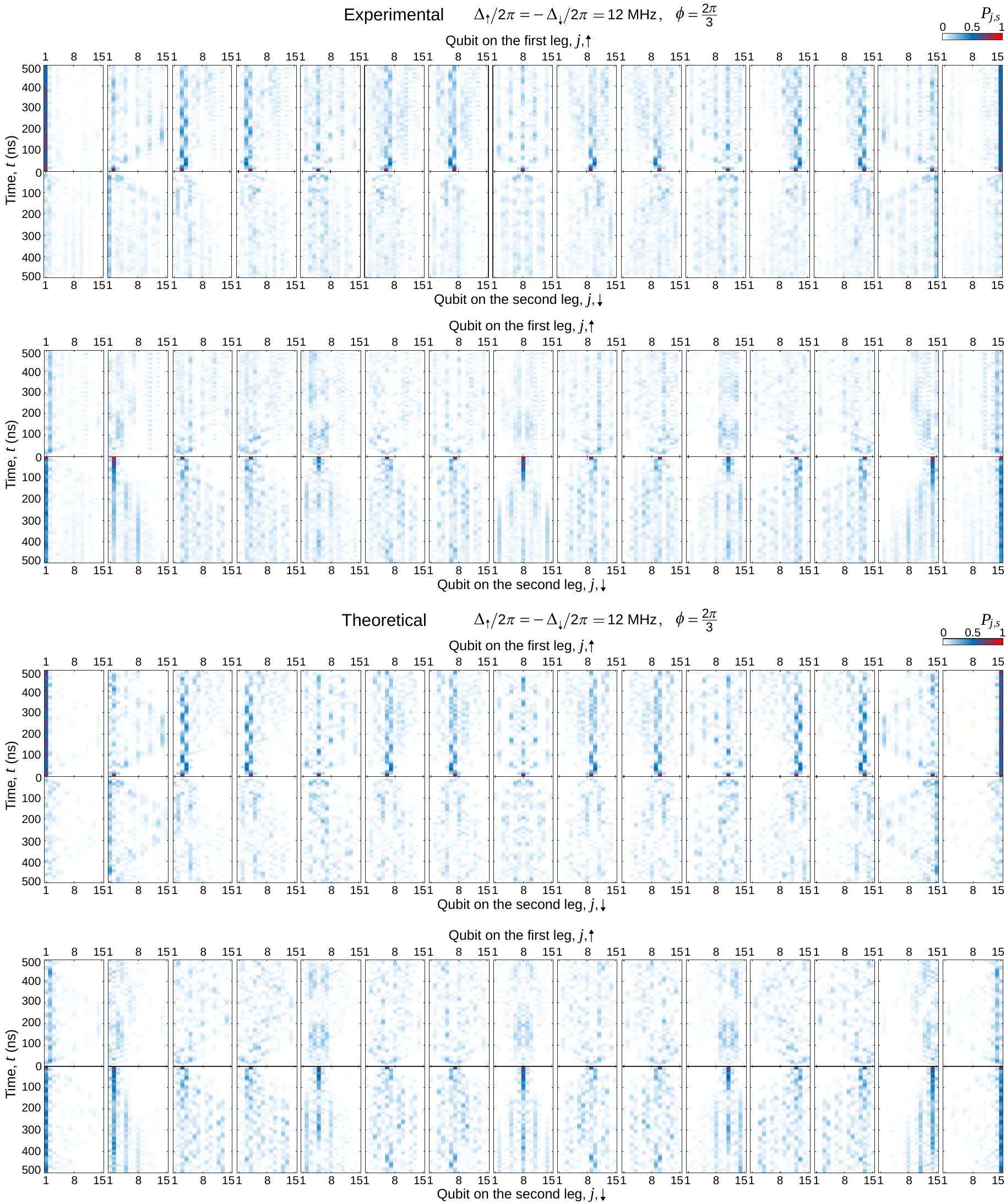}\\
    \caption{Time evolution of the excitation probability
  $P_{j,s}$ with $\Delta_{\uparrow}/2\pi =-\Delta_{\downarrow}/2\pi = 12$~MHz, $b=\frac{1}{3}$, and $\phi = \frac{2\pi}{3}$ after initially exciting each qubit on the
  30-qubit ladder. }\label{sfig:11}
  \end{figure*}

  \begin{figure*}[!ht]
    \centering
    \includegraphics[width=0.97\textwidth]{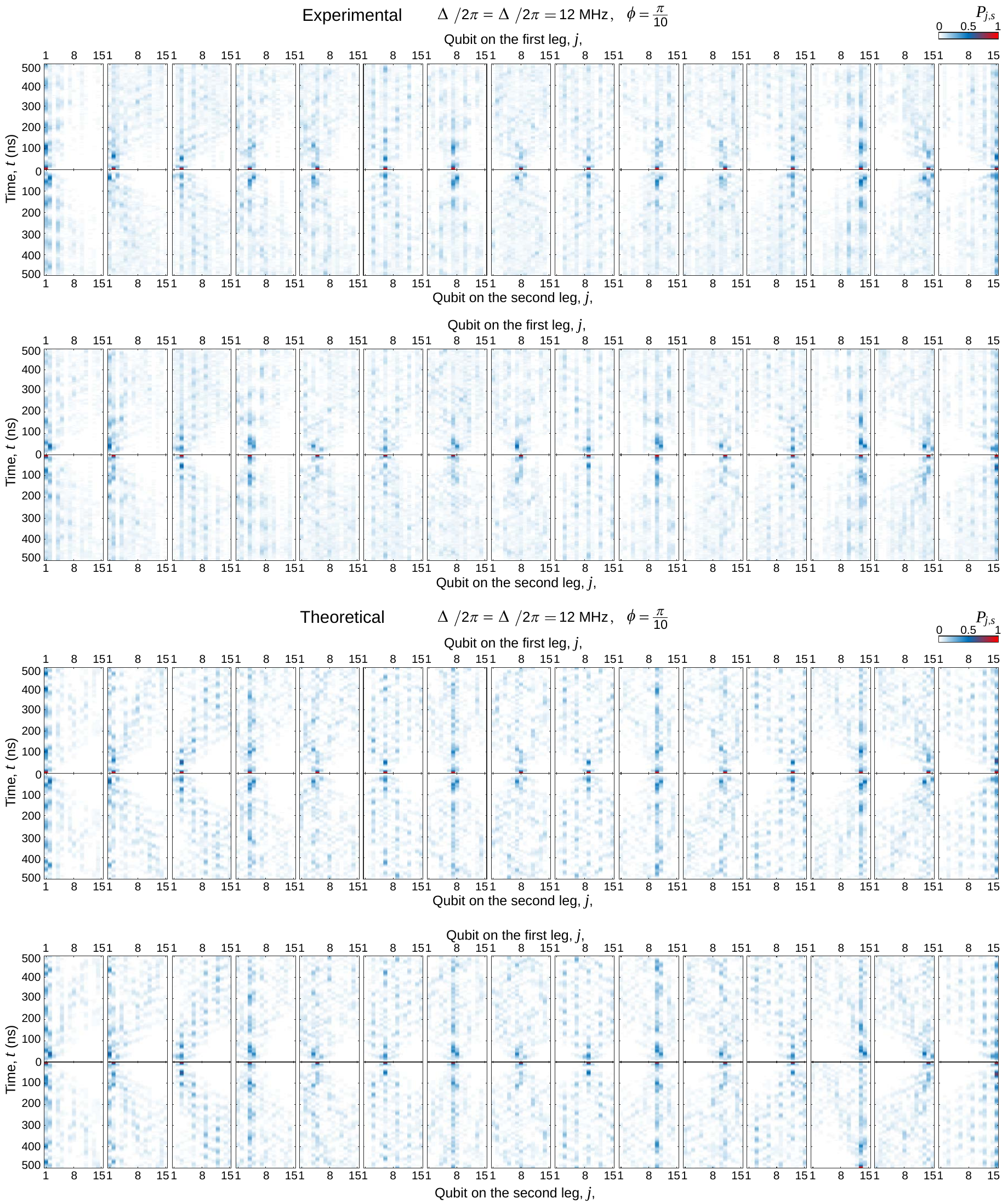}\\
    \caption{ Time evolution of the excitation probability
  $P_{j,s}$ with $\Delta_{\uparrow}/2\pi =\Delta_{\downarrow}/2\pi = 12$~MHz, $b=\frac{1}{3}$, and $\phi = \frac{\pi}{10}$ after initially exciting each qubit on the
  30-qubit ladder. }\label{sfig:12}
  \end{figure*}

  \begin{figure*}[!ht]
    \centering
    \includegraphics[width=0.97\textwidth]{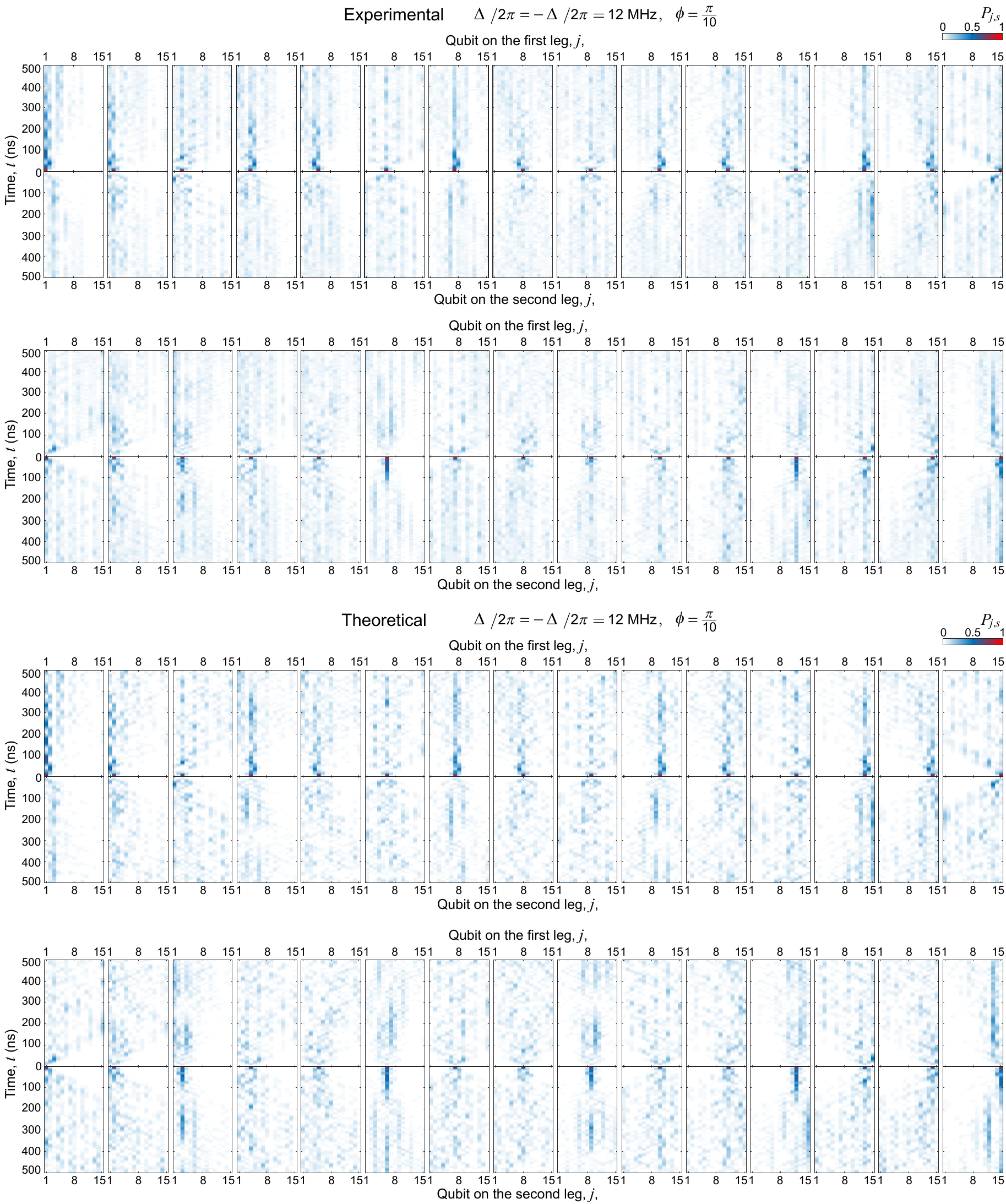}\\
    \caption{ Time evolution of the excitation probability
  $P_{j,s}$ with $\Delta_{\uparrow}/2\pi =-\Delta_{\downarrow}/2\pi = 12$~MHz, $b=\frac{1}{3}$, and $\phi = \frac{\pi}{10}$ after initially exciting each qubit on the
  30-qubit ladder. }\label{sfig:13}
  \end{figure*}

\begin{figure*}[t]
	\begin{center}
		\includegraphics[width=0.8\textwidth,clip=True]{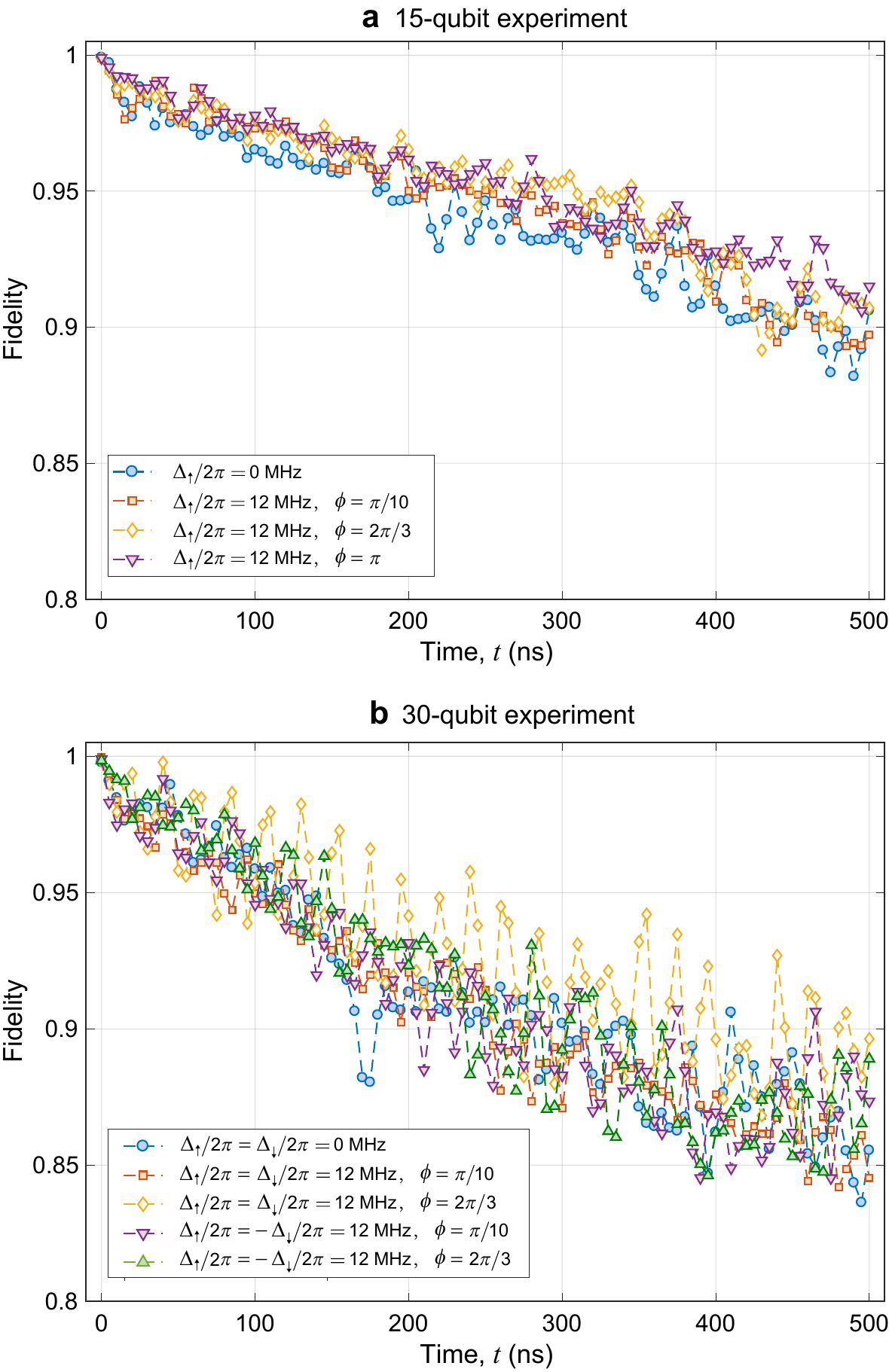}
		\caption{The fidelity for the measured and theoretical probability distributions of the 15-qubit (a) and 30-qubit (b) quantum walks.}
		\label{fig:fidelity}
	\end{center}
\end{figure*}

\begin{figure*}[t]
	\centering
	\includegraphics[width=0.8\textwidth]{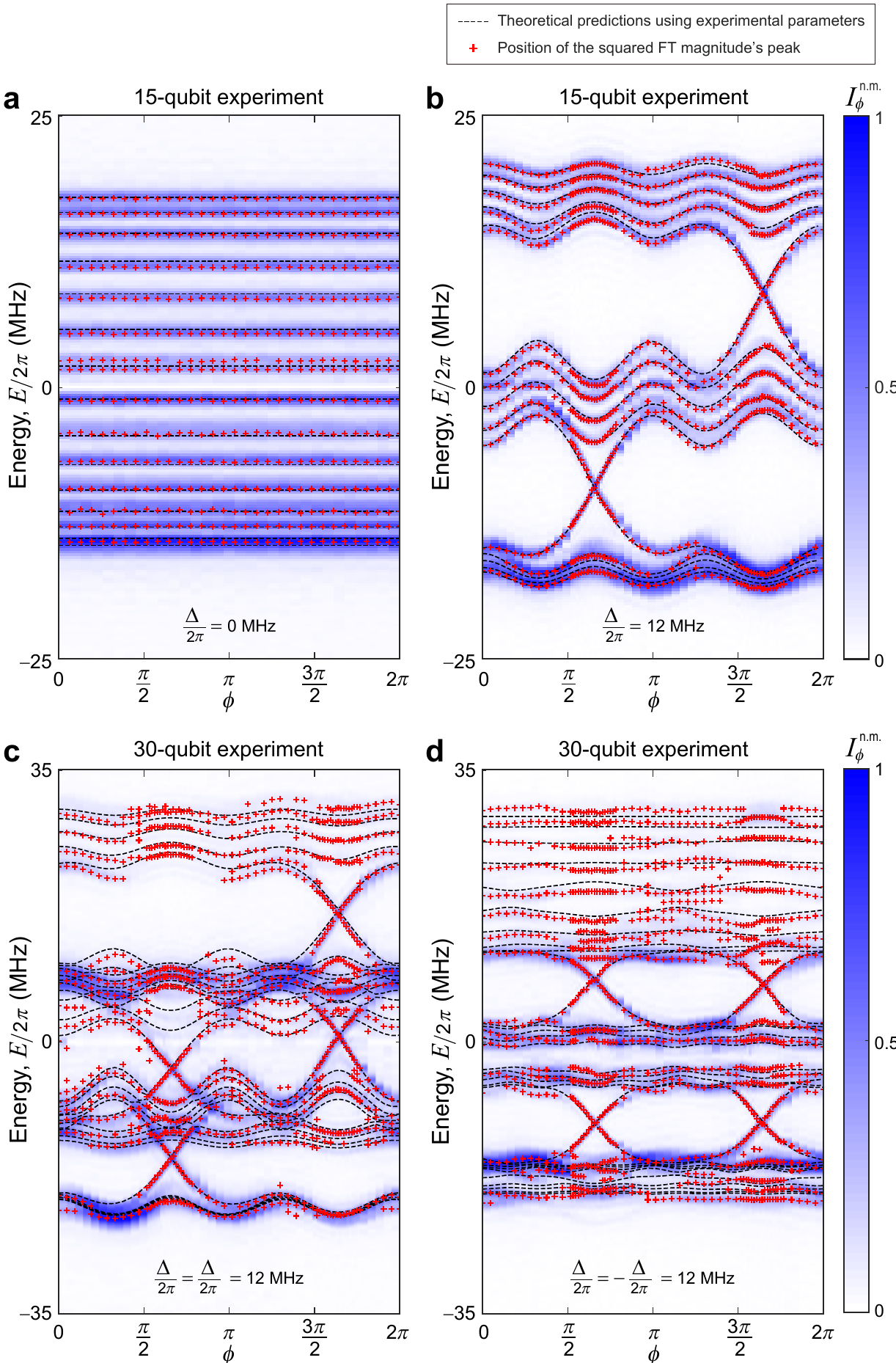}\\
	\caption{Comparison between the experimentally measured band structures by summing
  the FT magnitudes of all selected qubits (normalised data $I_\phi^{\textrm{n.m.}}$) with the theoretical predictions using experimental
  parameters. (a) is for the 15-qubit experiment with $\Delta_{\uparrow}/2\pi=0$~MHz.
  (b) is for the 15-qubit experiment with $\Delta_{\uparrow}/2\pi=12$~MHz and $b=\frac{1}{3}$.
  (c) is for the 30-qubit experiment with $\Delta_{\uparrow}/2\pi=\Delta_{\downarrow}/2\pi=12$~MHz and
  $b=\frac{1}{3}$.
  (d) is for the 30-qubit experiment with $\Delta_{\uparrow}/2\pi=-\Delta_{\downarrow}/2\pi=12$~MHz and
  $b=\frac{1}{3}$.
  }\label{sfig:14}
\end{figure*}

\begin{figure*}[t]
	\centering
	\includegraphics[width=0.97\textwidth]{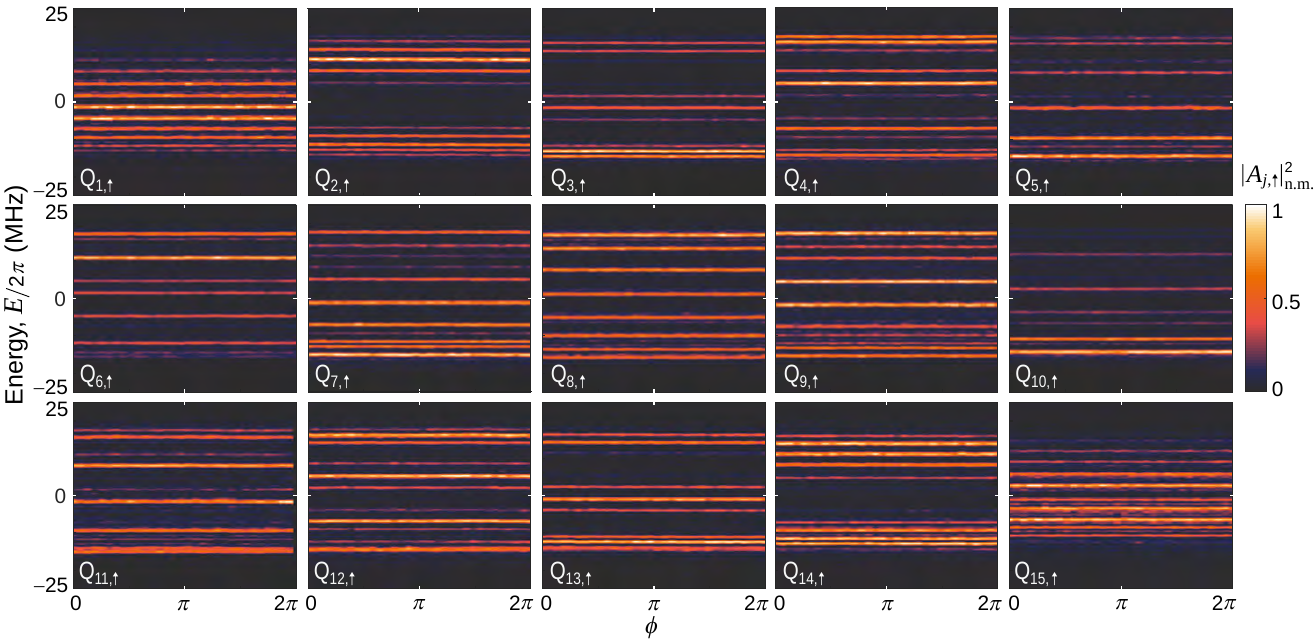}\\
	\caption{ Experimental data of the normalised squared FT magnitudes $|A_{j,\uparrow}|_{\textrm{n.m.}}^2$ when choosing
  Q$_{j,\uparrow}$ as the target qubit in the 15-qubit experiment with $\Delta_{\uparrow}/2\pi=0$~MHz.}\label{sfig:1}
\end{figure*}

\begin{figure*}[t]
	\centering
	\includegraphics[width=0.97\textwidth]{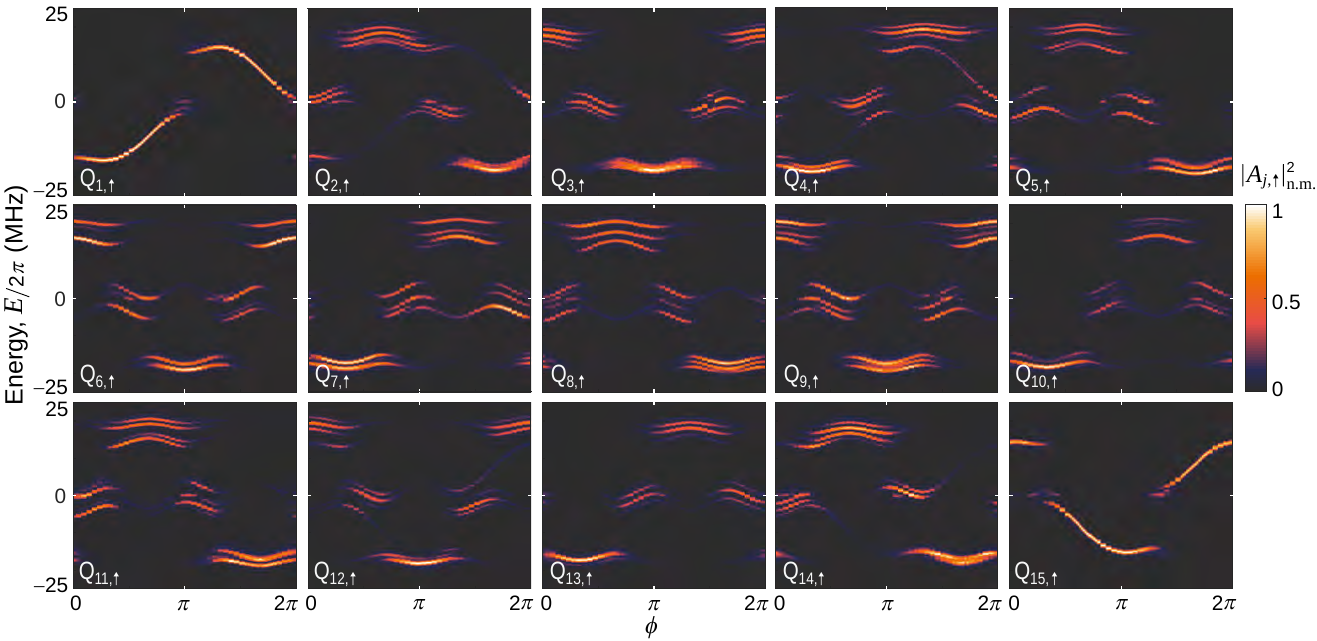}\\
	\caption{ Experimental data of the normalised squared FT magnitudes $|A_{j,\uparrow}|_{\textrm{n.m.}}^2$ when choosing
  Q$_{j,\uparrow}$ as the target qubit in the 15-qubit experiment with $b=\frac{1}{3}$ and $\Delta_{\uparrow}/2\pi=12$~MHz.}\label{sfig:2}
\end{figure*}

\begin{figure*}[t]
	\centering
	\includegraphics[width=0.95\textwidth]{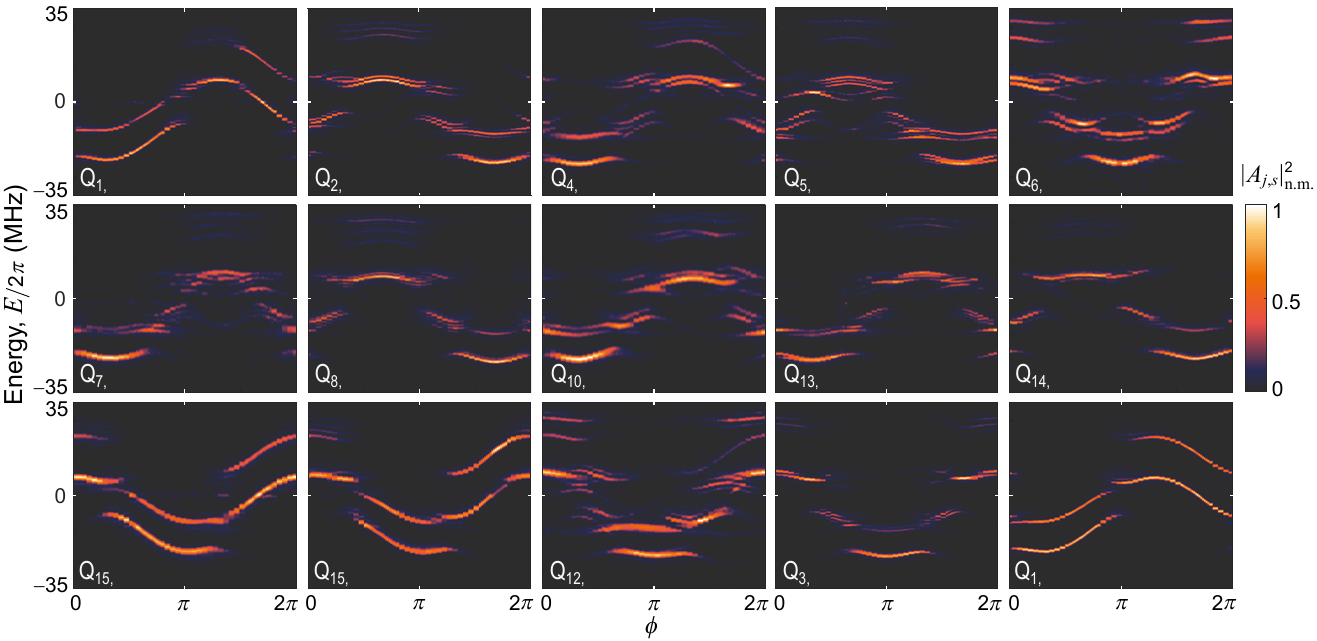}\\
	\caption{ Experimental data of the normalised squared FT magnitudes $|A_{j,s}|_{\textrm{n.m.}}^2$ when choosing
  Q$_{j,s}$ as the target qubit in the 30-qubit experiment with $\Delta_{\uparrow}/2\pi=\Delta_{\downarrow}/2\pi=12$~MHz and
  $b=\frac{1}{3}$.}\label{sfig:3}
\end{figure*}

\begin{figure*}[t]
	\centering
	\includegraphics[width=0.95\textwidth]{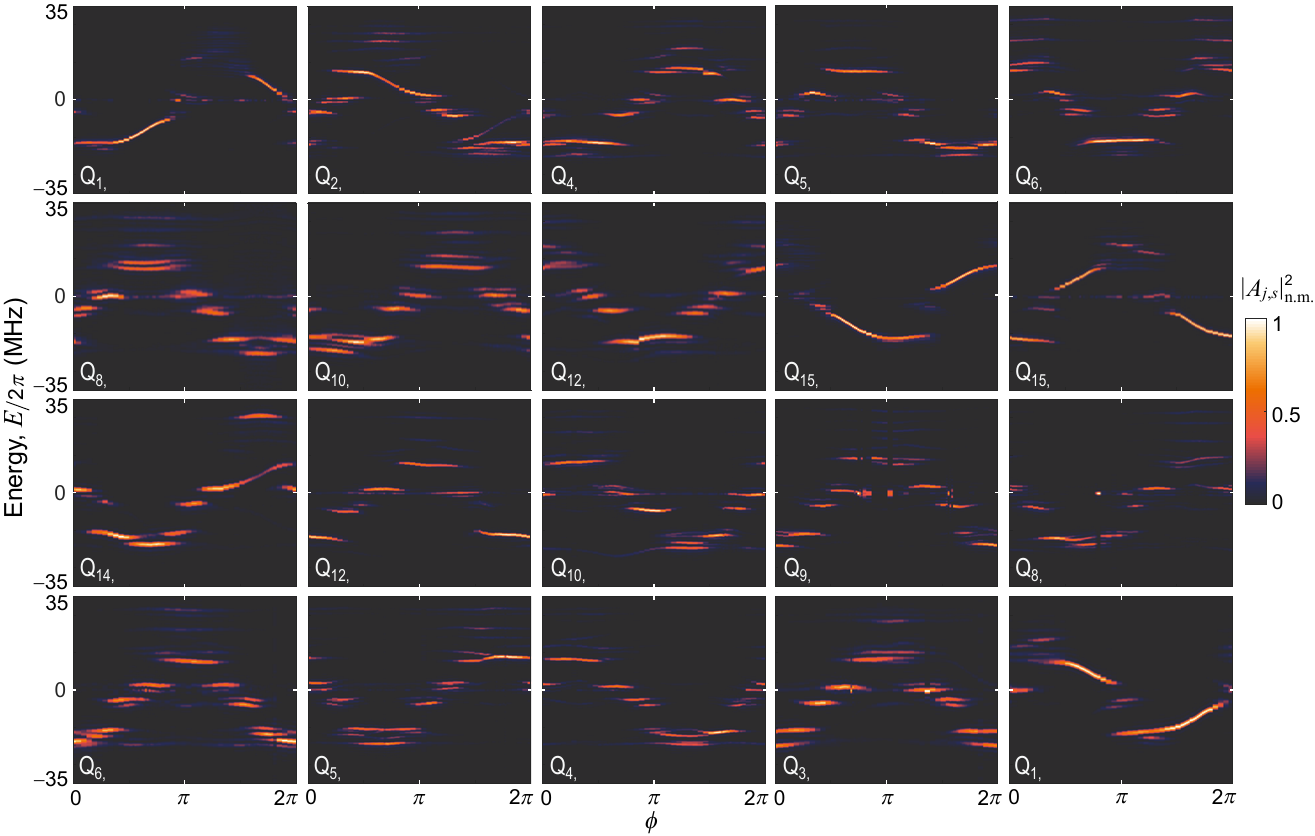}\\
	\caption{Experimental data of the normalised squared FT magnitudes $|A_{j,s}|_{\textrm{n.m.}}^2$ when choosing
  Q$_{j,s}$ as the target qubit in the 30-qubit experiment with $\Delta_{\uparrow}/2\pi=-\Delta_{\downarrow}/2\pi=12$~MHz and
  $b=\frac{1}{3}$. }\label{sfig:4}
\end{figure*}

\begin{figure*}[t]
	\centering
	\includegraphics[width=0.8\textwidth]{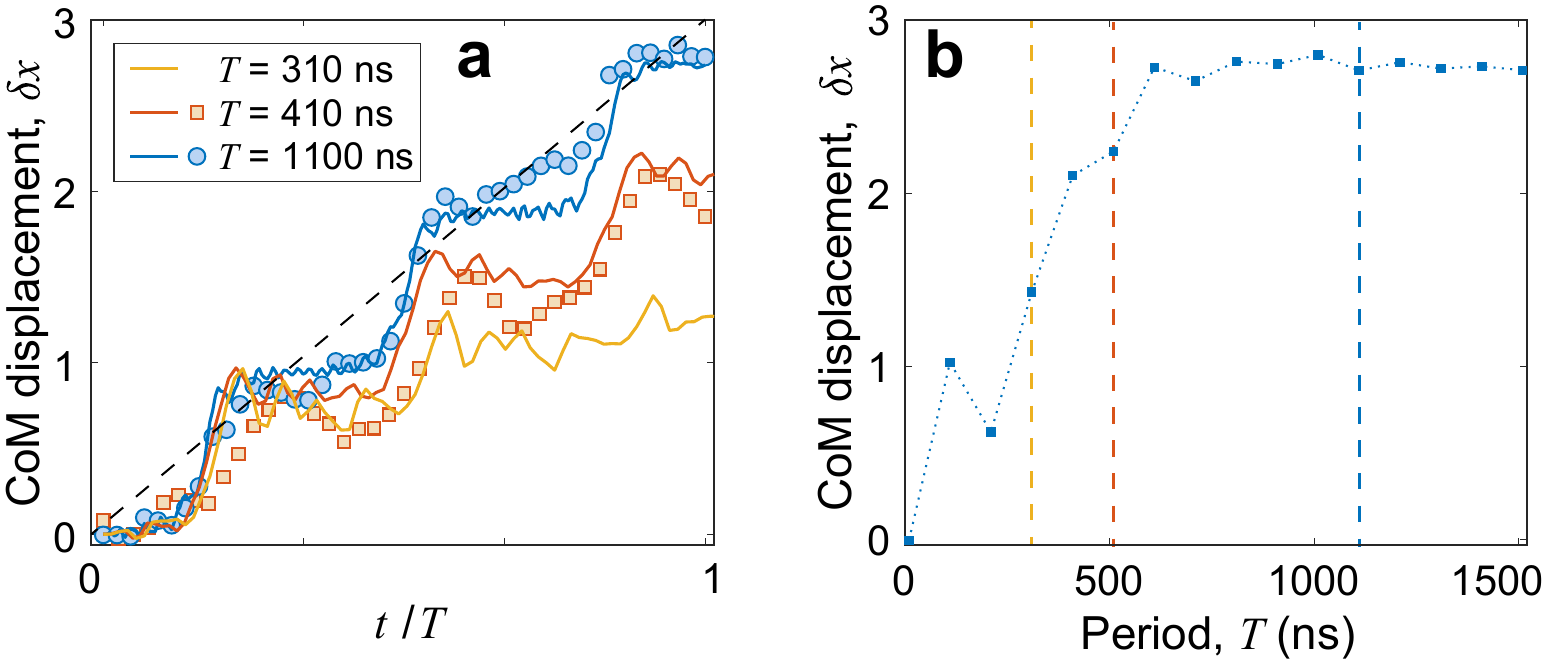}\\
	\caption{{\textbf{a,} Experimental data (circles and squares for different periods 
	$T=410$~ns and 1,100~ns, respectively) for the displacement 
	of the centre of mass (CoM) $\delta x$ versus time $t$ 
	within one pumping cycle, compared with the numerical simulations (solid curves)
	with different periods $T=310$~ns, 410~ns, and 1,100~ns. \textbf{b,} Numerical data for the displacement 
	of the CoM $\delta x$ in one pumping cycle versus different pumping periods $T$.}}\label{fig:rf1}
\end{figure*}

\begin{figure*}[t]
	\centering
	\includegraphics[width=0.95\textwidth]{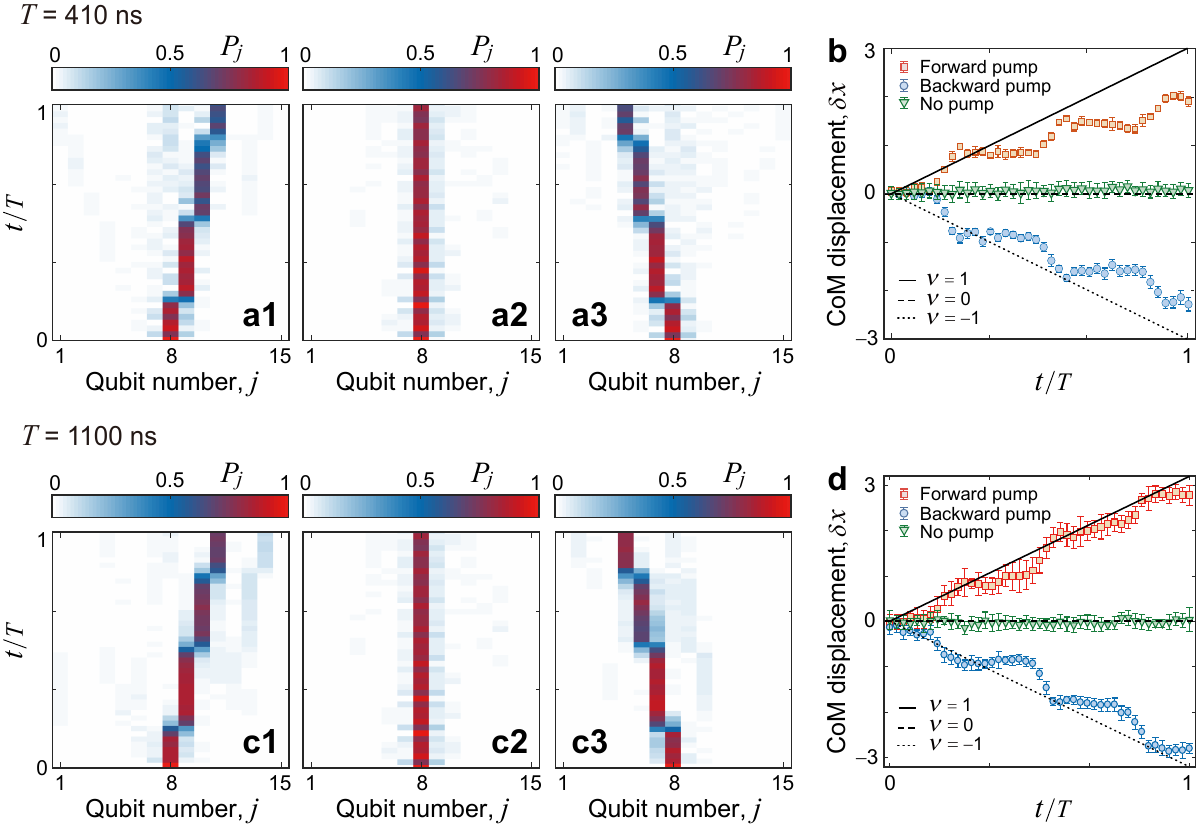}\\
	\caption{{\textbf{a1}--\textbf{a3}, Time evolution of an excitation initially prepared at the central qubit Q$_8$ when it is forward pumped (\textbf{a1}),
	not pumped (\textbf{a2}) and backward pumped (\textbf{a3}), respectively,  with $T=410$~ns and $\Delta/2\pi=36$~MHz for an initial $\phi_0=5\pi/3$.
	\textbf{b}, Displacement of the centre of mass (CoM) $\delta x$ versus time $t$ in one pumping cycle with period $T$ for the cases in (\textbf{a1}--\textbf{a3}). 
  The error bars are 1 SD, calculated from all groups of experimental results.
	\textbf{c1}--\textbf{c3}, Time evolution of an excitation initially prepared at the central qubit Q$_8$ when it is forward pumped (\textbf{c1}),
	not pumped (\textbf{c2}) and backward pumped (\textbf{c3}), respectively,  with $T=1,100$~ns and $\Delta/2\pi=36$~MHz for an initial $\phi_0=5\pi/3$.
	\textbf{d}, Displacement of the CoM $\delta x$ versus time $t$ in one pumping cycle with period $T$ for the cases in (\textbf{c1}--\textbf{c3}). The error bars are 1 SD, calculated from 10 groups of experimental measurement results.
	}}\label{fig:rf2}
\end{figure*}